\pdfoutput=1
\documentclass[11pt,twoside,a4paper,cmspaper,final,collab]{cms-tdr}
\def\svnVersion{01fd1cb-D}\def\svnDate{2021/08/30}\def\cmsCernNoTag{CERN-EP-2020-229}\def\cmsCernDate{\today}\def\cmsMessage{Published in Phys. Rev. D as \href{http://dx.doi.org/10.1103/PhysRevD.104.032009}{\doi{10.1103/PhysRevD.104.032009}.}}
\begin{document}\cmsNoteHeader{SMP-19-006}

\cmsNoteHeader{SMP-19-006}

\renewcommand{\cmsCollabName}{The CMS and TOTEM Collaborations}
\renewcommand{\cmslogo}{\includegraphics[height=2.33cm]{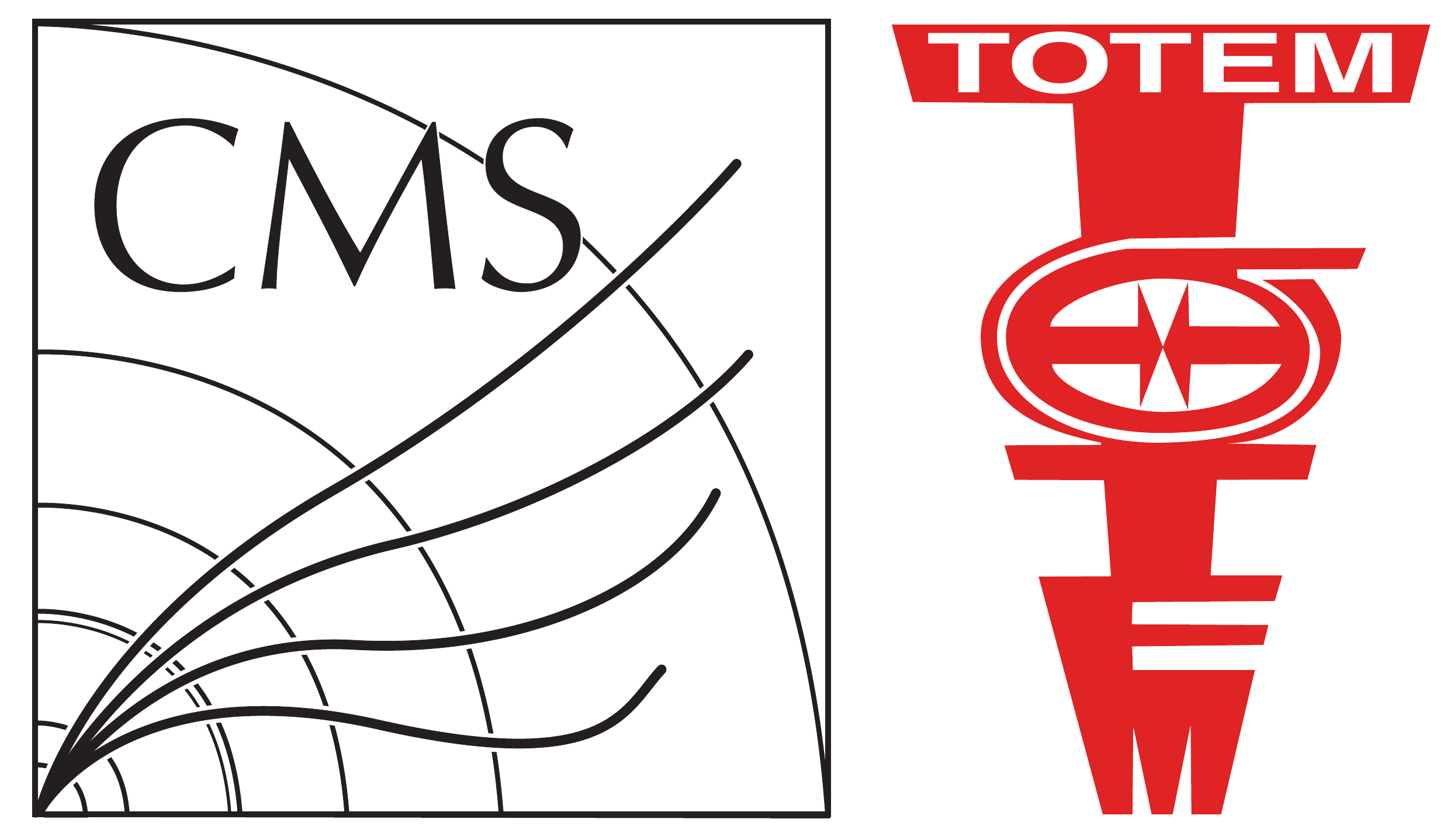}}
\renewcommand{\cmsTag}{CMS-\cmsNUMBER\\&TOTEM-2021-001\\}
\newlength\cmsTabSkip\setlength{\cmsTabSkip}{1ex}
\newlength\cmsFigWidth
\ifthenelse{\boolean{cms@external}}{\setlength{\cmsFigWidth}{\columnwidth}}{\setlength{\cmsFigWidth}{0.65\textwidth}}
\ifthenelse{\boolean{cms@external}}{\newcommand{\cmsLeft}{upper\xspace}}{\newcommand{\cmsLeft}{left\xspace}}
\ifthenelse{\boolean{cms@external}}{\newcommand{\cmsRight}{lower\xspace}}{\newcommand{\cmsRight}{right\xspace}}
\title{Hard color-singlet exchange in dijet events in proton-proton collisions at \texorpdfstring{$\sqrt{s} = 13\TeV$}{sqrt(s) = 13 TeV}}
\date{\today}

\abstract{
Events where the two leading jets are separated by a pseudorapidity interval devoid of particle activity, known as jet-gap-jet events, are studied in proton-proton collisions at $\sqrt{s} = 13\TeV$. The signature is expected from hard color-singlet exchange. Each of the highest transverse momentum (\pt) jets must have $\pt^\text{jet} > 40\GeV$ and pseudorapidity $1.4 < \abs{\eta^\text{jet}} < 4.7$, with $\eta^\text{jet1} \eta^\text{jet2} < 0$, where $\text{jet1}$ and $\text{jet2}$ are the leading and subleading jets in $\pt$, respectively. The analysis is based on data collected by the CMS and TOTEM experiments during a low luminosity, high-$\beta^*$ run at the CERN LHC in 2015, with an integrated luminosity of 0.66\pbinv. Events with a low number of charged particles with $\pt > 0.2 \GeV$ in the interval $\abs{\eta}<1$ between the jets are observed in excess of calculations that assume only color-exchange. The fraction of events produced via color-singlet exchange, $f_\text{CSE}$, is measured as a function of $\pt^\text{jet2}$, the pseudorapidity difference between the two leading jets, and the azimuthal angular separation between the two leading jets. The fraction $f_\text{CSE}$ has values of $0.4$--$1.0$\%. The results are compared with previous measurements and with predictions from perturbative quantum chromodynamics. In addition, the first study of jet-gap-jet events detected in association with an intact proton using a subsample of events with an integrated luminosity of 0.40\pbinv is presented. The intact protons are detected with the Roman pot detectors of the TOTEM experiment. The $f_\text{CSE}$ in this sample is $2.91 \pm 0.70\stat^{+ 1.08}_{- 1.01}\syst$ times larger than that for inclusive dijet production in dijets with similar kinematics.
}

\hypersetup{
pdfauthor={CMS Collaboration},
pdftitle={Hard color-singlet exchange in dijet events in proton-proton collisions at sqrt(s) = 13 TeV},
pdfsubject={CMS, TOTEM},
pdfkeywords={CMS, TOTEM, diffraction, BFKL, pomeron, resummation effects}}

\maketitle

\section{Introduction}\label{sec:intro}
\index{Introduction}

Quantum chromodynamics (QCD) is the established theory of strong interactions and it is especially successful at very short distances, where physical observables can be computed in a perturbative expansion in powers of the strong coupling, \alpS. However, there remain corners of phase space where predictions from perturbative QCD (pQCD) have yet to be confirmed. One such kinematic region is the high-energy limit of strong interactions, which is particularly important for better understanding the initial state in hadronic collisions and for studies of high-energy scattering~\cite{PDG,Akiba:2016ofq}.

In $2{\to}2$ parton scattering, the high-energy limit of QCD is mathematically represented by $\hat{s} \gg -\hat{t} \gg \Lambda^2_\text{QCD}$, where $\hat{s}$ is the square of the partonic center-of-mass energy, $\hat{t}$ is the square of the partonic four-momentum transfer, and $\Lambda_\text{QCD}$ is the energy scale below which QCD becomes strongly coupled. In this limit, some powers of $\alpS$ are multiplied by a large logarithm of $\hat{s}$ in the perturbative expansion, compensating for the smallness of $\alpS \ll 1$ such that $\alpS\ln(\hat{s}/\abs{\hat{t}})\lesssim 1$. Thus, the fixed-order perturbation theory approach is no longer valid. These logarithmically enhanced terms correspond to multiple-parton splittings that are strongly ordered in rapidity. The Balitsky--Fadin--Kuraev--Lipatov (BFKL) evolution equation resums these terms to all orders in $\alpS$ in the perturbative expansion~\cite{Kuraev:1977fs, Balitsky:1978ic,Lipatov:1985uk}, and its solutions are known up to next-to-leading logarithmic (NLL) accuracy~\cite{Fadin:1998py,Ciafaloni:1998gs}. In dijet production, the expected onset of BFKL dynamics is reached in configurations where the two jets are separated by a large rapidity interval. The BFKL radiation pattern is also expected to be important in the study of parton distribution functions (PDFs) of hadrons~\cite{Kuraev:1977fs, Balitsky:1978ic,Lipatov:1985uk}. In this context, the high-energy limit of QCD corresponds to the regime of very small values of the parton momentum fraction $x$ at low momentum transfer. The resummation of $\ln (1/x)$ terms to all orders in $\alpS$ predicts a power-law growth of gluon densities at small $x$.

At the CERN LHC, dedicated studies of BFKL dynamics include measurements of azimuthal angular ($\phi$) decorrelations between jets in forward-backward dijet configurations~\cite{Khachatryan:2016udy} and cross section measurements at large values of the rapidity difference between the jets~\cite{Aad:2011jz,Chatrchyan:2012pb}. Exclusive vector meson production at the LHC~\cite{Aaij:2014iea,Aaij:2015kea,Abelev:2012ba,TheALICE:2014dwa,Adam:2015gsa,Sirunyan:2018sav,Sirunyan:2019nog} can be treated within the BFKL framework, as discussed in Refs.~\cite{Bautista:2016xnp,Garcia:2019tne}. Measurements of inclusive jet or multijet cross sections at different center-of-mass energies show no significant deviations from predictions based on the Dokshitzer--Gribov--Lipatov--Altarelli--Parisi (DGLAP) evolution equations~\cite{dglap1,dglap2,dglap3}, where parton emissions are strongly ordered in transverse momentum ($\pt$), distinct from the BFKL ordering in rapidity, over a large region of phase space~\cite{Chatrchyan:2011qta,Chatrchyan:2012gwa,Chatrchyan:2012pb,Khachatryan:2015luy,
Khachatryan:2016hkr,Khachatryan:2016wdh,Khachatryan:2016mlc,
Sirunyan:2017jnl,Sirunyan:2018ffo,Aad:2010ad,Aad:2011jz,Aad:2011fc,Aad:2013tea,Aad:2014pua,
Aaboud:2017dvo,Aaboud:2018hie}. State-of-the-art global PDF fits highlight the importance of including resummation of small $x$ terms to all orders in $\alpS$ to describe inclusive deep inelastic scattering data collected by the DESY HERA experiments~\cite{Ball:2017otu}. A lesson from these studies is that BFKL dynamical effects associated with multiple parton splittings are very difficult to separate from other effects predicted by higher-order corrections in pQCD. More restrictive final-state studies, where other effects expected from pQCD are suppressed, may provide clearer indications of BFKL dynamics.

\begin{figure*}[tbp]
\centering
\hfill
\includegraphics[width=0.4\textwidth]{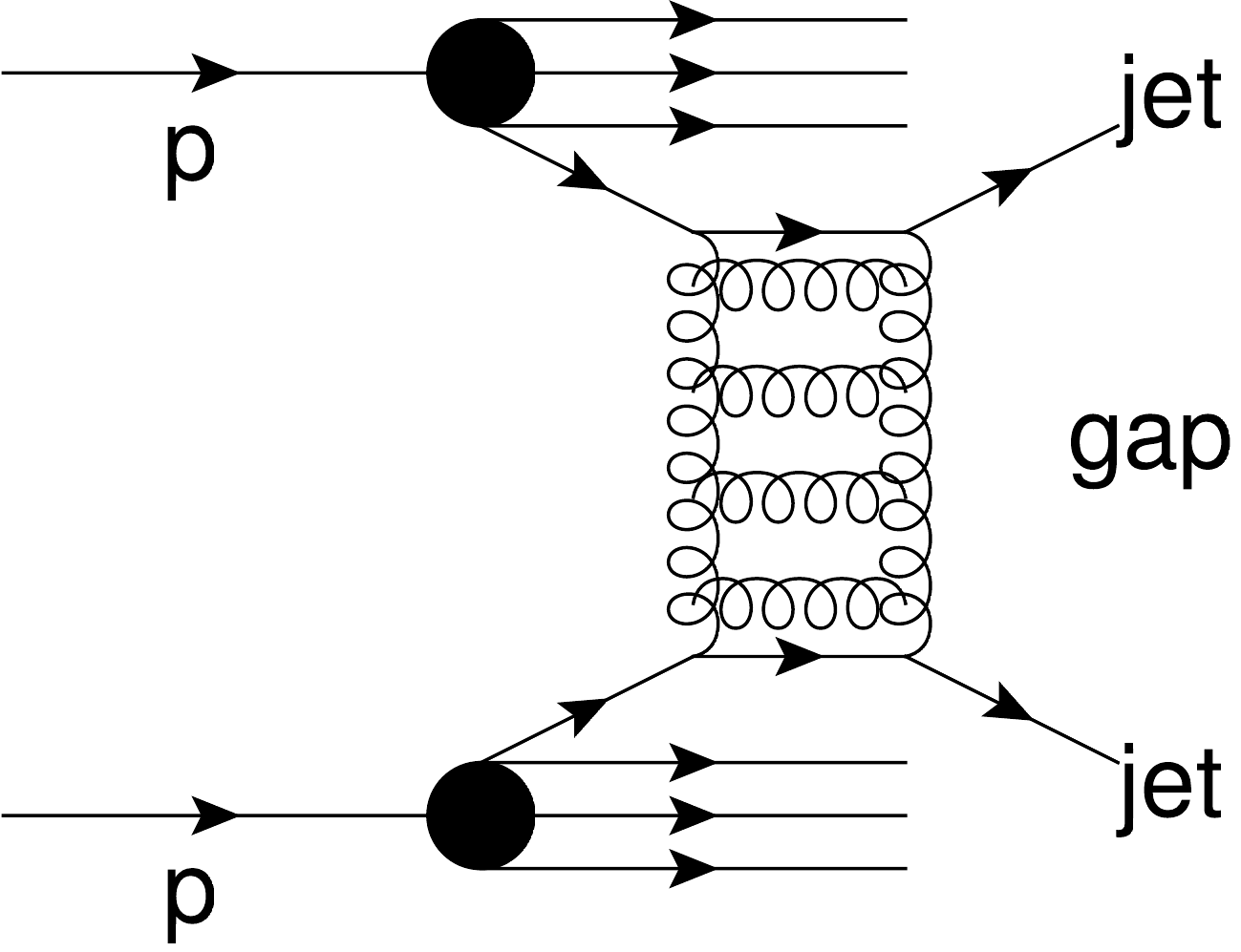}
\hspace{0.1\textwidth}
\includegraphics[width=0.4\textwidth]{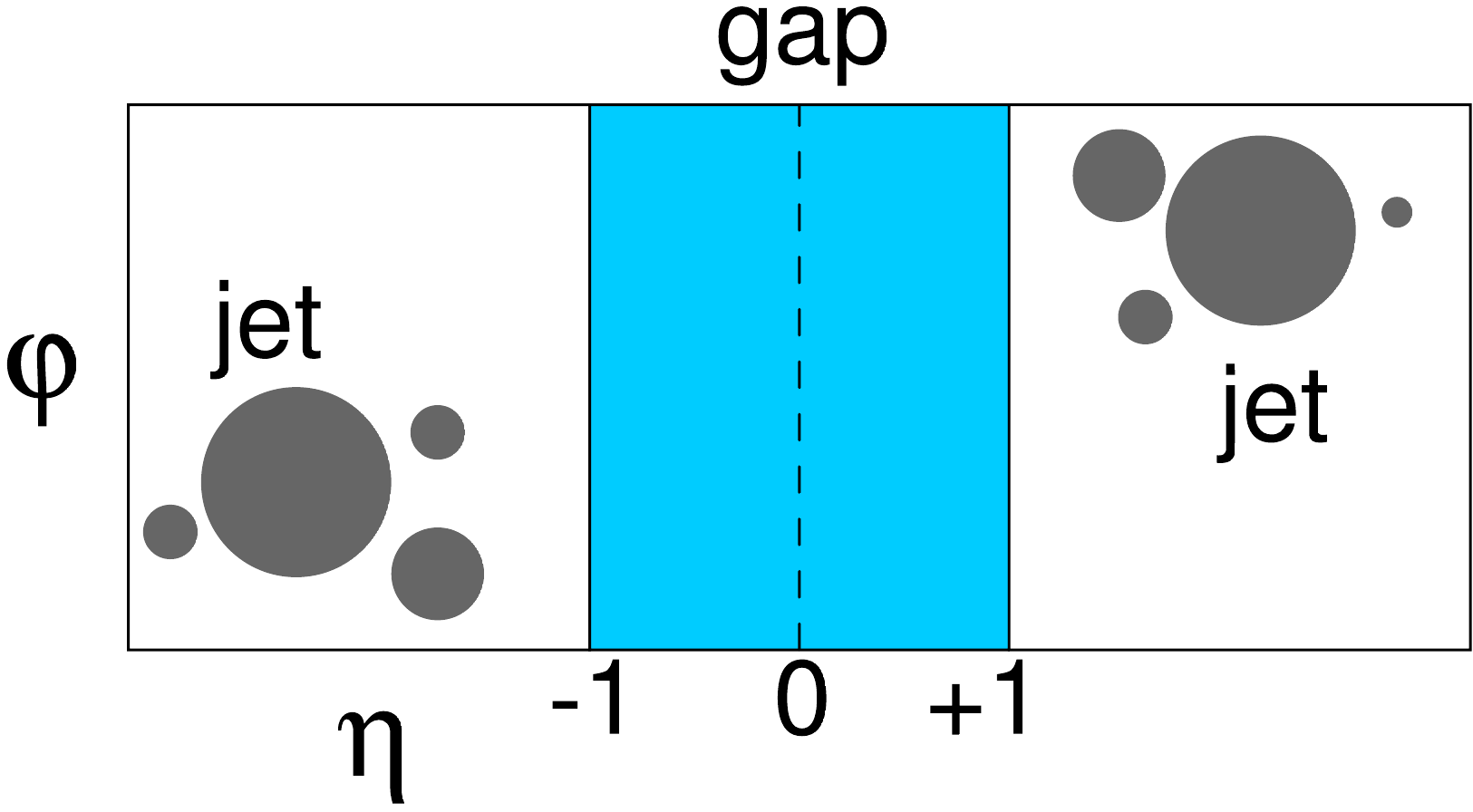}
\hfill
\caption{\label{jetgapjet_feynman} (Left) Schematic diagram of a jet-gap-jet event by hard color-singlet exchange in $\Pp\Pp$ collisions. The lines following the protons represent the proton breakup. (Right) Jet-gap-jet event signature in the $\eta$-$\phi$ plane. The filled circles represent final-state particles. The shaded rectangular area between the jets denotes the interval $\abs{\eta} < 1$ devoid of charged particles.}
\end{figure*}

A study of events is presented in proton-proton ($\Pp\Pp$) collisions with two jets separated by a large pseudorapidity ($\eta$) interval devoid of particle activity. These are known as Mueller--Tang jets~\cite{MUELLER1992123} or jet-gap-jet events. The jet-gap-jet events in this study are observed with the CMS detector. Previous studies of jet-gap-jet events have been carried out by the H1 and ZEUS Collaborations in dijet photoproduction in electron-proton collisions at the DESY HERA~\cite{h1,zeus}, by the CDF and D0 Collaborations in $\Pp\PAp$ collisions at center-of-mass energies $\sqrt{s} = 0.63$ and $1.8\TeV$ at the Fermilab Tevatron~\cite{d01,d02,d03,cdf1,cdf2,cdf3}, and by CMS at $7\TeV$ in $\Pp\Pp$ collisions at the CERN LHC~\cite{jgjCMS}. The pseudorapidity gap is indicative of an underlying $t$-channel hard color-singlet exchange~\cite{Bjorken, barone, donnachie, forshaw}. In the BFKL framework, hard color-singlet exchange is described by $t$-channel two-gluon ladder exchange between the interacting partons, as shown in Fig.~\ref{jetgapjet_feynman}, where the color charge carried by the exchanged gluons cancel, leading to a suppression of particle production between the final-state jets. This is known as perturbative pomeron exchange~\cite{Kuraev:1977fs, Balitsky:1978ic,Lipatov:1985uk}. Color-singlet exchange can occur in quark-quark, quark-gluon, and gluon-gluon scattering. Of these, gluon-gluon scattering is expected to be substantially favored as a result of the larger color charge of gluons~\cite{barone,donnachie,forshaw}. In contrast, in most collisions that lead to dijet production, the net color charge exchange between partons results in final-state particle production over wide intervals of rapidity between the jets. These color-exchange dijet events are referred to in this paper as ``background'' events. Dynamical effects predicted by the DGLAP evolution equations are largely suppressed in events with pseudorapidity gaps, since the predicted dijet production rate is strongly reduced by way of a Sudakov form factor~\cite{Bjorken,barone,donnachie,forshaw}. This factor, which accounts for the probability of having no additional parton emissions between the hard partons, is not necessary for BFKL pomeron exchange~\cite{MUELLER1992123}. The ratio of jet-gap-jet yields to inclusive dijet yields is sensitive to dynamical effects predicted by the BFKL evolution equations, as first suggested in Ref.~\cite{MUELLER1992123} and further studied in Refs.~\cite{Cox:1999dw, csp, Chevallier:2009cu, Kepka:2010hu,cspLHC}.

\begin{figure*}[tbp]
\centering
\hfill
\includegraphics[width=0.4\textwidth]{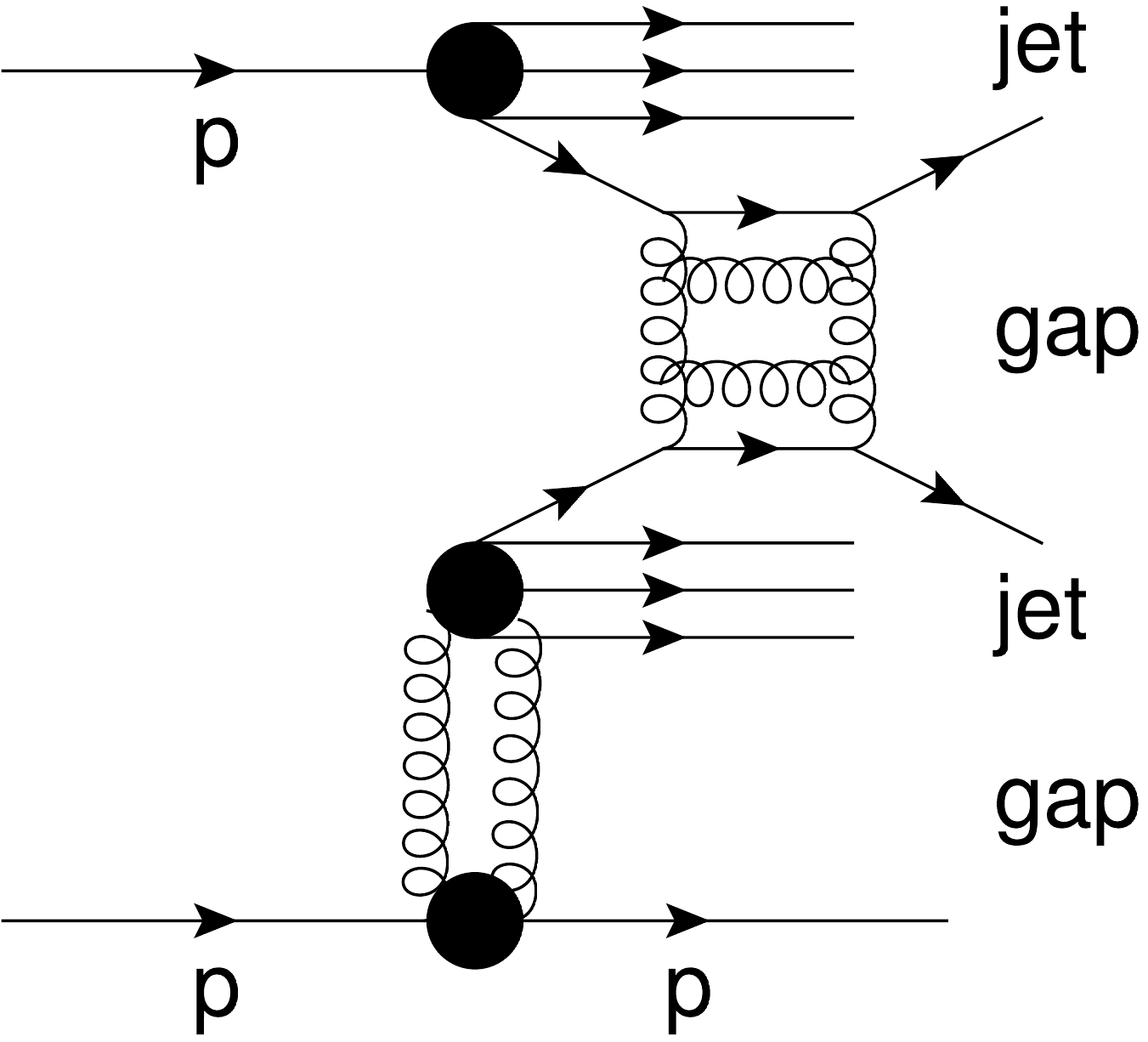}
\hspace{0.1\textwidth}
\includegraphics[width=0.41\textwidth]{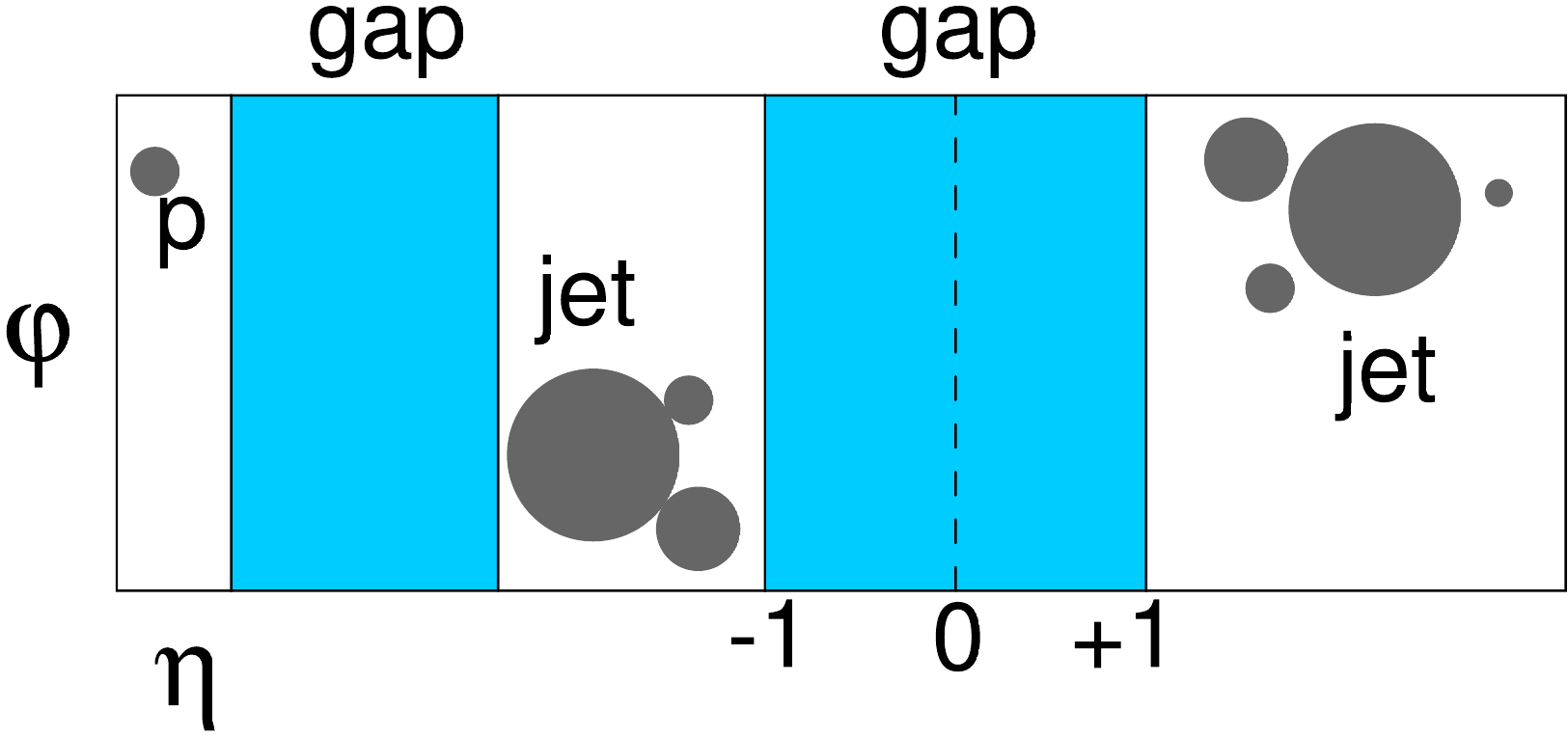}
\hfill
\caption{\label{jetgapjet_sd} (Left) Schematic diagram of a jet-gap-jet event by hard color-singlet exchange with an intact proton in $\Pp\Pp$ collisions. The jet-gap-jet is reconstructed in the CMS detector, while the intact proton is detected with one of the forward proton spectrometers of the TOTEM experiment. (Right) Proton-gap-jet-gap-jet event signature in the $\eta$-$\phi$ plane. The filled circles represent final-state particles. The shaded rectangular areas denote the central gap region $\abs{\eta} < 1$ devoid of charged particles and the forward gap that is inferred from the forward proton detection.}
\end{figure*}

The presence of soft rescattering effects between partons and the proton remnants modify the visible cross section of jet-gap-jet events. These soft interactions can induce the production of particles in the $\eta$ interval that would otherwise be devoid of particles. This results in a reduction of the number of events identified as having a jet-gap-jet signature. This reduction is parametrized using a multiplicative factor known as the rapidity gap survival probability, $\abs{\mathcal{S}}^2$. The survival probability is a process-dependent, nonperturbative quantity~\cite{Bjorken, bjorken_survival,survival_levin,khoze_survival_2013,gotsman_survival_2015,Khoze:2017sdd} that is expected to have values of the order of $\abs{\mathcal{S}}^2 = 1$--$10\%$ at LHC energies. This factor is often assumed to be largely independent of the dijet event kinematics~\cite{Bjorken}, although some nonperturbative models, such as the soft color interactions (SCI) model~\cite{csp,cspLHC}, suggest that this is not always the case. In particular, multiple-parton interactions (MPI) can further reduce the survival probability in dijet events with a central gap, as discussed in Refs.~\cite{csp,cspLHC,Babiarz:2017jxc}.

Soft rescattering effects can be suppressed in processes where one or both of the colliding protons remain intact after the interaction, such as in single- or central-diffractive dijet processes or in dijet photoproduction. These can be used to better separate events with a central gap between the jets, as discussed in Ref.~\cite{PhysRevD.87.034010}. Hence, parallel to the study of jet-gap-jet events in inclusive dijet production, a study of jet-gap-jet events with an intact proton, as shown in Fig.~\ref{jetgapjet_sd}, is also presented. Although no forward rapidity gap is required in the analysis, these events are referred to as ``proton-gap-jet-gap-jet'' throughout the paper, where the forward rapidity gap signature is inferred from the detection of the intact proton. This part of the analysis uses a subset of dijet events that, in addition, have intact protons detected with the forward proton spectrometers of the TOTEM experiment~\cite{totem1}. This diffractive event topology has not been previously measured.

The present study is based on low instantaneous luminosity data collected in $\Pp\Pp$ collisions at $\sqrt{s} = 13\TeV$ by the CMS and TOTEM experiments at the CERN LHC. These data were recorded with special LHC optics settings, $\beta^*=90$\unit{m}, where $\beta^*$ is the betatron amplitude function at the interaction point~\cite{LHC_TDR}. Data were recorded by CMS with an integrated luminosity of 0.66\pbinv; a subset of the data with 0.40\pbinv was collected jointly with the TOTEM experiment. The present analysis uses a similar event selection and central gap definition as the previous measurement by CMS at $7\TeV$~\cite{jgjCMS}. Each of the two highest $\pt$ jets must have $\pt^\text{jet}>40\GeV$ and $1.4<\abs{\eta^\text{jet}}<4.7$, and they must be in opposite hemispheres of the detector $\eta^\text{jet1}  \eta^\text{jet2} < 0$, where $\text{jet1}$ and $\text{jet2}$ denote the leading and subleading jets in $\pt$, respectively. The charged particle multiplicity ($N_\text{tracks}$) in the interval $\abs{\eta}<1$ between the two leading jets, where each charged particle must have $\pt>200\MeV$, is used to isolate color-singlet exchange dijet events from color-exchange dijet events. Jet-gap-jet events due to color-singlet exchange are characterized by a sharp excess at the lowest $N_\text{tracks}$ values above the expected contribution of color-exchange dijet events. The increase in $\sqrt{s}$ to $13\TeV$ provides improved conditions to study the hard color-singlet exchange process in an unexplored region of phase space. The increased sample size relative to the previous analysis at $7\TeV$ allows finer binning in the kinematic variables of interest and an improved precision in the determination of the fraction of dijet events produced via hard color-singlet exchange. Furthermore, the analysis based on CMS and TOTEM data provides a first investigation of dijet events with a central gap and an intact proton. This analysis can elucidate the role of soft parton exchanges in the creation and destruction mechanisms of pseudorapidity gaps in strong interactions~\cite{PhysRevD.87.034010}. The intact protons in the analysis have a fractional momentum loss ($\xi$) of up to $20$\%, with values of the square of the four-momentum transfer at the proton vertex ($t$) in the range between $-4$ and $-0.025\GeV^2$.

The paper is organized as follows. The CMS and TOTEM detectors are introduced in Section~\ref{sec:cms}. The data sample used in the analysis is described in Section~\ref{sec:data_sample}. The event selection requirements are presented in Section~\ref{sec:event_selection}. The central pseudorapidity gap and observable definitions are discussed in Sections~\ref{sec:jetgapjet} and \ref{sec:observable}, respectively. Section ~\ref{sec:background} gives a description of the background treatment used in the analysis. The systematic uncertainties are detailed in Section~\ref{sec:systematic_unc}. The results of the paper are shown in Section~\ref{sec:results}. A summary of the paper is found in Section~\ref{sec:summary}.

\section{The CMS and TOTEM detectors}\label{sec:cms}

The central feature of the CMS apparatus is a superconducting solenoid of 6\unit{m} internal diameter, providing a magnetic field of 3.8\unit{T}. Within the solenoid volume are a silicon pixel and strip tracker, a lead tungstate crystal electromagnetic calorimeter (ECAL), and a brass and scintillator hadron calorimeter (HCAL), each composed of a barrel and two endcap sections. Forward calorimeters extend the $\eta$ coverage provided by the barrel and endcap detectors. Muons are detected in gas-ionization chambers embedded in the steel flux-return yoke outside the solenoid.

The silicon tracker measures charged particles within the range $\abs{\eta} < 2.5$. It consists of 1440 silicon pixel and 15\,148 silicon strip detector modules. For nonisolated particles with $1 < \pt < 10\GeV$ and $\abs{\eta} < 1.4$, the track resolutions are typically 1.5\% in \pt and 25--90 (45--150)\mum in the transverse (longitudinal) impact parameter \cite{TRK-11-001}.

The particle-flow (PF) algorithm~\cite{PFnew} aims to reconstruct and identify each individual particle (physics-object) in an event, with an optimized combination of information from the various elements of the CMS detector. The energy of photons is obtained from the ECAL measurement. The energy of electrons is determined from a combination of the electron momentum at the primary interaction vertex as determined by the tracker, the energy of the corresponding ECAL cluster, and the energy sum of all bremsstrahlung photons spatially compatible with originating from the electron track. The energy of muons is obtained from the curvature of the corresponding track. The energy of charged hadrons is determined from a combination of their momentum measured in the tracker and the matching ECAL and HCAL energy deposits, corrected for the response function of the calorimeters to hadronic showers. Finally, the energy of neutral hadrons is obtained from the corresponding corrected ECAL and HCAL energies.

Tracks are reconstructed with the standard iterative algorithm of CMS~\cite{TRK-11-001}. To reduce the misidentification rate, tracks are required to pass standard CMS quality criteria, referred to as high-purity criteria. High-purity tracks satisfy requirements on the number of hits and the $\chi^2$ of the track-fit. The requirements are functions of the charged particle track \pt and $\eta$, as well as the number of layers with a hit. A more detailed discussion of the combinatorial track finding algorithm and the definition of high-purity tracks is reported in Ref.~\cite{TRK-11-001}. The reconstruction efficiency for high-purity tracks is about 75\% with $\pt > 200\MeV$. The candidate vertex with the largest value of summed physics-object $\pt^2$ is taken to be the primary $\Pp\Pp$ interaction vertex. In the vertex fit, each track is assigned a weight between $0$ and $1$, which reflects the likelihood that it genuinely belongs to the vertex. The number of degrees of freedom in the fit is strongly correlated with the number of tracks arising from the interaction region, as described in Ref.~\cite{TRK-11-001}.

\begin{figure*}[tbp]
\centering
\includegraphics[scale=0.95]{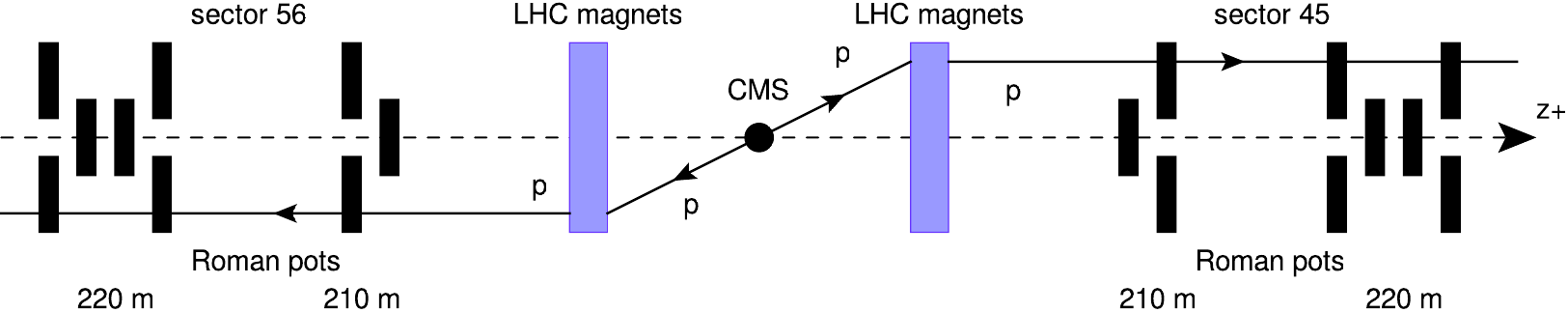}
\caption{\label{fig:cms_totem_config} Profile schematic of the CMS-TOTEM detector configuration during the 2015 run. The horizontal dashed line represents the beamline. The CMS detector is denoted by the filled circle in the center. The intact proton(s) are transported via the accelerator magnetic fields (violet light rectangles), eventually passing through the silicon detectors housed in the Roman pots (black dark rectangles) of the TOTEM experiment. Sectors 45 and 56 are located in the positive and negative $\eta$ regions in the CMS coordinate system, respectively.}
\end{figure*}

The jets are clustered using the infrared- and collinear-safe anti-\kt algorithm \cite{Cacciari:2008gp,Cacciari:2011ma}, with a distance parameter of $R = 0.4$. The clustering is performed with the \FASTJET package~\cite{Cacciari:2011ma}. The key feature of the anti-\kt algorithm is the resilience of the jet boundary with respect to soft radiation. This leads to cone-shaped hard jets. The jet momentum is determined as the vector sum of all particle momenta in the jet. The simulations show the CMS detector response is within $5$--$10$\% of the true hadron-level momentum over a wide range of the jet \pt and $\eta$. Jet energy corrections are derived from simulation to bring, on average, the measured jet energies to the known energies at the generator level~\cite{Khachatryan:2016kdb}. In situ measurements of the momentum balance in dijet, $\text{photon}$+$\text{jet}$, $\PZ$+$\text{jet}$, and multijet events are used to correct any residual differences in the jet energy scale in data and simulation~\cite{Khachatryan:2016kdb}. The jet energy resolution typically amounts to 15\% at 10\GeV, 8\% at 100\GeV, and 4\% at 1\TeV. A more detailed description of the CMS detector, together with a definition of the coordinate system used and the relevant kinematic variables, is described in Ref.~\cite{Chatrchyan:2008zzk}.

The proton spectrometer of the TOTEM experiment consists of two sets of telescopes, known as Roman pot (RP) stations~\cite{totem1} that are located close to the beamline. The arms are referred to as sectors 45 and 56 for positive and negative $\eta$, respectively. An RP that contains silicon strip detectors can approach the LHC beam to a distance of a few millimeters without affecting the LHC operation~\cite{totem1}. The RPs are used to detect protons deflected at scattering angles of only a few microradians relative to the beam. During the 2015 special run, there were two RP stations operating in each sector located at $\pm 210$\unit{m} and $\pm 220$\unit{m} relative to the interaction point. The configuration during 2015 is depicted in Fig.~\ref{fig:cms_totem_config}. The station at $210$\unit{m} has one unit of RPs, while the station at $220$\unit{m} has two units of RPs. Each unit has three RPs: one located above (``top''), one below (``bottom''), and one to one side (``horizontal'') of the LHC beam~\cite{totem1}. Before being detected, the trajectories of protons that have lost a small amount of their original momentum slightly deviate from the beam trajectory, with the deviation dependent on the momentum of the proton. The intact proton kinematics are reconstructed after modeling the transport of the protons from the interaction point to the RP location~\cite{totem1, optics}. With the $\beta^* = 90$\unit{m} conditions, small horizontal displacements of the forward proton tracks at the RPs are directly proportional to $\xi$. The detection of the forward protons also enables the reconstruction of $t$, which is related to the horizontal and vertical scattering angles of the proton track at the RPs~\cite{totem2,Niewiadomski:2008zz}. The resolution in $\xi$ is $0.008$ for $\xi \approx 0$ and $0.002$ for $\xi = 0.2$~\cite{totem2}. The RPs are aligned following the standard techniques developed by the TOTEM Collaboration~\cite{totem2}. The TOTEM detector is described in Refs.~\cite{totem1, totem2}.

\section{Data sample and trigger selection}\label{sec:data_sample}

The $\Pp\Pp$ collision data used in this analysis were collected in a combined special run by the CMS and TOTEM experiments in 2015 at $\sqrt{s} = 13\TeV$, when the LHC operated in a mode with low probability of overlapping $\Pp\Pp$ interactions in the same bunch crossing (pileup). With $\beta^* = 90$\unit{m} optics at the interaction point of CMS, there were about $0.05$--$0.10$ pileup interactions per event. Events were selected by trigger signals delivered simultaneously to the CMS and TOTEM detectors. The CMS orbit-counter reset signal, delivered to the TOTEM electronics at the start of the run, assures the time synchronization of the two experiments. The samples were combined offline by matching bunch crossing and orbit numbers, as in the previous CMS and TOTEM combined run at $\sqrt{s} = 8\TeV$~\cite{Sirunyan:2020ifc}. Since CMS and TOTEM collected data simultaneously for a fraction of the special run, the integrated luminosity for the CMS-TOTEM sample corresponds to 0.40\pbinv. The data were collected with an unprescaled inclusive dijet trigger. This trigger requires at least two leading jets ($\text{jet1}, \text{jet2}$), both with $\pt>32\GeV$ with $\abs{\eta}<5$~\cite{Khachatryan:2016bia}. The trigger is about $85$\% efficient for $\pt^\text{jet2} = 40\GeV$, and is fully efficient at $\pt^\text{jet2} > 55\GeV$, as measured with dijet events in a zero-bias sample collected using a random trigger in the presence of nonempty bunch crossings. Trigger efficiency effects largely cancel in the ratio of yields of events with a central gap, $f_\text{CSE}$, the main observable measured in this analysis, which is described in Section~\ref{sec:observable}. Thus, no efficiency correction is applied in the analysis. A subset of events of the zero-bias sample that contains forward proton information collected by the TOTEM experiment is used for systematic checks in the analysis.

\section{Event selection}\label{sec:event_selection}

\subsection{Dijet event selection}\label{subsec:dijet_selection}

The following selection requirements are used for the study of jet-gap-jet events within inclusive dijet events as well as for the analysis of jet-gap-jet events with an intact proton:

\begin{itemize}

\item Each of the two leading jets is required to have $\pt^\text{jet}> 40\GeV$. This selection maximizes the number of dijet events considered in the analysis, while ensuring high dijet reconstruction efficiency. The phase space explored in the present analysis is similar to that studied in the previous CMS measurement at $7\TeV$~\cite{jgjCMS}. There are no requirements on additional jets that may be produced in the collision.

\item The two leading jets are measured in opposite hemispheres of the CMS detector, $\eta^\text{jet1}   \eta^\text{jet2} < 0$, and must have $1.4 < \abs{\eta^\text{jet}} < 4.7$. This selection favors the phase space region for production of jet-gap-jet events.
Jets are reconstructed with the anti-\kt algorithm distance parameter $R = 0.4$ and the adopted jet $\eta$ range thus locates the jets at least one unit of $R$ away from the $\abs{\eta}<1$ region used to extract the multiplicity of charged particles.

\item The number of reconstructed primary vertices in the event is required to be at most one. This requirement is used to reject residual pileup interactions. For this analysis, a primary vertex is kept if it has at least two degrees of freedom as defined in Ref.~\cite{TRK-11-001}. Keeping events with no primary vertex retains forward-backward dijet configurations that have too few tracks to establish a primary vertex, as is likely for the jet-gap-jet topology.

\item The primary vertex, if present, is required to be located within a longitudinal distance of $24$\unit{cm} of the nominal interaction point of CMS.

\end{itemize}

There were 362\,915 dijet events satisfying these selection requirements.

\subsection{Intact proton selection}\label{subsec:forward_selection}

For the study of jet-gap-jet events with an intact proton (proton-gap-jet-gap-jet), in addition to the dijet event selection described in Section~\ref{subsec:dijet_selection}, the following selection requirements on the protons reconstructed in the RPs are also applied:

\begin{itemize}

\item At least one proton must be detected in either sector 45 or 56 RP stations.

\item The proton track must cross at least two overlapping RP units (\eg, top-top, bottom-bottom), to ensure quality proton reconstruction.

\item The $\xi$ reconstructed with the RP ($\xi_\Pp (\text{RP})$) must have values of $\xi_\Pp(\text{RP}) < 0.2$ and $t$ must have values of $-4 < t < -0.025\GeV^2$. These bounds are based on acceptance studies of the RPs.

\item The proton track impact location at the RP must satisfy the fiducial selection requirements $8<\abs{y(\text{RP})}<30$\unit{mm} and $0 < x(\text{RP}) < 20$\unit{mm} for vertical RPs, and $\abs{y(\text{RP})}<25$\unit{mm} and $7 < x(\text{RP}) < 25$\unit{mm} for horizontal RPs, where $x(\text{RP})$ and $y(\text{RP})$ denote the horizontal and vertical coordinates of the tracks in the plane transverse to the beamline at the RP. The beam position is at $x(\text{RP}) = y(\text{RP}) = 0$. This selection requirement ensures good proton reconstruction efficiency and acceptance within the RPs, and is based on acceptance studies of the RPs.

\end{itemize}

For the final selection requirement, the main goal is the removal of beam background events, which consist mostly of dijet events paired with uncorrelated beam halo particles or protons from residual pileup interactions. The beam halo is created by the interaction of beam particles with the collimation instrumentation or with residual gas in the vacuum chamber. To suppress these contributions, the following condition is applied:

\begin{itemize}
\item Events must satisfy $\xi_\Pp(\text{PF}) - \xi_\Pp(\text{RP}) < 0$, where $\xi_\Pp(\text{PF}) = \sum_{i} (E^i \pm p_z^i)/\sqrt{s}$ is the fractional momentum loss of the proton calculated with the PF candidates of CMS. Here, $E^i$ and $p_z^i$ are the energy and longitudinal momentum of the $i$-th PF candidate in the event, respectively. The positive or negative sign in the sum corresponds to the scattered proton moving towards the positive or negative $z$ direction in the CMS coordinate system, corresponding to the sector 45 or 56 directions, respectively. The PF candidates considered in the analysis have $\abs{\eta}<5.2$.
\end{itemize}

For the calculation of $\xi_\Pp (\text{PF})$, charged PF objects in $\abs{\eta}<2.5$ with a minimum $\pt> 200\MeV$ are considered. A minimum energy of $1.7$ and $1.2\GeV$ is used at $\abs{\eta}<1.4$ for neutral hadrons and photon candidates of the PF candidate collection, respectively. For neutral hadrons and photon candidates in the region $1.4<\abs{\eta}<2.5$, a respective minimum energy of $3.25$ and $3.00\GeV$ is used. In the forward region, $2.5<\abs{\eta}<5.2$, PF candidates with an energy greater than $5\GeV$ are selected. These $\eta$-dependent energy thresholds were optimized based on zero-bias data collected during the same run conditions as in the dijet data sample. This follows from a similar procedure used in the $7\TeV$ single-diffractive dijet analysis by CMS~\cite{Chatrchyan:2012vc} and the $8\TeV$ CMS-TOTEM study on diffractive dijet production~\cite{Sirunyan:2020ifc}.

Ideally, it is expected that the fractional momentum loss reconstructed with the central detector or the forward proton detectors should be the same, \ie, $\xi_\Pp(\text{PF}) = \xi_\Pp(\text{RP})$. However, because of reconstruction inefficiencies and acceptance limitations of the CMS detector, and the use of energy thresholds applied for each PF candidate reconstructed in CMS, these events satisfy instead the inequality $\xi_\Pp(\text{PF}) - \xi_\Pp(\text{RP})<0$, \ie, the fractional momentum loss is underestimated by the CMS detector. Therefore, the region $\xi_\Pp(\text{PF}) - \xi_\Pp(\text{RP}) > 0$ is dominated by events with uncorrelated forward protons that arise from pileup interactions or beam halo activity, since they do not have to satisfy the same bounds as the physical diffractive events. There is a residual contribution from these events in $\xi_\Pp(\text{PF}) - \xi_\Pp(\text{RP}) < 0$, which is subtracted from the data, as explained in Section~\ref{subsec:background_leadingproton}. The same selection requirement that targets the suppression of beam background contributions was also used in the measurement of single-diffractive dijet production at $\sqrt{s}=8\TeV$ by the CMS and TOTEM Collaborations~\cite{Sirunyan:2020ifc}.

There are $341$ and $336$ events satisfying the dijet and intact proton selection requirements in sectors 45 and 56, respectively.

\section{Central gap between the jets}\label{sec:jetgapjet}

Jet-gap-jet events arising from color-singlet exchange cannot be identified on an event-by-event basis since color-exchange dijet events can also have central gaps through fluctuations in the particle activity between the two jets. Nevertheless, the color-singlet exchange dijet process is expected to lead to an increase in the number of dijet events at the lowest particle multiplicities over those expected to arise from color exchange.

In this analysis, the charged particle activity between the two leading jets is used to characterize the pseudorapidity gap between the jets. The multiplicity of charged particles, $N_\text{tracks}$, is defined as the number of reconstructed charged particle tracks between the two leading jets, where each charged particle is in the interval $\abs{\eta}<1$ and has $\pt > 200\MeV$. The measured relative $\pt$ uncertainty of each charged particle is required to be smaller than $10$\%; this reduces the contribution from badly reconstructed or low-quality tracks. Reconstructed charged particle tracks satisfy the high-purity criteria of CMS described in Ref.~\cite{TRK-11-001}. The central gap is defined as the absence of charged particle production for $\abs{\eta}<1$, which is the same definition used in the previous study at $\sqrt{s}=7\TeV$~\cite{jgjCMS}.

The fixed pseudorapidity gap region $\abs{\eta}<1$ is the same as that employed in previous measurements by the CDF, D0, and CMS Collaborations ~\cite{d02,d03,cdf2,cdf3,jgjCMS}, which facilitates the comparison of the findings of the present analysis with those previously reported at lower $\sqrt{s}$. Since $\abs{\eta^\text{jet1,2}} > 1.4$ and $\eta^\text{jet1}\eta^\text{jet2} < 0$, the separation between the jet axes starts at about 3 units in $\eta$, which is the minimum gap width typically used in studies of diffractive reactions in high energy physics. At the same time, the $\eta$ region is large enough to allow for a controlled subtraction of color-exchange dijet contributions. There are $1650$ jet-gap-jet candidate events with $N_\text{tracks} = 0$ in the sample. Although it is expected that jet-gap-jet events should only yield $N_\text{tracks} = 0$, events with multiplicities up to $N_\text{tracks} = 2$ occur in the signal when jet constituents are emitted at wide angles into the $\abs{\eta}<1$ region, as discussed in Section~\ref{sec:background}.

For a central gap definition based on neutral hadrons or photons, the corresponding \pt thresholds cannot be lowered to the $200\MeV$ scale as with charged particle tracks. The noise level $\pt$ thresholds are $0.5$ and $2\GeV$ for photons and neutral particles at central pseudorapidities, respectively, which leads to a looser definition of an $\eta$ interval devoid of particle activity. Consequently, neutral hadrons and photons are not used in the definition of the central gap in this analysis.

When an intact proton is included, the same definition of the central gap between the jets described above is used. The forward gap is inferred from the direct detection of the scattered proton, \ie, no calorimeter-based rapidity gap is applied. A total of $11$ events are found with $N_\text{tracks} = 0$ for dijet events with an intact proton.

\begin{figure*}[tbp]
\centering
\includegraphics[width=0.45\textwidth]{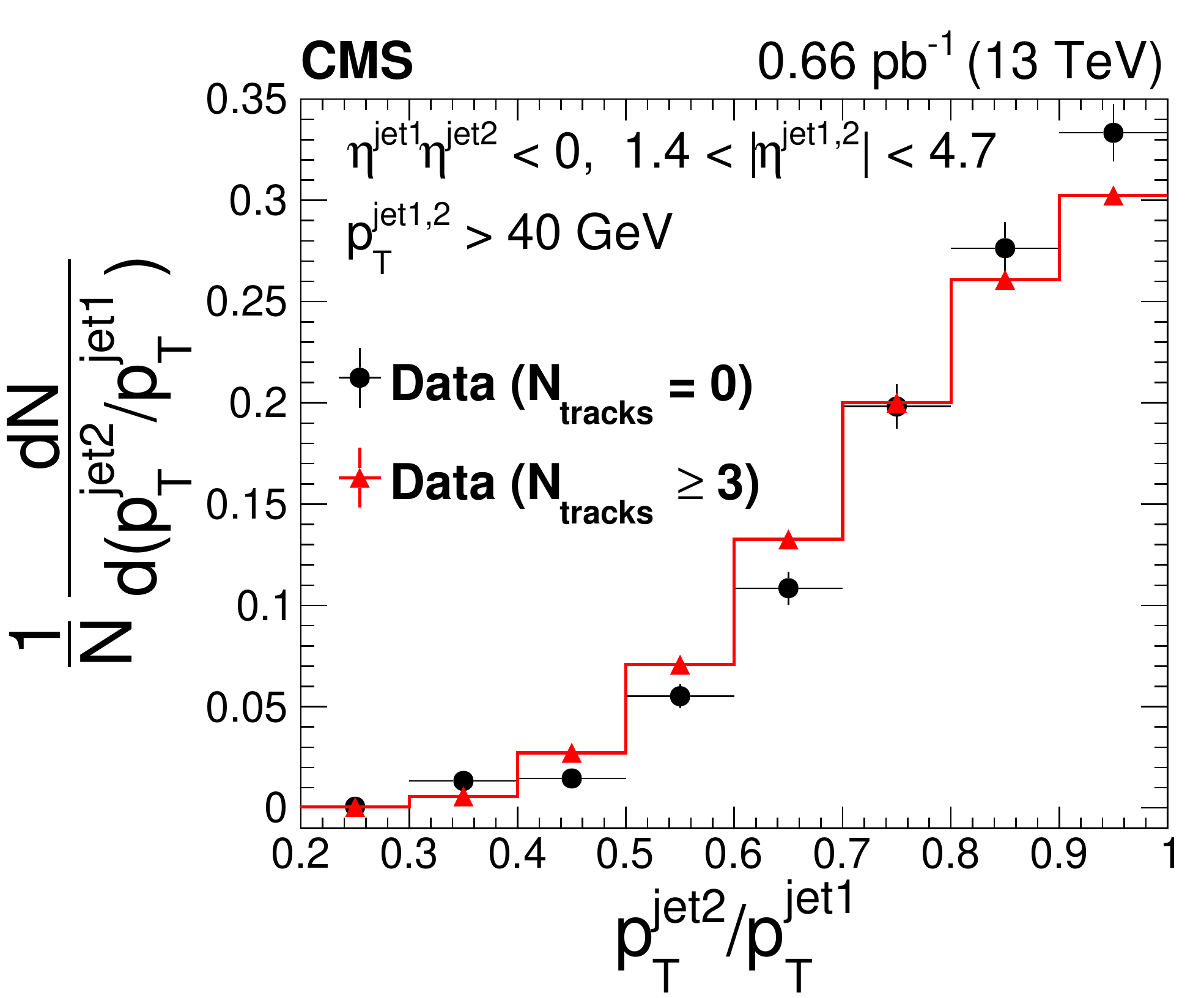}
\includegraphics[width=0.45\textwidth]{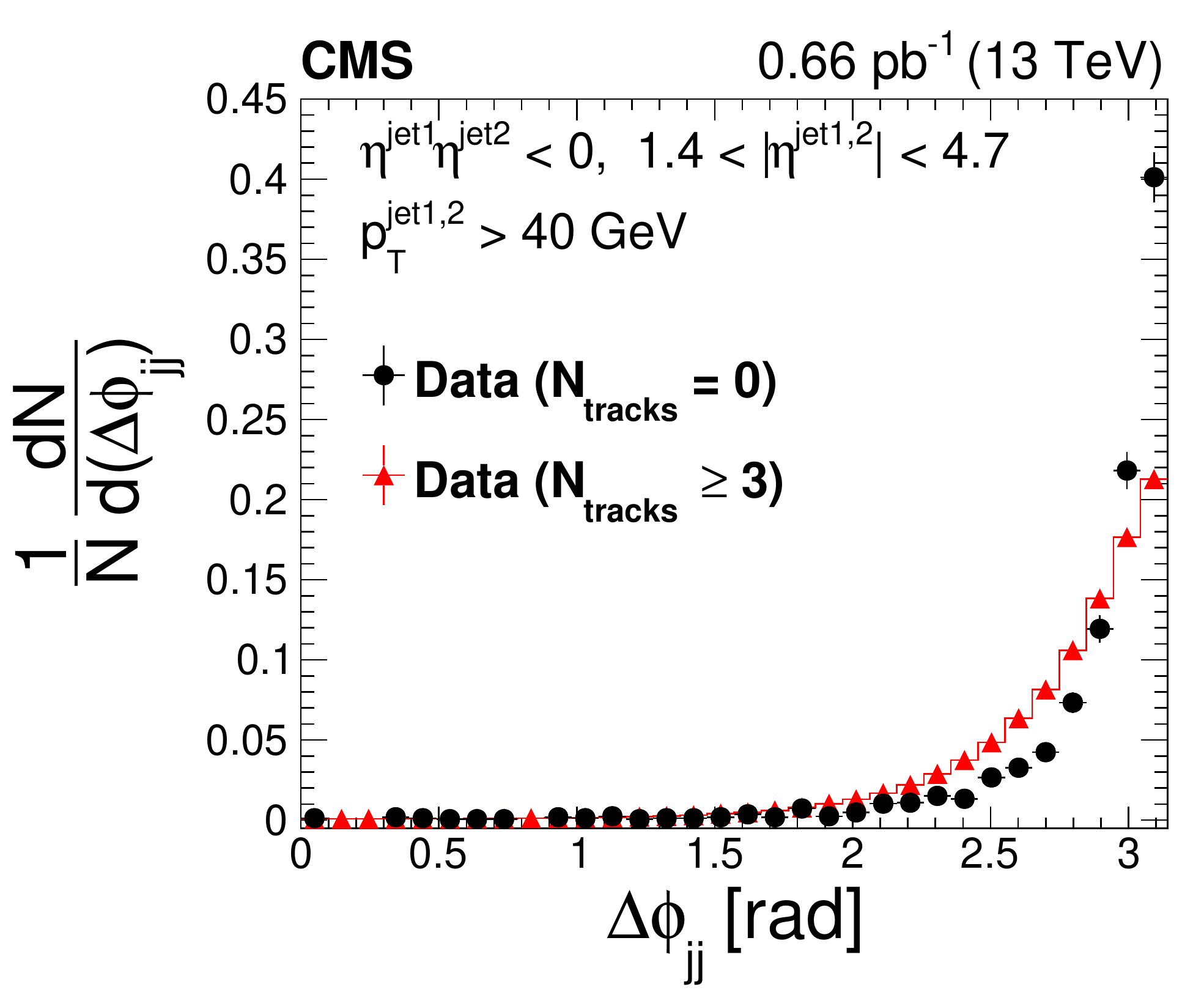}
\includegraphics[width=0.45\textwidth]{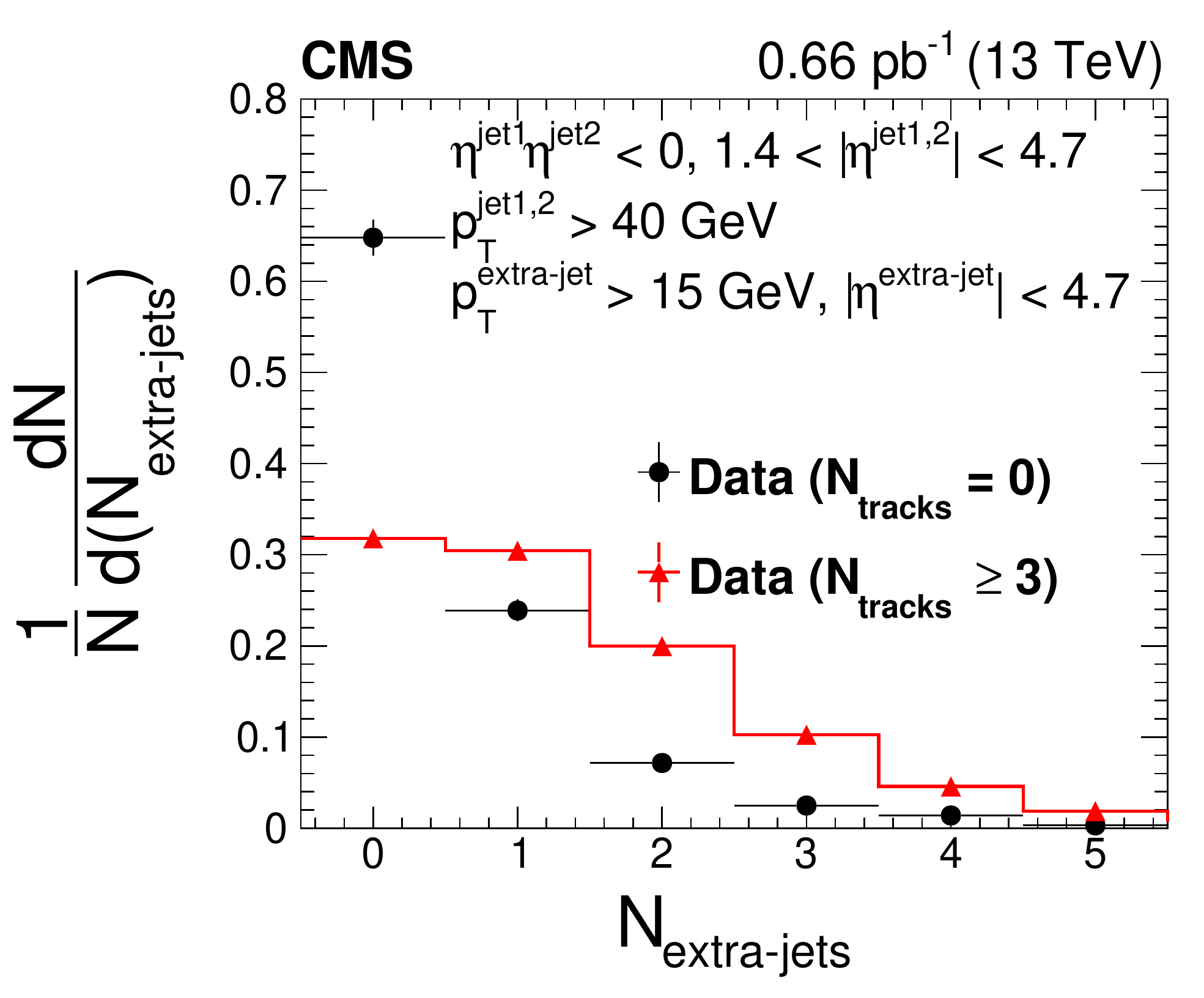}
\caption{ \label{momentum_balance} Distributions of the ratio of the subleading jet to leading jet transverse momenta $\pt^\text{jet2}/\pt^\text{jet1}$ (left panel), the azimuthal angular separation between the two leading jets $\Delta\phi_\text{jj}$ (right panel), and the number of additional jets $N_\text{extra-jets}$ with $\pt^\text{extra-jet} > 15\GeV$ (lower panel), for jet-gap-jet candidates with $N_\text{tracks} = 0$ in $\abs{\eta}<1$ (black circle) and color-exchange dijet candidates $N_\text{tracks} \geq 3 $ in $\abs{\eta}<1$ (red triangle). The vertical bars represent the statistical uncertainties, which are smaller than the marker for some data points. The horizontal bars represent the bin width. The distributions are normalized to unity.}
\end{figure*}

Features of the dijet sample enriched in jet-gap-jet events are presented in Fig.~\ref{momentum_balance}. Events with $N_\text{tracks} = 0$ are dominated by jet-gap-jet events, whereas events with $N_\text{tracks} \geq 3$ are dominated by color-exchange dijet events. Jet-gap-jet candidates have the two leading jets strongly correlated in their transverse momenta, as shown in the upper panels of Fig.~\ref{momentum_balance}. This is characteristic of the nearly elastic parton-parton hard scattering process that initiated the jet production. The jet multiplicity, where each extra jet has $\pt^\text{extra-jet} > 15\GeV$ and $\abs{\eta^\text{extra-jet}}<4.7$, is shown in the lower panel of Fig.~\ref{momentum_balance}. Most of the jet-gap-jet event candidates consist of two-jet events, whereas color-exchange dijet events feature multiple jets.

\section{Color-singlet exchange fraction}\label{sec:observable}

Ideally, hard color-singlet exchange events should have only $N_\text{tracks} = 0$. Occasionally, however, charged particles created during the fragmentation process are produced at large angles with respect to the jet boundary, such that they are emitted into the $\abs{\eta}<1$ region. This leads to spillage of the color-singlet exchange signal events into the neighboring multiplicity counts. Therefore, the jet-gap-jet contributions are extracted for multiplicities up to $N_\text{tracks} = 2$. The integration interval $N_\text{tracks} < 3$ is optimized based on the background studies described in Section~\ref{sec:background}, where the excess of events over the expected number of color-exchange dijet events at low multiplicities is observed to stabilize, within the statistical uncertainties, at $N_\text{tracks} < 3$.

The number of dijet events with $N_\text{tracks} < 3$ is denoted by $N^\text{F}$, the number of dijet events with no underlying color-singlet exchange with $N_\text{tracks} < 3$ by $N_\text{non-CSE}^\text{F}$, and the total number of dijet events by $N$. The yields $N^\text{F}$ and $N$ are extracted directly via event counting, whereas $N_\text{non-CSE}^\text{F}$ requires modeling of color-exchange dijet events, which is discussed in Section~\ref{sec:background}.

The fraction of color-singlet exchange dijet events is given by
\begin{equation}\label{eqn:fcse}
f_\text{CSE} = \frac{N^\text{F} - N_\text{non-CSE}^\text{F}}{N},
\end{equation}
\noindent and is measured as a function of kinematic variables of interest. Because $f_\text{CSE}$ is a ratio of yields, jet reconstruction uncertainties approximately cancel. The $f_\text{CSE}$ fraction can be measured as a function of:

\begin{itemize}

\item The pseudorapidity separation of the two leading jets, $\Delta\eta_\text{jj} \equiv \abs{\eta^\text{jet1} - \eta^\text{jet2}}$.

\item The subleading jet transverse momentum, $\pt^\text{jet2}$.

\item The azimuthal angular separation between the two leading jets, $\Delta\phi_\text{jj} \equiv \abs{\phi^\text{jet1} - \phi^\text{jet2}}$.

\end{itemize}

The fraction $f_\text{CSE}$, measured as a function of $\Delta\eta_\text{jj}$, is particularly sensitive to predictions based on perturbative calculations within the BFKL framework~\cite{Cox:1999dw, csp, Chevallier:2009cu, Kepka:2010hu,cspLHC}, since it is directly related with the resummation of large logarithms of energy. The fraction $f_\text{CSE}$, as a function of $\pt^\text{jet2}$, can be compared with phenomenology studies that predict a weak dependence of this fraction on $\pt^\text{jet2}$ based on BFKL calculations~\cite{Cox:1999dw, csp, Chevallier:2009cu, Kepka:2010hu}. This $\pt^\text{jet2}$ dependence also compares better with previous measurements by D0~\cite{d02,d03} and CMS~\cite{jgjCMS}. The fraction $f_\text{CSE}$, as a function of $\Delta\phi_\text{jj}$, is sensitive to deviations from the back-to-back topology of jet-gap-jet events caused by higher-order perturbative QCD corrections, \eg, those induced by higher order corrections to the impact factors, which are related to the coupling of the perturbative pomeron to quarks and gluons~\cite{hentschinski1,hentschinski2}. The $f_\text{CSE}$ is extracted in bins of the kinematic variables of interest with ranges specified in Tables~\ref{tab:fcse_delta_eta}--\ref{tab:fcse_delta_phi} of Section~\ref{sec:results}.

For the measurement with intact protons, $f_\text{CSE}$ is the ratio of the number of proton-gap-jet-gap-jet events to the number of standard diffractive dijet events. In this case, signal events are extracted in the first two multiplicity bins, $N_{\text{tracks}} < 2$. The integration region of $N_\text{tracks}<2$ is optimized based on the background studies described in Section~\ref{subsec:background_leadingproton}, where an excess of events over background expectations is observed up to $N_\text{tracks} <2$, and on the lower mean multiplicity found in data in events with intact protons. Because of the limited sample size, a measurement as a function of kinematic variables is not possible. Thus, the respective $f_\text{CSE}$ is extracted using the entire sample of events with the intact proton.

\section{Background treatment}\label{sec:background}

Two independent, data-based techniques are used to describe the contribution of color-exchange dijet events in the lowest multiplicity bins. The first method relies on a data sample independent of the nominal sample, whereas the second method relies on a parametrization of particle multiplicity distributions in hadronic collisions. These techniques avoid model-dependent treatment of the underlying event activity, hadronization effects, and other effects that impact the description of particle activity between the jets that are embedded in Monte Carlo events.

\subsection{Background for jet-gap-jet events}\label{subsec:background_inclusive}

In the first approach, a separate $N_\text{tracks}$ distribution is obtained from a sample of events where the two leading jets are reconstructed on the same side of the CMS detector ($\eta^\text{jet1}   \eta^\text{jet2} > 0$) with jets satisfying the requirements $1.4 < \abs{\eta^\text{jet}} < 4.7$ and $\pt^\text{jet}>40\GeV$. The independent sample of events where jets are produced on the same side is referred to as ``SS dijet sample.'' The nominal sample, where jets are reconstructed on opposite sides of the detector ($\eta^\text{jet1}   \eta^\text{jet2} < 0$), is denoted by ``OS dijet sample.'' To suppress single-diffractive dijet contributions in the SS sample (dijet production with a forward pseudorapidity gap), which could affect the shape of the multiplicity distribution at very low multiplicities, at least one calorimeter tower with a minimum energy of $5\GeV$ above the calorimeter noise level in the forward region opposite to the dijet system within $3<\abs{\eta}<5.2$ is required. This SS method for estimating the color-exchange contributions in jet-gap-jet analyses has been used by the CDF and CMS Collaborations~\cite{cdf1,cdf2,cdf3,jgjCMS}.

\begin{figure*}[ht!]
\centering
\includegraphics[width=0.49\textwidth]{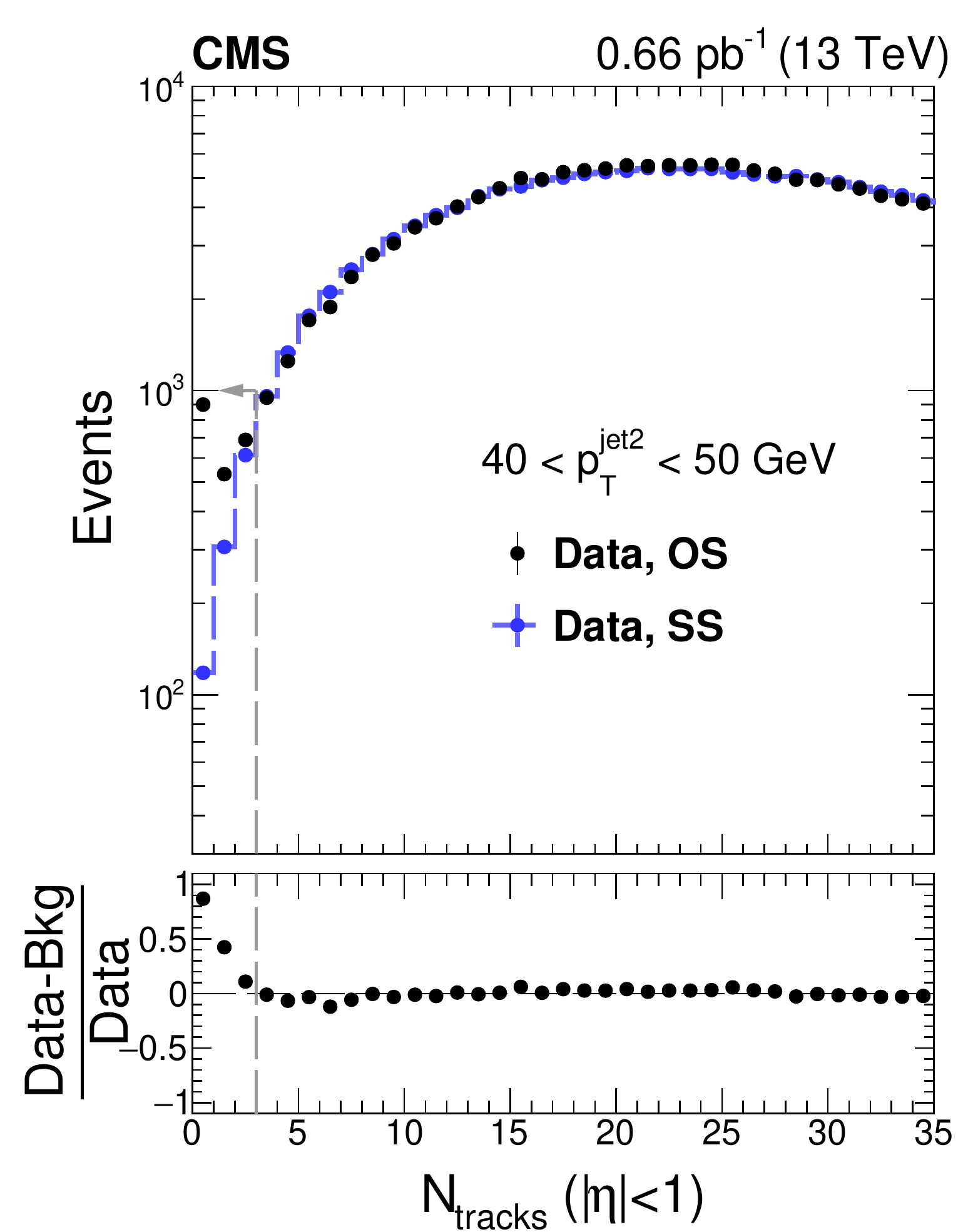}
\includegraphics[width=0.49\textwidth]{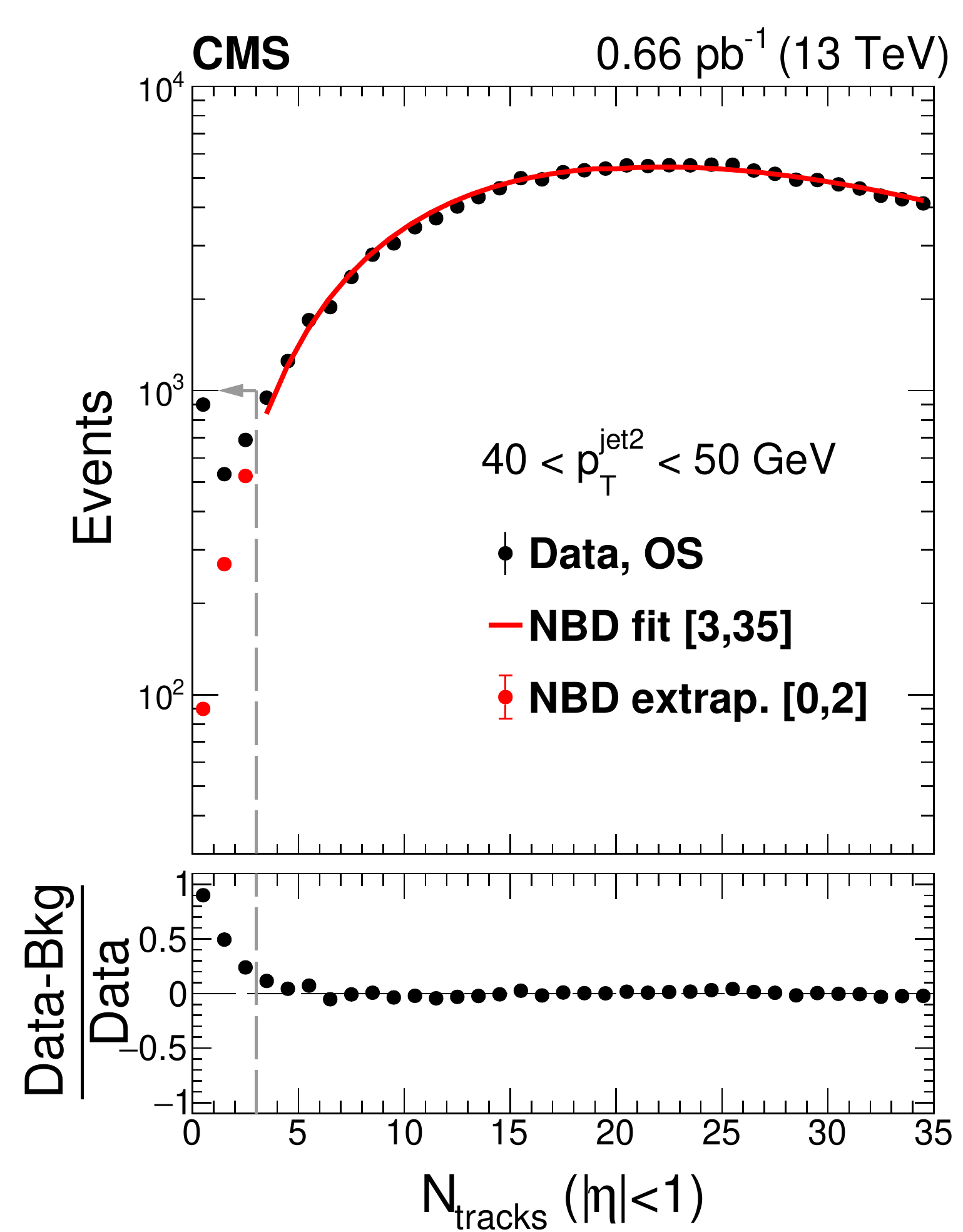}
\caption{\label{fig:multiplicities} Charged particle multiplicity distribution $N_\text{tracks}$ in the $\abs{\eta}<1$ region for charged particle tracks with $\pt > 200 \MeV$ for opposite side (OS) dijet events satisfying $\eta^\text{jet1} \eta^\text{jet2}<0$ with $40 < \pt^\text{jet2} < 50\GeV$. The vertical bars, which represent statistical uncertainties, are smaller than the markers for most data points. Results from color-exchange dijet background estimation based on the same side (SS) dijet events and the negative binomial distribution (NBD) function fit are shown on the left and right panels, respectively. The NBD function is fit in the interval $3 \leq N_\text{tracks} \leq 35$, and extrapolated to $N_\text{tracks} = 0$. The dashed-line arrow represents the jet-gap-jet signal region used in the analysis, $N_\text{tracks} \leq 2$. The vertical bars of the NBD extrapolation points, which are smaller than the markers, represent the uncertainty in the extrapolation based on the fit parameter uncertainties extracted in the $3 \leq N_\text{tracks} \leq 35$ interval. The fraction $f_\text{CSE}$ corresponds to the ratio of the excess of events at low multiplicities relative to the integrated number of events, as described in the text.}
\end{figure*}

Although the multiplicity distribution of charged particles in $\abs{\eta}<1$ has a similar shape in the SS and the OS dijet samples, the SS dijet sample has a lower mean $N_\text{tracks}$ than the OS sample. To compensate for this difference and obtain a better superposition of the $N_\text{tracks}$ distributions of the SS and the OS dijet samples for multiplicities of $N_\text{tracks} > 2$, the $\eta$ region for the SS dijet sample is adjusted. The adjustment is estimated by matching the mean multiplicity of the distributions of the SS and OS samples by varying the pseudorapidity gap width in the SS sample. The optimal $\eta$ interval for the SS dijet sample is $\abs{\eta} < 1.2$, consistent with findings by the CDF and CMS Collaborations at lower $\sqrt{s}$~\cite{cdf1,cdf2,cdf3,jgjCMS}. The multiplicity distribution in the SS sample is then normalized to the one of the OS dijet sample in an interval dominated by color-exchange dijet events, $3 \leq N_\text{tracks} \leq 40 $. The number of events of the SS sample in the first multiplicity bins $N_\text{tracks} < 3$ becomes the estimated number of color-exchange events contributing in the color-singlet exchange signal region. This is illustrated in the left panel of Fig.~\ref{fig:multiplicities} for one of the bins in the region $40 < \pt^\text{jet2} < 50\GeV$ used in the analysis. An excess of OS dijet events at low multiplicities above the expected color-exchange events is observed, which is interpreted as the contribution of hard color-singlet exchange dijet events. The $f_\text{CSE}$ fraction is observed to stabilize up to $N_\text{tracks} = 2$ with the SS method, within the statistical uncertainties, and thus this is the integration region used for the $f_\text{CSE}$ extraction in the analysis. For events at low nonzero $N_\text{tracks}$, strong correlations in $\eta$-$\phi$ between the charged particles and the jets are observed. This suggests that events with low nonzero $N_\text{tracks}$ are due to charged particle constituents of the jet falling into the $\abs{\eta}<1$ region.
 
The second method used to estimate the color-exchange background relies on a fit to the $N_\text{tracks}$ distribution with a negative binomial distribution (NBD) function. This distribution is used to describe $N_\text{tracks}$ distributions with underlying color charge exchanges in hadronic collisions~\cite{Giovannini1986,Ghosh:2012xh}, as first reported by the UA5 Collaboration~\cite{ua5, ua5failstart} at $\sqrt{s} = 540\GeV$. The NBD functional form has also been used to describe $\Pp\Pp$ collision data at several $\sqrt{s}$ values by the ALICE Collaboration~\cite{alice}. The NBD function is less successful in describing the high multiplicity tails of $N_\text{tracks}$ distributions for $\sqrt{s}$ larger than $900\GeV$~\cite{ua5failstart,alice}, and requires the use of more complex phenomenological parametrizations necessary for very wide multiplicity intervals. For the study of jet-gap-jet events, a single NBD function fit is sufficient, since the main focus is at low $N_\text{tracks}$. The NBD method for estimating the color-exchange contributions in jet-gap-jet analyses has been used by the D0 and CMS Collaborations~\cite{d01,d02,d03,jgjCMS}.

The NBD function is fit in the interval $3 \leq N_\text{tracks} \leq 35$, which is expected to be dominated by color-exchange dijet events. The range of $3 \leq N_\text{tracks} \leq 35$ also compares better to the $7\TeV$ analysis, since the shape of the $N_\text{tracks}$ distribution is similar. The NBD function is extrapolated to $N_\text{tracks} = 0$ to estimate the contribution of color-exchange dijet background counts. This is illustrated in the right panel of Fig.~\ref{fig:multiplicities} for one of the bins, $40 < \pt^\text{jet2} < 50\GeV$, used in the analysis. As with the SS method, an excess at low $N_\text{tracks}$ over the NBD extrapolation is observed. The fraction $f_\text{CSE}$ is observed to stabilize by integrating the excess up to $N_\text{tracks} = 2$ with the NBD method, within the statistical uncertainties, and hence this is the integration region used to extract $f_\text{CSE}$. The estimated color-exchange dijet yield in the signal region is stable with respect to variations of the starting and ending points of the fit region, as verified explicitly by changing the fit interval to $3 \leq N_\text{tracks} \leq 25$, $3 \leq N_\text{tracks} \leq 45$, or $4 \leq N_\text{tracks} \leq 35$. The shape of the $N_\text{tracks}$ distribution is very similar for events with low $\Delta\eta_\text{jj}$ (more central dijets) compared with those with large $\Delta\eta_\text{jj}$ (very forward-backward dijet configurations). This is because, for the majority of the events, the gap region is far from the edges of the jets due to the $\abs{\eta^\text{jet}}>1.4$ requirement, which reduces the contamination of soft radiation from the jet.

The NBD method is used to extract the main results in the analysis, since it computes the fraction $f_\text{CSE}$ as a function of the kinematic variables of interest. It also provides for a more direct comparison with the previous measurement by CMS at $\sqrt{s} = 7\TeV$~\cite{jgjCMS}, where the main results are extracted with an NBD function fit in similar $N_\text{tracks}$ intervals. The SS method is used for systematic checks in the analysis. The SS method overestimates the contribution of color-exchange dijet events by about $15$\% relative to the results extracted with the NBD method in $40 < \pt^\text{jet2} < 50\GeV$, and by about $1$--$5$\% for larger values of $\pt^\text{jet2} > 50\GeV$. These differences are taken as a systematic uncertainty.

The performance of the NBD method is tested on the $N_\text{tracks}$ distribution of the SS dijet sample by performing the NBD fit in the range $3 \leq N_\text{tracks} \leq 35$. The extrapolation of the fit results to $N_\text{tracks} = 0$ agrees with the SS data. As an additional check, a subset of the OS dijet sample characterized by the presence of a third leading jet with $\pt^\text{jet3} > 15\GeV$ and $\abs{\eta^\text{jet3}} < 1$ is analyzed. This selection yields a trijet sample enriched in color-exchange events. The NBD function fit describes correctly the $N_\text{tracks}$ distribution of this trijet sample, further confirming the validity of the NBD approach.

The $f_\text{CSE}$ fractions are extracted from the data using dijet yields uncorrected for detector effects. No unfolding of the data is necessary, since reconstruction, resolution, and migration effects cancel in the ratio of yields in $f_\text{CSE}$. The number of color-singlet exchange dijet events in the numerator of Eq.~(\ref{eqn:fcse}) does not depend on track reconstruction inefficiencies; the latter only influence the color-exchange dijet events in the denominator of Eq.~(\ref{eqn:fcse}), which are subtracted in the analysis. Simulation events show that the results do not change within the statistical uncertainties if hadron-level or detector-level variables are used. This was also true for the $7\TeV$ CMS paper~\cite{jgjCMS}.

For these simulation studies, inclusive dijet events (with no hard color-singlet exchange contributions) were simulated using the leading order (LO) {\PYTHIA}8 Monte Carlo event generator~\cite{pythia8} (version 8.212) with the PDF set NNPDF2.3LO~\cite{NNPDF1,NNPDF2}. The {\PYTHIA}8 generator relies on a parton showering algorithm for resummation of soft and collinear gluon emissions at leading-logarithm accuracy, and on the Lund string fragmentation model for hadronization effects~\cite{LundString}. The underlying event tune CUETP8M1~\cite{CUETP8M1} is used, together with initial- and final-state radiation effects. Hard color-singlet exchange events are simulated with the {\HERWIG}6 Monte Carlo event generator~\cite{herwig} (version 6.520) with the PDF set CTEQ6L1~\cite{cteq6l1}. The {\HERWIG}6 generator simulates events with hard color-singlet exchange between two partons following predictions based on simplified leading-logarithm BFKL calculations. Hadronization effects in {\HERWIG}6 are based on the cluster fragmentation model~\cite{clusterFragmentation}. The \textsc{Jimmy} package~\cite{jimmy} is used to supplement MPI. A detailed simulation of the CMS detector response is performed with the \GEANTfour toolkit~\cite{geant}. The reconstruction of these simulated events uses the same algorithms as the data. Stable particles, whose decay length is greater than $20$\unit{mm}, are used for jet reconstruction and measurement of the charged particle multiplicity distribution between the jets in these studies. The hadron-level results on $f_\text{CSE}$ are compared with those obtained when considering the detector response, and agree within the statistical uncertainties, provided that the signal extraction is performed at most at $N_\text{tracks} < 3$. The $f_\text{CSE}$ values in simulation are matched to those in data for these studies. For a check of the background subtraction methods used in the analysis, the $f_\text{CSE}$ values calculated with {\PYTHIA}8 (color-exchange dijet events, no jet-gap-jet signal) are compared, and found consistent with those extracted using the SS or NBD methods, within the statistical uncertainties.
 
\subsection{Background for proton-gap-jet-gap-jet events}\label{subsec:background_leadingproton}

\begin{figure}[tbp]
\centering
\includegraphics[width=0.49\textwidth]{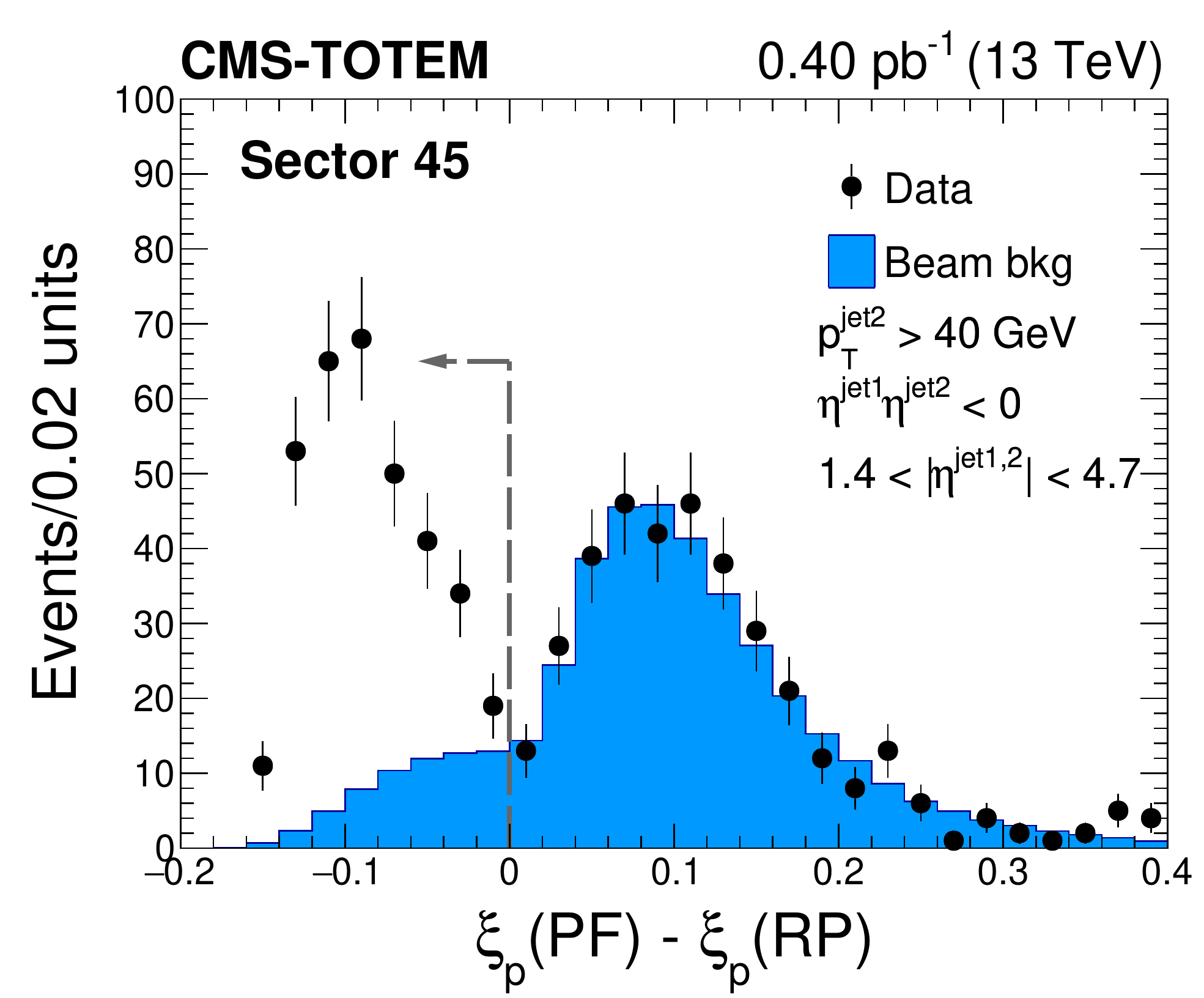}
\includegraphics[width=0.49\textwidth]{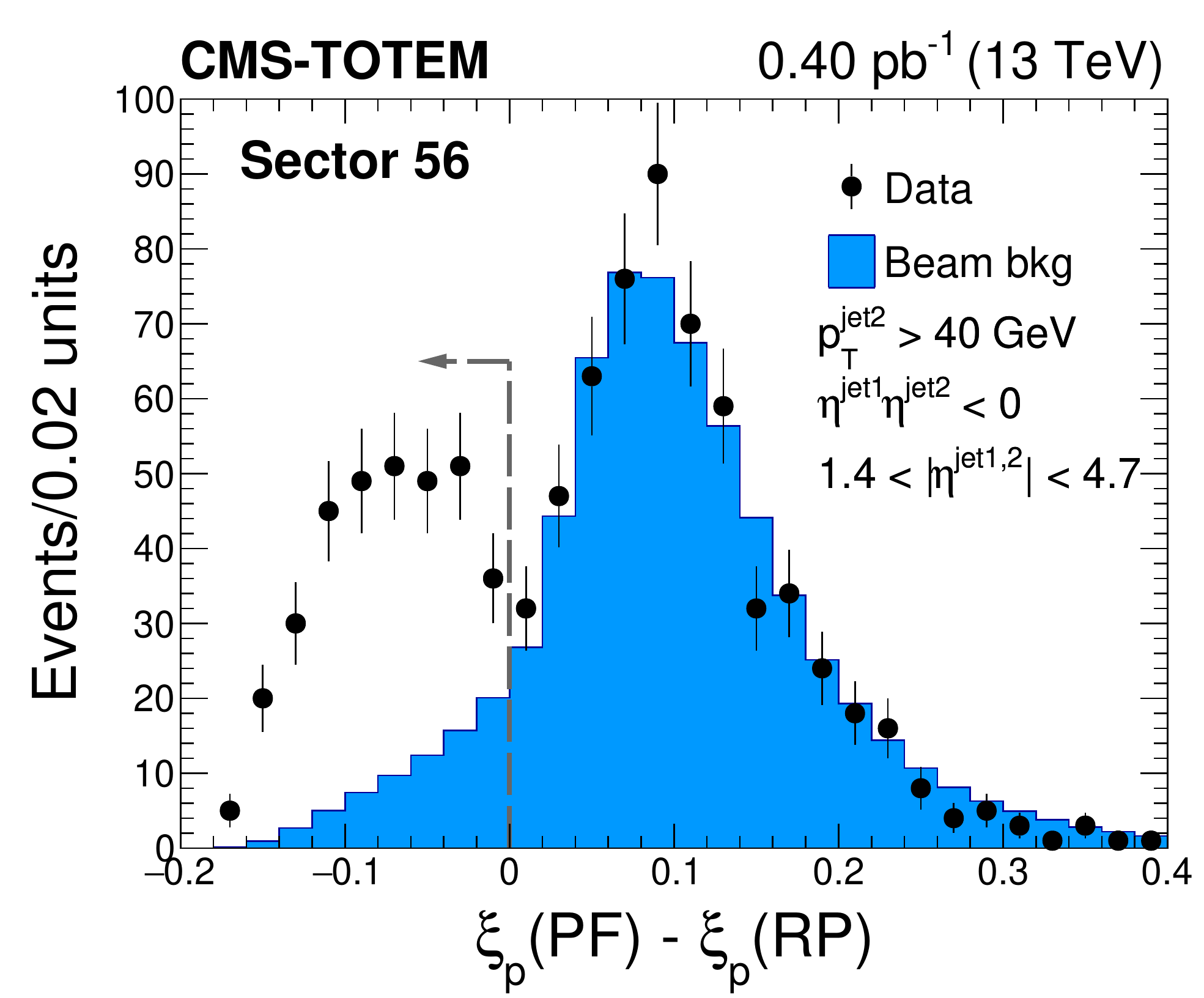}
\caption{\label{fig:xi_difference}  Distribution of $\xi_\Pp(\text{PF}) - \xi_\Pp(\text{RP})$ in sectors 45 (left) and 56 (right) in data, where $\xi_\Pp(\text{PF})$ and $\xi_\Pp(\text{RP})$ denote the fractional momentum loss of the proton reconstructed with the particle-flow (PF) candidates of CMS and the Roman pots (RP) of TOTEM, respectively. The vertical bars indicate statistical uncertainties only. The estimated background contamination (beam background events) is represented by the filled histogram, and is estimated from the data, as described in the text. The statistical uncertainties of the beam background histograms are smaller than the histogram lines. No central gap is required for this plot. The dashed-line arrow represents the requirement applied in the analysis to remove most of the beam background contribution.}
\end{figure}

In the sample with intact protons, the contribution of protons from pileup interactions and beam halo activity must be subtracted. The residual contamination that survives the selection requirement $\xi_\Pp(\text{PF}) - \xi_\Pp(\text{RP}) < 0$, as noted in Section~\ref{subsec:forward_selection}, is estimated using an event mixing procedure that mimics the beam background contribution in the nominal sample, as described below.

\begin{figure*}[ht!]
\centering
\includegraphics[width=0.49\textwidth]{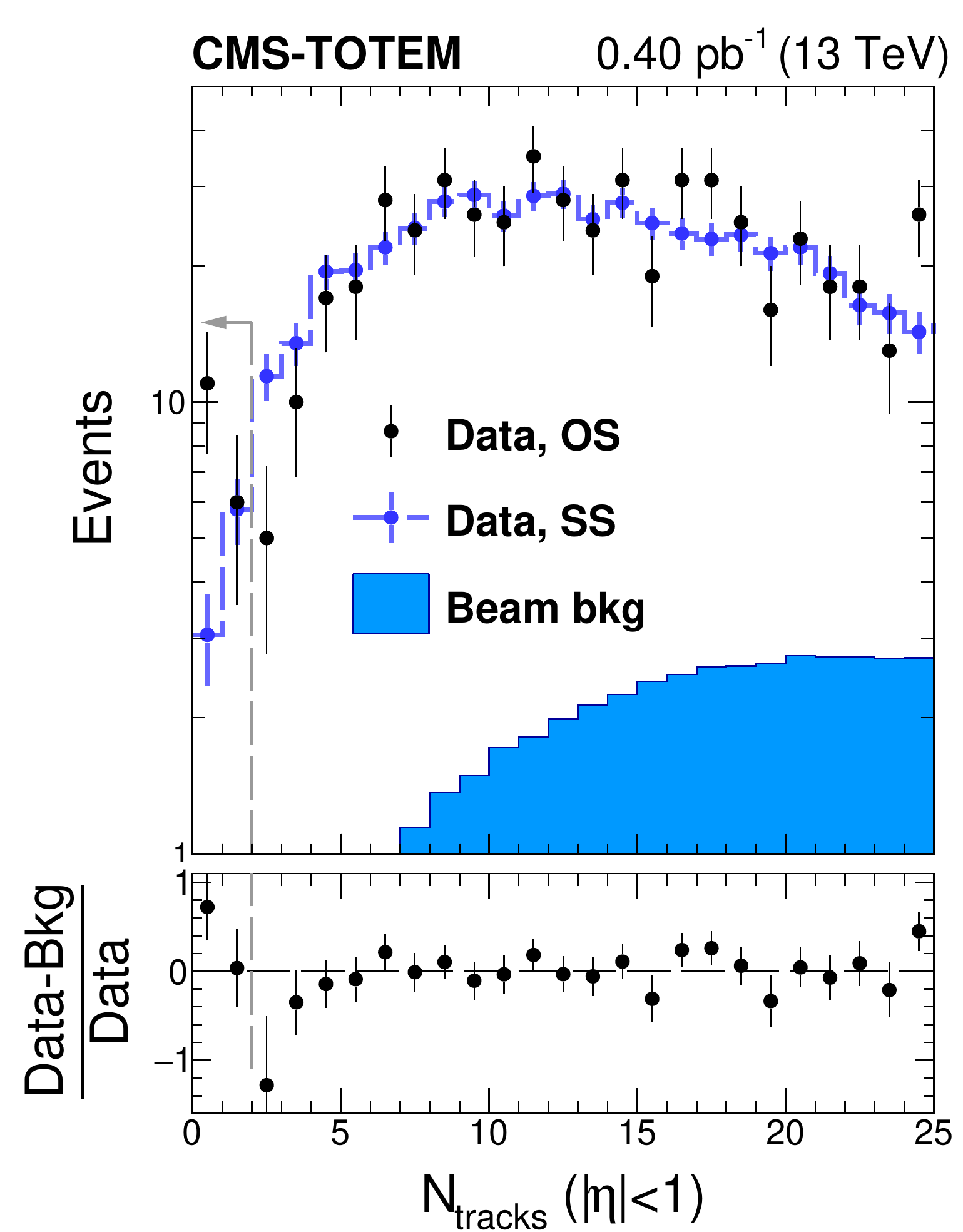}
\includegraphics[width=0.49\textwidth]{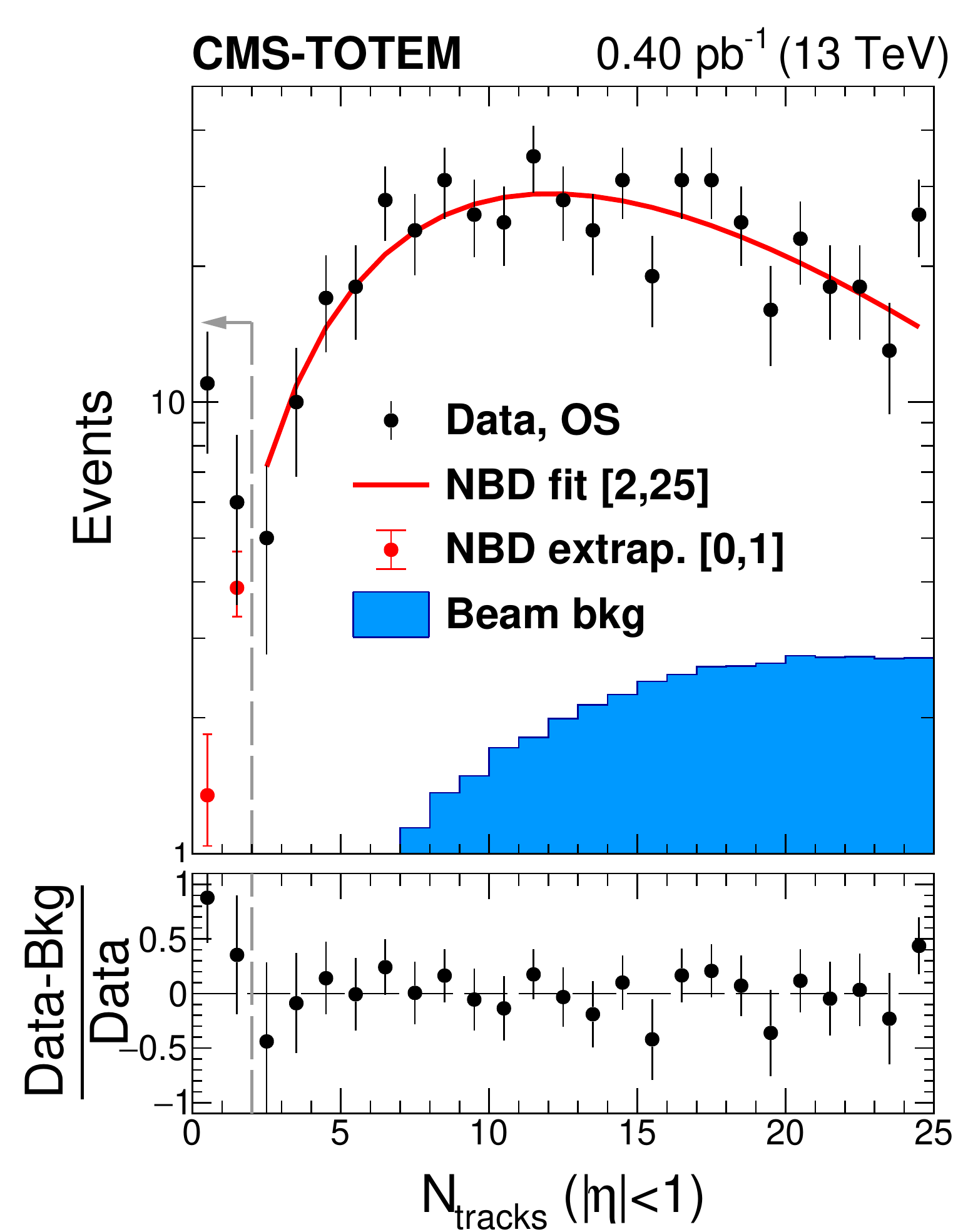}
\caption{\label{fig:multiplicity_onearm} Charged particle multiplicity distribution in the $\abs{\eta}<1$ region after the dijet and proton selection. Opposite side (OS) dijet events satisfy $\eta^\text{jet1} \eta^\text{jet2}<0$. The vertical bars represent the statistical uncertainties. The filled histogram represents the residual beam background contamination. The contribution of standard diffractive dijet events that feature a central gap is modeled with the same side (SS) dijet events (left) and with the negative binomial distribution (NBD) function fit (right), as described in the text. The NBD function is fit in the interval $2 \leq N_\text{tracks} \leq 25$, and extrapolated to $N_\text{tracks} = 0$. The dashed-line arrow represents the region $N_\text{tracks} < 2$ used for signal extraction in the analysis. The vertical bars of the NBD extrapolation points represent the uncertainty in the extrapolation based on the fit parameter uncertainties extracted in the $2 \leq N_\text{tracks} \leq 25$ interval.}
\end{figure*}

Events from the inclusive dijet sample are paired with uncorrelated protons from events in the zero-bias data sample. The dijet events should satisfy the same event selection requirement described in Section~\ref{sec:event_selection}. The number of events from this event mixing procedure is normalized to data with $\xi_\Pp(\text{PF}) - \xi_\Pp(\text{RP}) > 0$, which is dominated by beam background events. Then, the number of events with $\xi_\Pp(\text{PF}) - \xi_\Pp(\text{RP}) < 0$ is the estimated number of beam background events present in the nominal sample. The results of this procedure are presented in Fig.~\ref{fig:xi_difference}. Beam background contamination constitutes $18.6$ and $21.5$\% of the sample in sectors 45 and 56, respectively. Similar procedures have been used in Refs.~\cite{Abbott:1999km,Affolder:2000vb,Affolder:2001zn,PhysRevD.77.052004,
Aaltonen:2012tha,Andreev:2015cwa,Sirunyan:2020ifc}. The distribution of $N_\text{tracks}$ from beam background, shown in Fig.~\ref{fig:multiplicity_onearm}, is determined from the event mixing procedure. A larger number of events in the $\xi_\Pp(\text{PF}) - \xi_\Pp(\text{RP}) > 0$ region is observed for intact protons detected in sector 56. This is indicative of a larger beam background contamination in sector 56 in comparison to sector 45. A higher beam background activity in sector 56 has been previously observed by CMS and TOTEM in the $8\TeV$ single-diffractive dijet measurement~\cite{Sirunyan:2020ifc}.

Standard single-diffractive dijet events can yield a central gap between the jets by fluctuations in $N_\text{tracks}$, analogous to the fluctuations of color-exchange dijet events in inclusive dijet production. The methods introduced in Section~\ref{subsec:background_inclusive} are used to estimate these contributions with modifications that account for differences in the sample with intact protons. Generally, the $N_\text{tracks}$ is lower in events with an intact proton than in inclusive dijet production events. For events passing the dijet and forward proton selection requirements, the mean multiplicity in the $\abs{\eta}<1$ region is $\langle N_\text{tracks} \rangle \approx 17$, compared to the larger $\langle N_\text{tracks} \rangle \approx 28$ in inclusive dijet production. This is consistent with the overall suppression of spectator parton interactions and lower energy available for production of particles in single-diffractive events. Since the $N_\text{tracks}$ distributions in sectors 45 and 56 are similar in shape, the $N_\text{tracks}$ values from the two sectors are summed for the analysis.

The first approach is the SS method. For the analysis with intact protons, the definition of the SS dijet sample introduced in Section~\ref{subsec:background_inclusive} cannot be used. The mean $\eta$ of the jets is not centered at zero in single-diffractive events. This is because single-diffractive dijet events are intrinsically boosted along the beam direction, in a direction opposite to the scattered proton. Thus, in considering single-diffractive dijet events located in the same hemisphere of the CMS detector, the $N_\text{tracks}$ in the $\abs{\eta}<1.2$ region is directly influenced by the intrinsic boost effects. To account for this, the $N_\text{tracks}$ distribution of the SS dijet sample is instead measured in intervals of $-2.0 < \eta < 0.4$ or $-0.4 < \eta < 2.0$ for protons detected in sector 45 or 56, respectively. These intervals are determined based on the mean jet $\eta$ in the data for events with an intact proton in sectors 45 and 56, which corresponds to boosts of about $0.8$ units in negative and positive $\eta$, respectively. The two leading jets are located on the same side relative to these $\eta$ intervals, \ie, $\eta^\text{jet} < -2.2$ or $ \eta^\text{jet} > 0.6$ for intact protons in sector 45 and $\eta^\text{jet} < -0.6$ or $ \eta^\text{jet} > 2.2$ for protons in sector 56. The location of the jet axes is $0.2$ units away from the $\eta$ interval, as in the construction of the SS dijet sample of Section~\ref{subsec:background_inclusive}. The resulting $N_\text{tracks}$ distribution of the SS dijet sample matches that of the OS sample at moderate multiplicities after these adjustments. The $N_\text{tracks}$ distribution of the SS dijet sample is normalized to that of the nominal sample in the range $2 \leq N_\text{tracks} \leq 40$. The number of events of the SS dijet sample in the lowest multiplicity bins is then used to estimate the standard single-diffractive dijet production at low multiplicities $N_\text{tracks} \leq 1$, as shown in Fig.~\ref{fig:multiplicity_onearm}. An excess of events over the expected background counts is observed, which is attributed to the presence of proton-gap-jet-gap-jet events.

The second approach is based on the NBD method introduced in Section~\ref{subsec:background_inclusive}. The NBD function is fit in the interval $2 \leq N_\text{tracks} \leq 25$, and is then extrapolated to $N_\text{tracks} = 0$ to estimate the contribution of standard diffractive dijet events that feature a central gap, as seen in Fig.~\ref{fig:multiplicity_onearm}. The upper bound at $N_\text{tracks} = 25$ is chosen to include the lower mean $N_\text{tracks}$ of the dijet sample with intact protons, and, at the same time, to avoid the contribution by beam background contamination that dominates at high multiplicities. The NBD is fit before beam background subtraction. The result is the same if the fit is carried out after the beam background subtraction, which has an effect on the extracted $f_\text{CSE}$ of less than $2\%$. An excess over the NBD extrapolation results is observed in the data, which provides for an interpretation in terms of proton-gap-jet-gap-jet events. The NBD method is used to extract the main results in the analysis, which facilitates a comparison with the jet-gap-jet results extracted in inclusive dijet production. Because of the lower mean value of $N_\text{tracks}$ and the smaller width of the $N_\text{tracks}$ distribution, the NBD fit extrapolation is more sensitive in jet-gap-jet events with an intact proton than in inclusive dijet events. This is quantified as part of the systematic uncertainties in the $f_\text{CSE}$ extraction.

\section{Systematic uncertainties}\label{sec:systematic_unc}

\subsection{Systematic uncertainties in the study of jet-gap-jet events}\label{subsec:systematic_unc_cms}

The sources of systematic uncertainties for the $f_\text{CSE}$ fraction measurement are:

\textit{Jet energy scale:} The \pt of each jet is varied with $\pt\to \pt \pm \delta \pt(\pt,\eta)$, where $\delta \pt(\pt,\eta)$ is the jet energy scale uncertainty as a function of the jet \pt and $\eta$. The new jet collection is reordered in $\pt$, and the analysis is repeated. The difference in the extracted fraction $f_\text{CSE}$ relative to the results found with the nominal jet energy corrections is a measure of the associated systematic uncertainty. The resulting relative uncertainty is $0.5$--$6.0$\%.

\textit{Track quality:}  The selection criteria used to define high-purity tracks are loosened and the difference in $f_\text{CSE}$ with respect to the nominal selection is taken as the associated systematic uncertainty. The loose quality criteria correspond to the minimum requirements yielding well-reconstructed tracks in the CMS detector, as described in Ref.~\cite{TRK-11-001}. The corresponding uncertainty in $f_\text{CSE}$ is $1.5$--$8.0$\%.

\textit{Charged particle \pt threshold:} Charged particles with $\pt < 200\MeV$ are not considered in identifying a central gap. To study the sensitivity of the results to this threshold, the analysis is repeated with \pt thresholds of $150$ and $250\MeV$ for particles with $\abs{\eta}<1$. The corresponding relative differences in the measured $f_\text{CSE}$ fractions are $1.1$--$5.8$\% and are assigned as systematic uncertainties.

\textit{Background subtraction method}: The background determined using the SS method is compared with the adopted NBD background approach, and the difference is the associated systematic uncertainty. This reflects the imperfect knowledge of the $N_\text{tracks}$ distributions for color-exchange dijet events. At lower $\pt^\text{jet2}$ values, with $40 < \pt^\text{jet2} < 50\GeV$, the relative systematic uncertainty is $14.6$\%, whereas for larger values, $\pt^\text{jet2} > 80\GeV$, it is $2$--$5$\%.

\textit{NBD fit parameters:} The NBD function has three free parameters, including an overall normalization. The color-exchange dijet yields in the signal region are recalculated by varying the NBD fit parameters within their uncertainties. Correlations between the fit parameters are included in this procedure. The maximal differences relative to the nominal results are a measure of the associated systematic uncertainty. These calculations result in a relative uncertainty of less than $2.6$\% in the extracted $f_\text{CSE}$.

\textit{Functional form of the fit:} To quantify the systematic uncertainty associated with the functional form chosen to fit the $N_\text{tracks}$ distribution at large multiplicities, the $N_\text{tracks}$ distribution is fit instead with a double NBD function (a sum of two NBD functions) to extract $f_\text{CSE}$. The double NBD function has been found to be an alternative empirical parametrization of charged particle multiplicities in hadronic collisions at various $\sqrt{s}$, particularly for very wide $N_\text{tracks}$ intervals~\cite{Ghosh:2012xh,ua5failstart,alice}. The symmetrized difference of the $f_\text{CSE}$ extracted with the double NBD fit with respect to the nominal $f_\text{CSE}$ fraction is taken as the respective systematic uncertainty. The relative uncertainty in the extracted $f_\text{CSE}$ is $2$--$7$\%.

A summary of the systematic uncertainties is presented in Table~\ref{tab:systematic_uncertainties}. The systematic uncertainties are added in quadrature and the total bin-specific systematic uncertainty varies between $7$ and $23$\%.

As mentioned in Section~\ref{sec:jetgapjet}, no neutral particles are used in the definition of the pseudorapidity gap because of the relatively large $\pt$ thresholds above the calorimeter noise for neutral hadrons and photons. Most dijet events with low $N_\text{tracks}$ in the region $\abs{\eta}<1$ have little, if any, neutral particle activity in that region. Simulation studies that include the detector response, based on the samples described in Section~\ref{subsec:background_inclusive}, suggest that the neutral hadron and photon activity observed in data originate from the emission of jet constituents into the $\abs{\eta}<1$ region, together with residual contributions of the calorimeter noise. The $f_\text{CSE}$ fractions remain mostly unaffected if the contribution of neutral particles at central $\eta$ is included in the analysis. In particular, if the vector \pt sum of the neutral hadrons and photons for $\abs{\eta}<1$ is required to be less than $15\GeV$, the results for $f_\text{CSE}$ are the same, within the statistical uncertainties of $f_\text{CSE}$. This is consistent since the color-exchange dijet background is already subtracted in the determination of $f_\text{CSE}$.

\begin{table*}[htbp]
\centering
\topcaption{\label{tab:systematic_uncertainties} Relative systematic uncertainties in percentage for the measurements of $f_\text{CSE}$ in jet-gap-jet and proton-gap-jet-gap-jet events. The jet-gap-jet results summarize the systematic uncertainties in bins of the kinematic variables of interest $\pt^\text{jet2}$, $\Delta\eta_\text{jj}$, and $\Delta\phi_\text{jj}$. When an uncertainty range is given, the range of values is representative of the variation found in $f_\text{CSE}$ in bins of the kinematic variables of interest.}
\begin{scotch}{ccccc}
\multirow{2}{*}{Source}                       & \multicolumn{3}{c}{{ Jet-gap-jet (\%)}}                            & \multirow{2}{*}{Proton-gap-jet-gap-jet (\%)} \\
                                              & $\Delta\eta_\text{jj}$ & $\pt^\text{jet2}$ & $\Delta\phi_\text{jj}$ &                                         \\ \hline
Jet energy scale                              & $1.0$--$5.0$          & $1.5$--$6.0$            & $0.5$--$3.0$              & $0.7$                                     \\
Track quality                        & $6.0$--$8.0$          & $5.4$--$8.0$            & $1.5$--$8.0$               & $8$                                     \\
Charged particle \pt threshold & $2.0$--$5.8$          & $1.6$--$4.0$            & $1.1$--$5.8$               & $11$                                    \\
Background subtraction method                 & $4.7$--$15$         & $2$--$15$             & $12$                   & 28                                    \\
NBD fit parameters                             & $0.8$--$2.6$          & $0.6$--$1.7$            & $0.1$--$0.6$               & $7.0$ \\
Functional form of the fit                            & $2$--$7.3$          & $1.4$--$8.0$            & $0.6$--$7.8$               & $11.5$                                     \\
NBD fit interval                              & \NA                & \NA                  & \NA                     & $12$                                    \\
Calorimeter energy scale                      & \NA                & \NA                  & \NA                     & $5.0$                                     \\
Horizontal dispersion                         & \NA                & \NA                  & \NA                     & $6.0$                                     \\
Fiducial selection requirements               & \NA                & \NA                  & \NA                     & $2.6$                                     \\
Total                                         & $7$--$23$         & $9$--$15$           & $12$--$18.5$             & $35$                                    \\
\end{scotch}
\end{table*}

\subsection{Systematic uncertainties in the study of proton-gap-jet-gap-jet events}

In addition to the sources of systematic uncertainties described in Section~\ref{subsec:systematic_unc_cms}, the following sources of systematic uncertainties that affect the extraction of $f_\text{CSE}$ in proton-gap-jet-gap-jet events are considered:

\textit{NBD fit interval:}  Because of the lower mean $N_\text{tracks}$ and the limited sample size, the NBD fit extrapolation is more sensitive to the fit interval in events with an intact proton than in inclusive dijet production. The color-exchange dijet background for intervals of $2 \leq N_\text{tracks} \leq 15$ and $2 \leq  N_\text{tracks} \leq 35$ is evaluated. The difference of the extracted $f_\text{CSE}$ value for these intervals relative to that for the nominal interval $2 \leq N_\text{tracks} \leq 25$ is taken as the associated systematic uncertainty. Based on these studies an uncertainty of $12$\% is assigned to the extracted $f_\text{CSE}$. The difference of the measured $f_\text{CSE}$ value using the fit interval $3 \leq N_\text{tracks} \leq 25$ relative to the nominal fit interval is negligible.

\textit{Calorimeter energy scale:} Beam background contributions are suppressed via the requirement $\xi_\Pp (\text{PF}) - \xi_\Pp(\text{RP}) < 0$ in the analysis. Since $\xi_\Pp(\text{PF})$ is constructed from the PF candidates of the CMS experiment, it is affected by the energy calibration uncertainties of each PF candidate. The impact on $\xi_\Pp(\text{PF})$ is estimated by varying the energy of the PF candidates conservatively by $\pm 10$\%~\cite{PFnew}. The corresponding relative difference in the extracted $f_\text{CSE}$ value is $5$\%, and is included as the associated systematic uncertainty.

\textit{Horizontal dispersion:} The determination of $\xi_\Pp(\text{RP})$ depends on the LHC optics parametrization in the transport matrix, which connects the kinematics of the proton at the interaction point with those measured at the RPs. The horizontal dispersion term in the transport matrix directly affects the measurement of $\xi_\Pp(\text{RP})$~\cite{totem1}. The associated systematic uncertainty is estimated by conservatively scaling the value of $\xi_\Pp(\text{RP})$ by $\pm 10$\%, and repeating the analysis. The $f_\text{CSE}$ has an uncertainty of $6$\%.

\textit{Fiducial selection requirements for $x(\text{RP})$--$y(\text{RP})$ coordinates at the RPs:} The vertical and horizontal fiducial requirements are varied by $0.2$ and $1$\unit{mm}, respectively. The relative difference of the $f_\text{CSE}$ result with respect to that obtained with the nominal fiducial $x(\text{RP})$--$y(\text{RP})$ requirements is less than $2.6$\%, and is assigned as the corresponding systematic uncertainty.

The systematic uncertainties are summarized in Table~\ref{tab:systematic_uncertainties}. The systematic uncertainties related to the jet reconstruction and central gap definition are larger in the proton-gap-jet-gap-jet study. The total systematic uncertainty is calculated as the quadratic sum of the individual contributions, and it has a value of 35\%.

\section{Results}\label{sec:results}

\subsection{Results for jet-gap-jet events in inclusive dijet production}

\begin{figure*}[htbp!]
\centering
\includegraphics[width=0.49\textwidth]{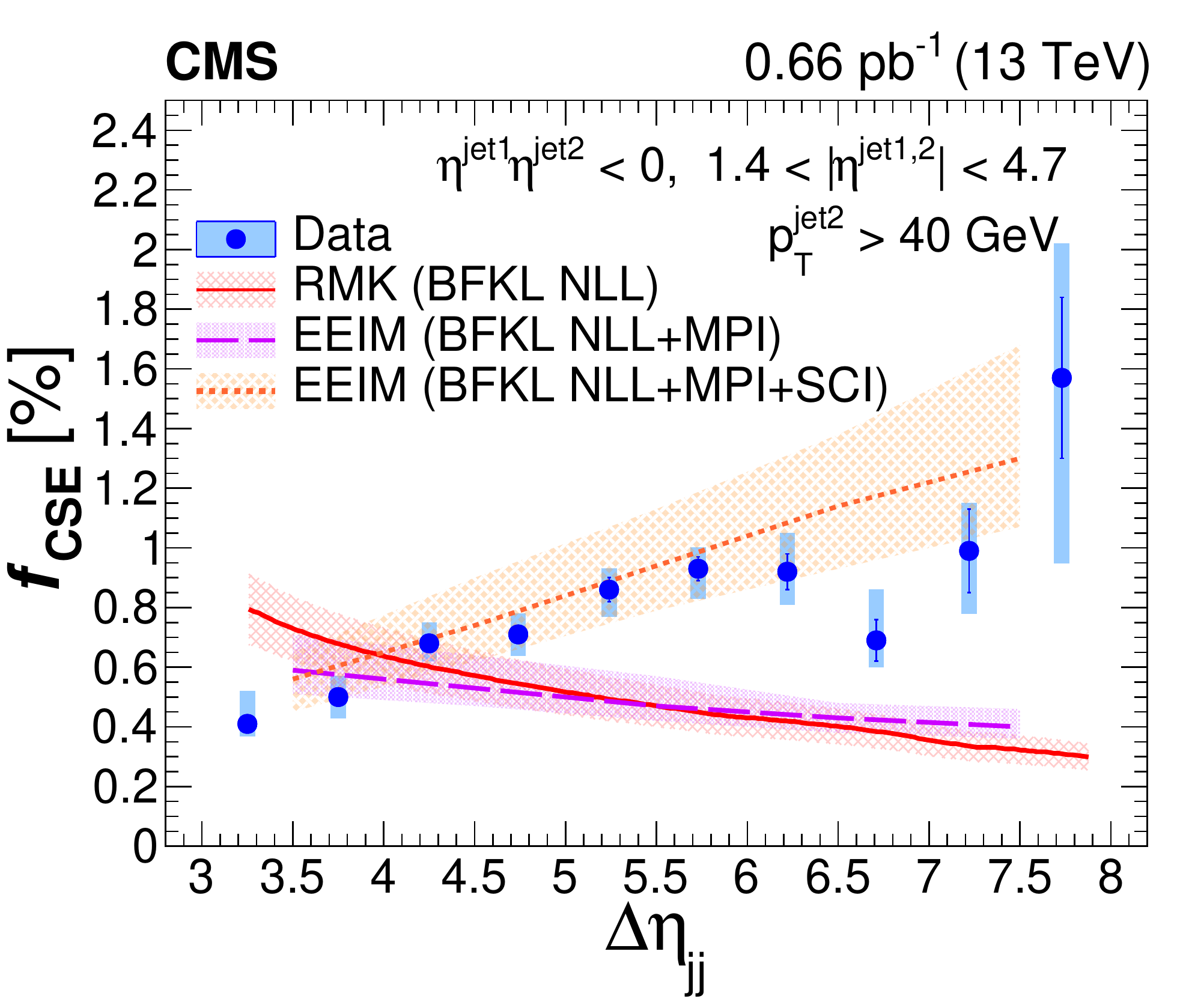}
\includegraphics[width=0.49\textwidth]{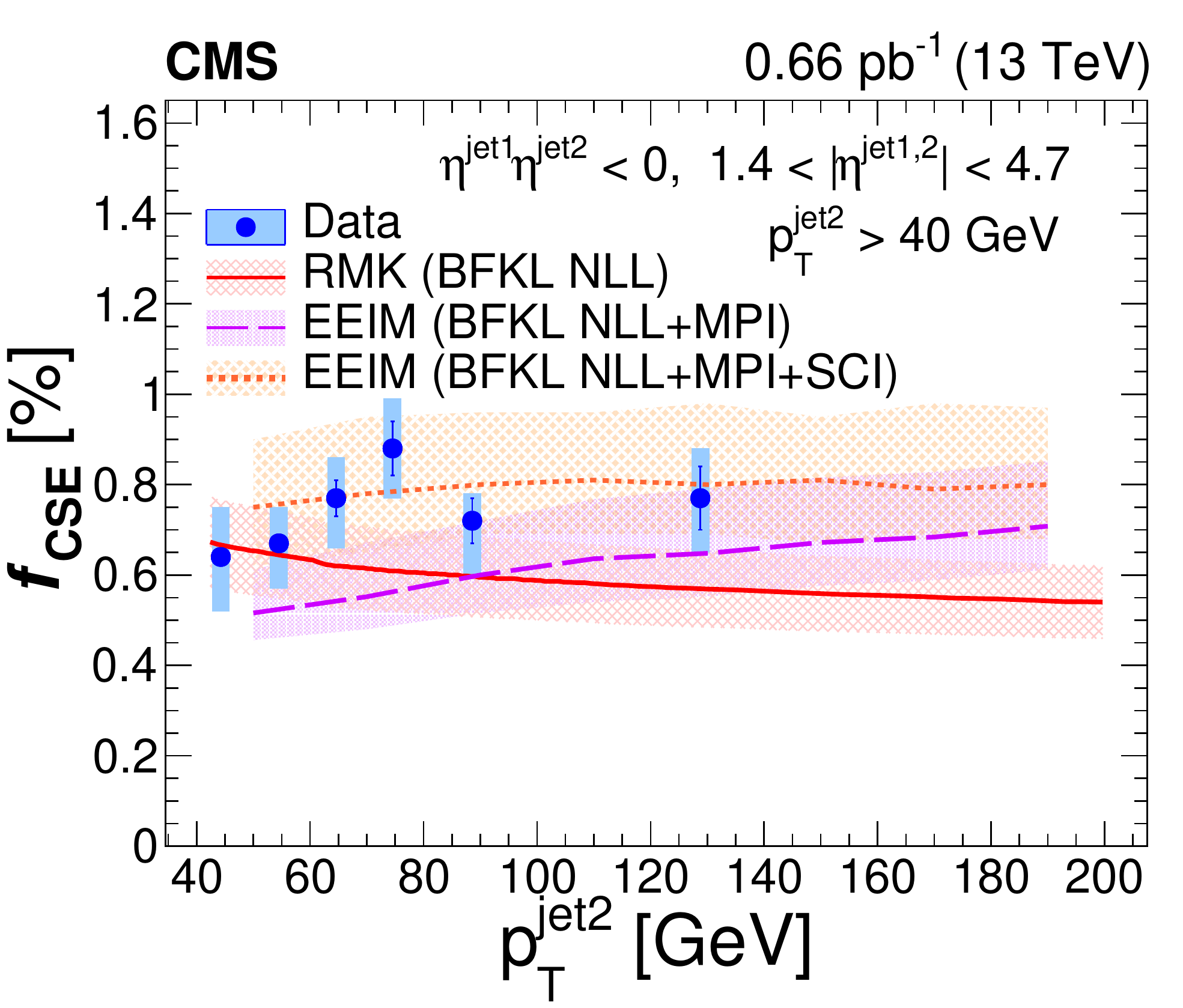}
\includegraphics[width=0.49\textwidth]{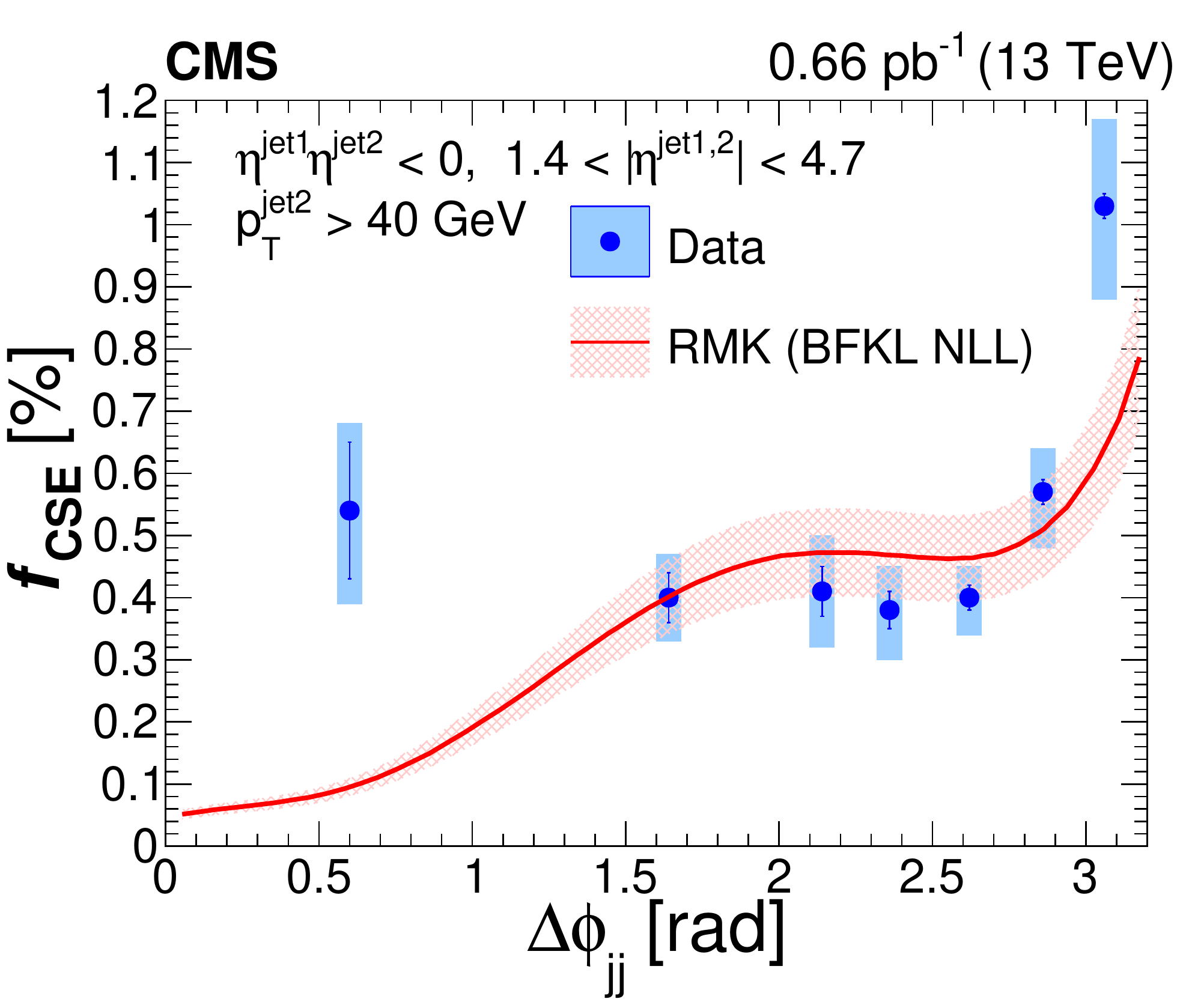}
\caption{ \label{fig:fcse_delta_eta} Fraction of color-singlet exchange dijet events, $f_\text{CSE}$, measured as a function of $\Delta\eta_\text{jj}$, $\pt^\text{jet2}$, and $\Delta\phi_\text{jj}$ in $\Pp\Pp$ collisions at $\sqrt{s} = 13\TeV$. The vertical bars represent statistical uncertainties, while boxes represent the combination of statistical and systematic uncertainties in quadrature. The results are plotted at the mean values of $\Delta\eta_\text{jj}$, $\pt^\text{jet2}$, and $\Delta\phi_\text{jj}$ in the bin. For a given plot of $f_\text{CSE}$ versus a kinematic variable of interest ($\pt^\text{jet2}$, $\Delta\eta_\text{jj}$, or $\Delta\phi_\text{jj}$), the other kinematic variables are integrated over their allowed range. The red solid curve corresponds to theoretical predictions based on the RMK model~\cite{Chevallier:2009cu, Kepka:2010hu} with gap survival probability of $\abs{\mathcal{S}}^2 = 10$\%. The EEIM model~\cite{csp,cspLHC} predictions with MPI-only contributions and $\abs{\mathcal{S}}^2 = 1.2$\% or MPI+SCI are represented by the purple dashed and orange dotted curves, respectively. The bands around the curves represent the associated theoretical uncertainties. The EEIM model has only small contributions far from back-to-back jets since no hard NLO $2\to 3$ processes are included, and thus predictions are not shown for the lower panel of $f_\text{CSE}$ versus $\Delta\phi_\text{jj}$.}
\end{figure*}

The measured fractions $f_\text{CSE}$ are presented in Fig.~\ref{fig:fcse_delta_eta} and Tables \ref{tab:fcse_delta_eta}--\ref{tab:fcse_delta_phi}. As a function of $\Delta\eta_\text{jj}$, the $f_\text{CSE}$ fraction shows a uniform increase from $0.4$ to $1.0$\% for $\Delta\eta_\text{jj}$ between $3$ and $6$ units. Within the experimental uncertainties, $f_\text{CSE}$ is about $0.7$\%, and shows little, if any, dependence on $\pt^\text{jet2}$. As a function of $\Delta\phi_\text{jj}$ between the two leading jets, the $f_\text{CSE}$ fraction exhibits a peak near $\Delta\phi_\text{jj} = \pi$ with a value of $1$\%, which suggests that jet-gap-jet events are more strongly correlated in the transverse plane than inclusive dijet events. A constant value of about $0.4\%$ is found for $\Delta\phi_\text{jj} < 2.8$; this implies that color-singlet exchange dijet events decorrelate at a similar rate as color-exchange dijet events in this interval.

The present results are compared with BFKL-based theoretical calculations of Royon, Marquet, Kepka (RMK)~\cite{Chevallier:2009cu, Kepka:2010hu} and Ekstedt, Enberg, Ingelman, Motyka (EEIM)~\cite{csp,cspLHC}, the results of which are shown in Fig.~\ref{fig:fcse_delta_eta}. The RMK and EEIM model predictions include dominant NLL corrections to the BFKL evolution of the parton-level cross section using LO impact factors. The RMK predictions are supplemented with a gap survival probability of $\abs{\mathcal{S}}^2 = 10$\%, whose value is used to match the $f_\text{CSE}$ values observed in data. The RMK predictions use an updated parametrization of the BFKL NLL amplitudes that include the larger phase space available at LHC energies~\cite{Trzebinski:2015dna}, which are then implemented in the {\HERWIG}6 generator~\cite{Kepka:2010hu}. The theoretical uncertainties in the RMK prediction are due to renormalization and factorization scale uncertainties in the BFKL calculation~\cite{Kepka:2010hu}. The EEIM predictions include soft rescattering effects based on MPI or MPI supplemented with SCI. In the EEIM approach, the spectator parton activity originating from perturbative gluons is modeled in terms of initial- and final-state parton showers, MPI, and hadronization processes, as implemented in {\PYTHIA}6~\cite{csp,cspLHC}. The SCI accounts for possible gap destruction effects caused by color exchange with negligible momentum transfer that rearrange the color field created in the $\Pp\Pp$ collision~\cite{csp}. The free parameters of the SCI model are fit to describe the previous $7\TeV$ measurement by CMS~\cite{cspLHC}. The remaining nonperturbative corrections are either modeled with a survival probability of $\abs{\mathcal{S}}^2=1.2$\% to match the $f_\text{CSE}$ value found in data (purple dashed line in Fig.~\ref{fig:fcse_delta_eta}) or with SCI (orange dotted line in Fig.~\ref{fig:fcse_delta_eta}). The theoretical uncertainties in the EEIM model predictions are dominated by the cutoff $\pt$ scale used for MPI in the simulation.

\begin{table}[htbp]
\centering
\topcaption{\label{tab:fcse_delta_eta} Measured values of the fraction of color-singlet exchange events $f_\text{CSE}$ in bins of the pseudorapidity difference between the two leading jets $\Delta\eta_\text{jj}$. The first column indicates the $\Delta\eta_\text{jj}$ intervals and the last column represents the measured fraction. The first and second uncertainties correspond to the statistical and systematic components, respectively. The results are integrated over the allowed $\pt^\text{jet2}$ and $\Delta\phi_\text{jj}$ values. The mean values of $\Delta\eta_\text{jj}$ in the bin are given in the middle column.}
\renewcommand{\arraystretch}{1.3}
\begin{scotch}{c c c}
$\Delta\eta_\text{jj}$ & $\langle \Delta\eta_\text{jj} \rangle$ & $f_\text{CSE}$ {[}\%{]}              \\ \hline 
$3.0$--$3.5$      &   $3.24$    & $0.41 \pm 0.02^{+0.11}_{-0.04}$  \\ 
$3.5$--$4.0$      &   $3.75$    & $0.50 \pm 0.02^{+0.07}_{-0.07}$  \\ 
$4.0$--$4.5$      &   $4.25$    & $0.68 \pm 0.02^{+0.07}_{-0.06}$  \\ 
$4.5$--$5.0$      &   $4.74$    & $0.71 \pm 0.03^{+0.06}_{-0.06}$  \\
$5.0$--$5.5$      &   $5.24$     & $0.86 \pm 0.04^{+0.06}_{-0.08}$  \\
$5.5$--$6.0$      &   $5.73$   & $0.93 \pm 0.04^{^+0.06}_{-0.09}$ \\
$6.0$--$6.5$      &   $6.22$    & $0.92 \pm 0.06^{+0.11}_{-0.09}$  \\
$6.5$--$7.0$      &   $6.71$    & $0.69 \pm 0.07^{+0.15}_{-0.05}$  \\
$7.0$--$7.5$      &   $7.22$    & $0.99 \pm 0.14^{+0.07}_{-0.15}$  \\
$7.5$--$8.0$      &   $7.73$    & $1.57 \pm 0.27^{+0.35}_{-0.56}$  \\ 
\end{scotch}
\end{table}

\begin{table}[htbp]
\centering
\topcaption{\label{tab:fcse_pt2} Measured values of the fraction of color-singlet exchange events $f_\text{CSE}$ in bins of the subleading jet transverse momentum $\pt^\text{jet2}$. The first column indicates the $\pt^\text{jet2}$ bin intervals and the last column represents the measured fraction. The first and second uncertainties correspond to the statistical and systematic components, respectively. The results are integrated over the allowed $\Delta\eta_\text{jj}$ and $\Delta\phi_\text{jj}$ values. The mean values of $\pt^\text{jet2}$ in the bin are given in the middle column.}
\renewcommand{\arraystretch}{1.3}
\begin{scotch}{c c c}
$\pt^\text{jet2}$ [\GeVns{}] & $\langle \pt^\text{jet2} \rangle$ [\GeVns{}] &$f_\text{CSE}$ {[}\%{]}                \\ \hline 
$40$--$50$        &  $44.3$   & $0.64\pm 0.01^{+0.11}_{-0.12}$   \\ 
$50$--$60$        &  $54.5$   & $0.67\pm 0.02^{+0.08}_{-0.10}$  \\
$60$--$70$        &  $64.6$   & $0.77\pm 0.04^{+0.08}_{-0.10}$  \\
$70$--$80$        &  $74.5$   & $0.88\pm 0.06^{+0.09}_{-0.09}$ \\
$80$--$100$       &  $88.6$   & $0.72\pm 0.05^{+0.04}_{-0.11}$  \\
$100$--$200$      &  $128.8$   & $0.77\pm 0.07^{+0.09}_{-0.10}$ \\
\end{scotch}
\end{table}

\begin{table}[htbp!]
\centering
\topcaption{\label{tab:fcse_delta_phi} Measured values of the fraction of color-singlet exchange events $f_\text{CSE}$ in bins of the azimuthal angular difference between the two leading jets $\Delta\phi_\text{jj}$. The first column indicates the $\Delta\phi_\text{jj}$ bin intervals and the last column represents the measured fraction. The first and second uncertainties correspond to the statistical and systematic components, respectively. The results are integrated over the allowed $\pt^\text{jet2}$ and $\Delta\eta_\text{jj}$ values. The mean values of $\Delta\phi_\text{jj}$ in the bin are given in the middle column.}
\renewcommand{\arraystretch}{1.3}
\begin{scotch}{c c c}
$\Delta\phi_\text{jj}$ & $\langle \Delta\phi_\text{jj} \rangle$ & $f_\text{CSE}$ {[}\%{]}                \\ \hline 
$0.00$--$1.00$      &       $0.60$              & $0.54 \pm 0.11^{+0.09}_{-0.10}$ \\
$1.00$--$2.00$       &         $1.64$           & $0.40 \pm 0.04^{+0.06}_{-0.06}$ \\
$2.00$--$2.25$      &         $2.14$         & $0.41 \pm 0.04^{+0.08}_{-0.08}$ \\
$2.25$--$2.50$    &        $2.36$          & $0.38 \pm 0.03^{+0.06}_{-0.07}$ \\
$2.50$--$2.75$     &       $2.62$          & $0.40 \pm 0.02^{+0.05}_{-0.06}$ \\
$2.75$--$3.00$        &      $2.86$          & $0.57 \pm 0.02^{+0.07}_{-0.09}$ \\
$3.00$--$\pi$        &     $3.06$             & $1.03 \pm 0.02^{+0.14}_{-0.15}$ \\
\end{scotch}
\end{table}

According to both the RMK and EEIM model calculations, $f_\text{CSE}$ should have a weak dependence on $\pt^\text{jet2}$. Within the uncertainties, this feature is consistent with the observed $f_\text{CSE}$ values. The predictions by RMK and EEIM (with MPI only) yield a decreasing $f_\text{CSE}$ with increasing $\Delta\eta_\text{jj}$. This is in disagreement with the data, which show a $f_\text{CSE}$ that generally grows with larger $\Delta\eta_\text{jj}$. The EEIM model predictions, when supplemented with SCI, correctly describe $f_\text{CSE}$ as a function of $\Delta\eta_\text{jj}$ within the uncertainties. The predictions of the RMK model for $f_\text{CSE}$ as a function of $\Delta\phi_\text{jj}$ are consistent with the data within the uncertainties for medium angular separations $1<\Delta\phi_\text{jj}<3$, but underestimate the experimental result by about $10$\% near $\Delta\phi_\text{jj} = \pi$. The model significantly underestimates the observed $f_\text{CSE}$ for small angular separations with $\Delta\phi_\text{jj} < 1$. The EEIM model uses LO $2{\to}2$ hard processes resulting in back-to-back hard jets, such that $\Delta\phi_\text{jj} \approx \pi$, with only small deviations due to the leading logarithmic parton showers, but no hard next-to-LO (NLO) $2\to 3$ processes causing larger deviations.

Present calculations include partial corrections at NLO in $\alpS$ within the BFKL framework, namely resummation of large logarithms of energy at NLL accuracy using LO impact factors. Higher-order corrections to impact factors are known to have significant effects in the description of similar processes, such as Mueller--Navelet jets~\cite{Colferai:2010wu}. Recently, major progress has been made in the calculation of NLO impact factors for the jet-gap-jet process~\cite{hentschinski1,hentschinski2}. These corrections have yet to be included in the BFKL theoretical calculations to complete the NLO analysis of the jet-gap-jet process.

\begin{figure}[tb]
\centering
\includegraphics[width=\cmsFigWidth]{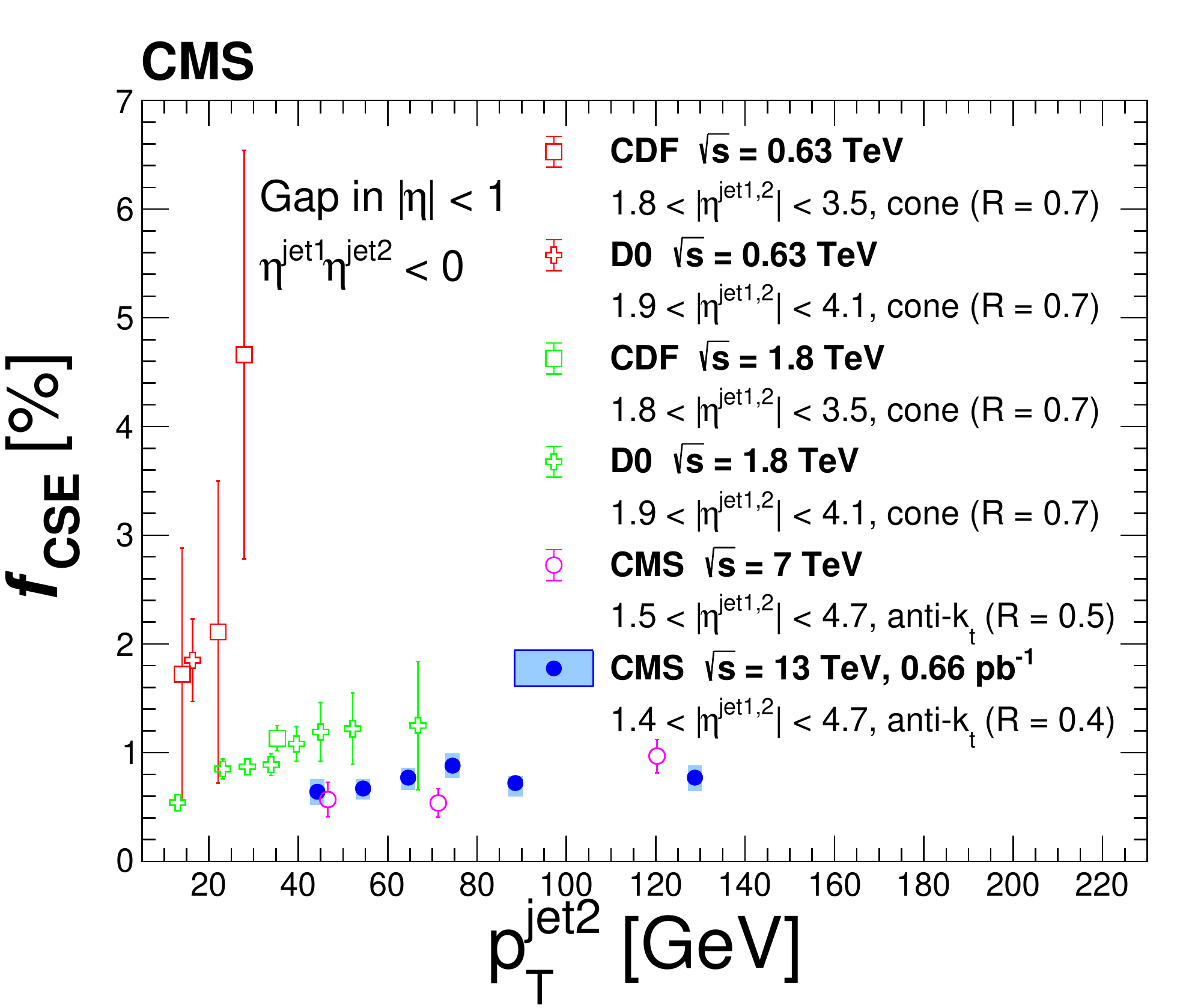}
\caption{ \label{fig:fcse_ptaverage_comparison} Fraction of color-singlet exchange dijet events, $f_\text{CSE}$, measured as a function of $\pt^\text{jet2}$ by the D0 and CDF Collaborations~\cite{d03,cdf2,cdf3} at $\sqrt{s} = 0.63$ (red open symbols) and $1.8\TeV$ (green open symbols), by the CMS Collaboration~\cite{jgjCMS} at $7\TeV$ (magenta open symbols), and the present results at $13\TeV$ (filled circles). The vertical bars of the open symbols represent the total experimental uncertainties. The vertical bars of the $13\TeV$ measurement represent the statistical uncertainties, and boxes represent the combination of statistical and systematic uncertainties in quadrature. The central gap is defined by means of the particle activity in the $\abs{\eta}<1$ interval in these measurements, as described in the text. The jet $\pt$ and $\eta$ requirements of the previous measurements are specified in the legend of the plot. No phase space extrapolations are made in plotting this figure.}
\end{figure}

In Fig.~\ref{fig:fcse_ptaverage_comparison}, the current results are compared with previous measurements of $f_\text{CSE}$ with a central gap in $\abs{\eta}<1$ by the D0 and CDF Collaborations at the Tevatron in $\Pp\PAp$ collisions at $\sqrt{s}=0.63$ and $1.8\TeV$~\cite{d02,d03,cdf2,cdf3}, and by the CMS Collaboration in $\Pp\Pp$ collisions at $7\TeV$~\cite{jgjCMS}. There are differences in the phase space volumes populated by the two leading jets, jet clustering algorithms, and distance parameters, which are described in the next paragraphs. Simulation studies that rely on hadron-level particle distributions, based on the samples described in Section~\ref{subsec:background_inclusive}, indicate that the choice of the jet reconstruction algorithm (cone or anti-\kt algorithms) has a negligible effect on the shape of the charged particle multiplicity distribution between the jets. The value of the distance parameter $R$ influences the charged particle multiplicity distribution shape of jet-gap-jet signal events. For large values of $R$, it is less likely for charged particle constituents of the jet to populate the central $\abs{\eta}<1$ region since the jet axes are further away from the edges of the gap region. This yields a sharper jet-gap-jet signal excess at $N_\text{tracks} = 0$ for large jet distance parameter. At small distance parameter $R$, there is more spillage of charged particles into the gap region, since the jet axes can approach the edge of the $\abs{\eta}<1$ interval more closely. The shape of the multiplicity distribution of color-exchange dijet events remains mostly unaffected by the size of $R$. In these simulation studies, these effects are negligible provided that $f_\text{CSE}$ is extracted over the first multiplicity bins $N_\text{tracks} < 3$, as is done in this measurement.

\begin{figure}[tb]
\centering
\includegraphics[width=\cmsFigWidth]{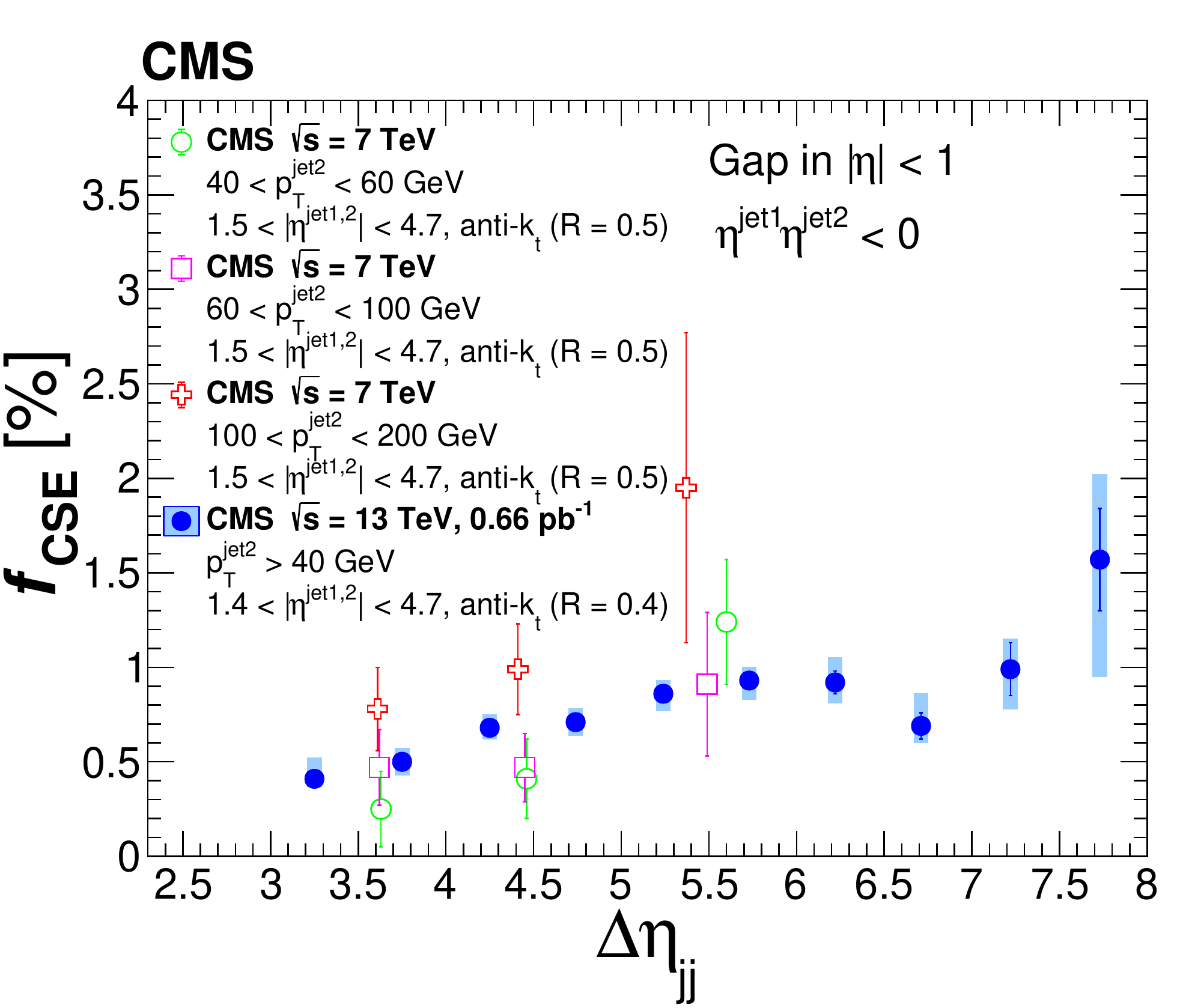}
\caption{ \label{fig:fcse_eta_comparison_cms7TeV} Fraction of color-singlet exchange dijet events, $f_\text{CSE}$, measured as a function of $\Delta\eta_\text{jj}$ by CMS at $7\TeV$~\cite{jgjCMS} and the present measurement at $13$\TeV. The $7$\TeV measurement was performed in three bins of  $\pt^\text{jet2} = 40$--$60$, $60$--$100$, and $100$--$200\GeV$, which are represented by the open circle, open square, and open cross symbols, respectively. The present $13\TeV$ results are represented by the filled circles. The vertical bars of the $7\TeV$ measurement represent the total experimental uncertainties. The vertical bars of the $13\TeV$ measurement represent the statistical uncertainties, and boxes represent the combination of statistical and systematic uncertainties in quadrature.}
\end{figure}

The study by the D0 Collaboration~\cite{d03} uses the calorimeter tower multiplicity distribution in $\abs{\eta}<1$, where each calorimeter tower has transverse energy $\ET>200\MeV$. The $0.63$ and $1.8\TeV$ studies consider jets with $\ET^\text{jet} > 12\GeV$ and $1.9 < \abs{\eta^\text{jet}}<4.1$. The CDF Collaboration measured jet-gap-jet events at $0.63$ and $1.8\TeV$~\cite{cdf2,cdf3}. The $N_\text{tracks}$ value in the region $\abs{\eta}<1$ with $\ET >300\MeV$ is used in the CDF analyses. Each of the two leading jets has $1.8<\abs{\eta^\text{jet}}<3.5$, with $\ET^\text{jet}>8\GeV$ and $>20$\GeV for the $0.63$ and $1.8\TeV$ studies, respectively. The jets are clustered using the cone algorithm with $R = 0.7$ for both CDF and D0 studies. The measurement by CMS at $7\TeV$ is done in three bins of $\pt^\text{jet2} = 40$--$60$, $60$--$100$, and $100$--$200\GeV$~\cite{jgjCMS}. The jets are clustered with the anti-\kt algorithm with $R = 0.5$ in the previous CMS study. Each of the two leading jets has $1.5 < \abs{\eta^\text{jet}} < 4.7$, and the signal extraction is based on the $N_\text{tracks}$ distribution with $\pt > 200\MeV$ in $\abs{\eta}<1$.

In Fig.~\ref{fig:fcse_ptaverage_comparison}, the D0 and CDF Collaborations find that $f_\text{CSE}$ decreases by a factor of $2.5 \pm 0.9$~\cite{d03} and $3.4 \pm 1.2$~\cite{cdf3}, respectively, when $\sqrt{s}$ increases from $0.63$ to $1.8\TeV$. Similarly, the results by the CMS experiment at $7\TeV$ show a $f_\text{CSE}$ that decreases by a factor of around $2$ with respect to the $1.8\TeV$ results at the Tevatron~\cite{jgjCMS}. The observed energy dependence of the previous measurements is generally attributed to a larger number of soft parton interactions with increasing $\sqrt{s}$, which enhances the probability of the gap being destroyed. The $13\TeV$ results show there is no further decrease of the $f_\text{CSE}$ values relative to the $7\TeV$ results, within the uncertainties. This could be an indication that the rapidity gap survival probability stops decreasing at the center-of-mass energies probed at the LHC for the jet-gap-jet process.

The present measurement of $f_\text{CSE}$ expands the reach in $\Delta\eta_\text{jj}$ covered in the earlier $7\TeV$ CMS measurement~\cite{jgjCMS}, as seen in Fig.~\ref{fig:fcse_eta_comparison_cms7TeV}. The measurement of $f_\text{CSE}$ as a function of $\Delta\eta_\text{jj}$ at $7\TeV$ is carried out in three bins of $\Delta\eta_\text{jj} = 3$--$4$, $4$--$5$, and $5$--$7$ units for each bin of $\pt^\text{jet2}$. The dependence of $f_\text{CSE}$ as a function of $\Delta\eta_\text{jj}$ at $13\TeV$ confirms the trend observed by CMS at $7\TeV$ and extends the range previously explored towards large values of $6.5 < \Delta\eta_\text{jj} < 8$.

\subsection{Results for jet-gap-jet events with an intact proton}\label{sec:results_gjgj}

\begin{figure}[htbp!]
\centering
\includegraphics[width=0.49\textwidth]{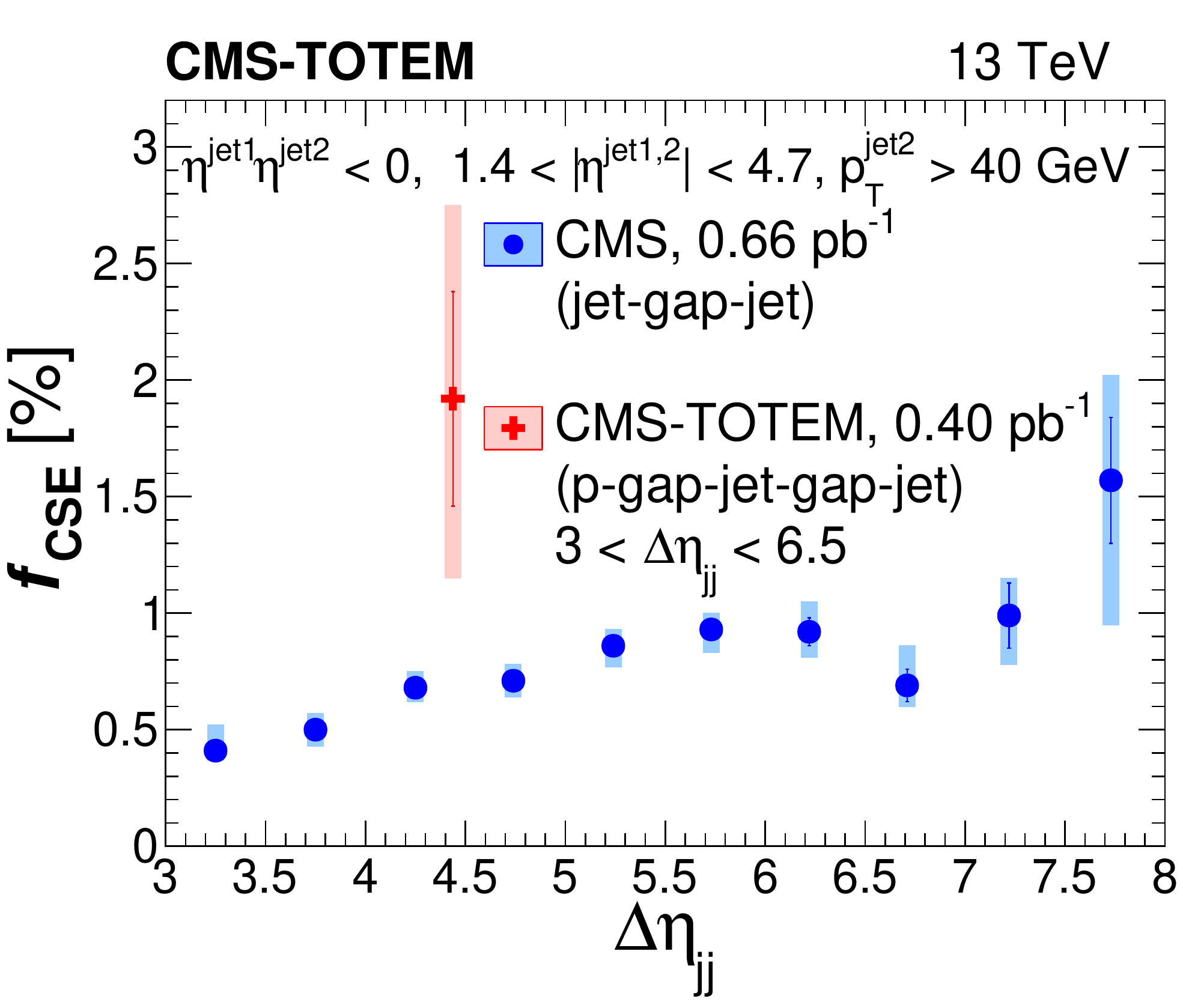}
\includegraphics[width=0.49\textwidth]{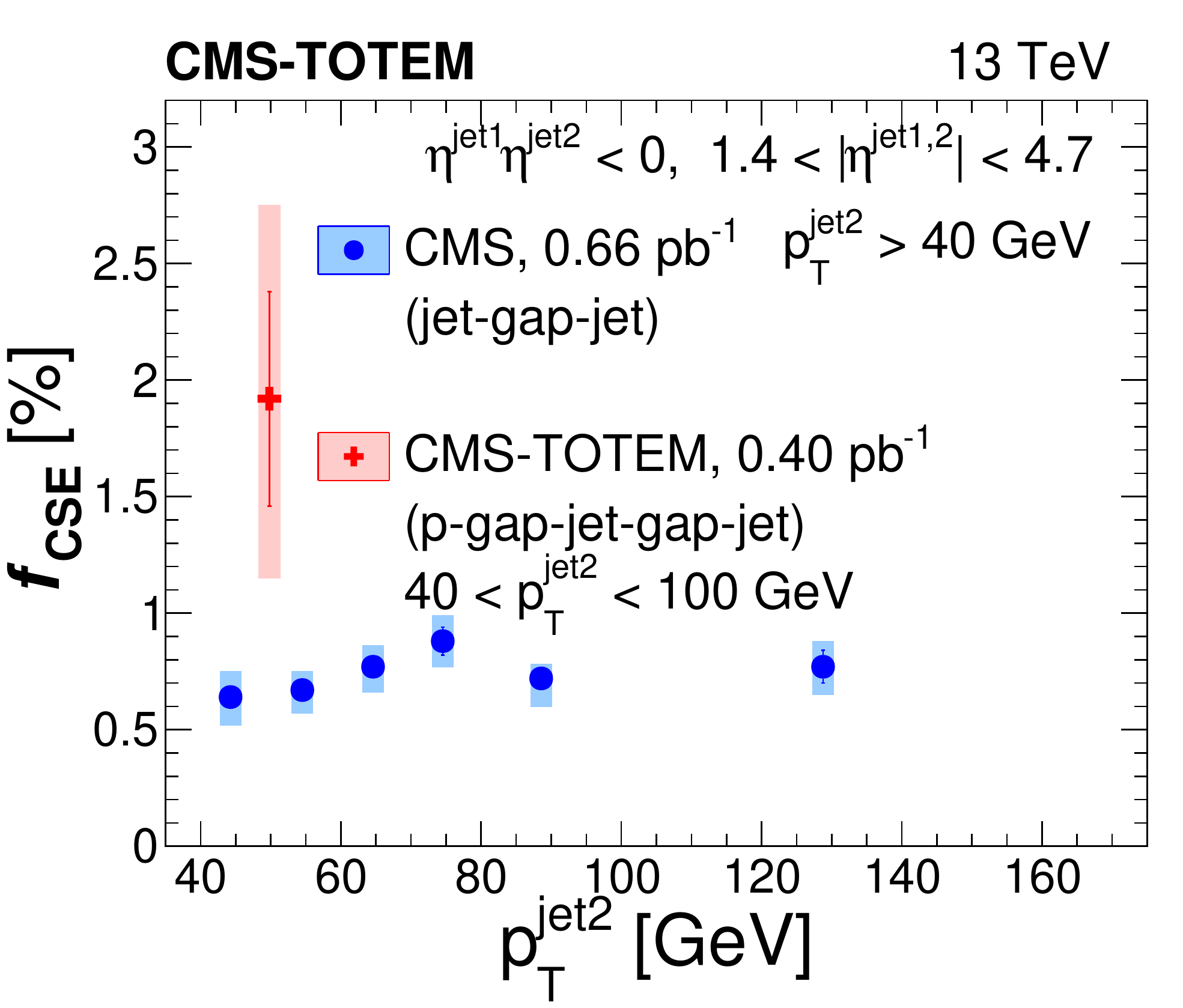}
\caption{\label{fig:gapfraction_proton} Fraction of hard color-singlet exchange dijet events $f_\text{CSE}$, measured as a function of $\Delta\eta_\text{jj}$ (\cmsLeft) and $\pt^\text{jet2}$ (\cmsRight) extracted in inclusive dijet event production (labeled CMS, represented by the blue circle markers) and in dijet events with an intact proton at $13\TeV$ (labeled CMS-TOTEM, represented by the red cross marker). The vertical bars represent the statistical uncertainties, and boxes represent the combination of statistical and systematic uncertainties in quadrature. The CMS results are plotted at the mean values of $\Delta\eta_\text{jj}$ and $\pt^\text{jet2}$ in the bin. Similarly, the CMS-TOTEM result is plotted at the mean value of $\Delta\eta_\text{jj}$ and $\pt^\text{jet2}$ in the CMS-TOTEM combined sample. The $40 < \pt^\text{jet2} < 100\GeV$ and $3.0 < \Delta\eta_\text{jj} < 6.5$ ranges below the CMS-TOTEM legend represent the dijet phase space covered by events with an intact proton with the present sample size, rather than a selection requirement, as described in the text.}
\end{figure}

The fraction $f_\text{CSE}$ in events with intact protons is $f_\text{CSE} = [1.92 \pm 0.46\stat ^{+0.69}_{-0.62 } \syst ]$\%. Although the dijet events with an intact proton cover the same phase space as those in the inclusive dijet analysis, most of the events used in the study populate the regions $3.0 < \Delta\eta_\text{jj} < 6.5$ and $40 < \pt^\text{jet2} < 100\GeV$ because of the limited sample size of events with intact protons. The fraction $f_\text{CSE}$ in events with an intact proton is $2.91 \pm 0.70$ \stat $^{+ 1.08}_{- 1.01} \syst$ times larger than that extracted for inclusive dijet production, where the two leading jets have similar kinematics to events with an intact proton, \ie, $40 < \pt^\text{jet2} < 100\GeV$ and $3.0 < \Delta \eta_\text{jj} < 6.5$ for jet-gap-jet events considered in the aforementioned double ratio calculation. The $f_\text{CSE}$ ratio in the latter jet-gap-jet subsample has a value of $f_\text{CSE} = [0.66 \pm 0.01\stat ^{+0.06}_{-0.09}\syst]$\%. Correlations of systematic uncertainties associated with jet reconstruction and central gap definition are included when evaluating the uncertainties in the double ratios. Statistical and systematic uncertainties in the double ratio are largely dominated by the uncertainties in the CMS-TOTEM $f_\text{CSE}$ measurement. The CMS-TOTEM results, when compared with the CMS results extracted in inclusive dijet production, suggest that the relative abundance of dijet events with a central gap is larger in events with an intact proton. This is illustrated in Fig.~\ref{fig:gapfraction_proton}, where the results for $f_\text{CSE}$ are presented as a function of $\Delta\eta_\text{jj}$ and $\pt^\text{jet2}$.

The larger $f_\text{CSE}$ value in events with an intact proton may reflect a reduced spectator parton activity in reactions with an intact proton in comparison to the soft parton activity present in interactions where the proton breaks up. In the latter, there can be soft parton exchanges between the proton remnants and partons produced in the collision, which can destroy the central gap signature between the final-state jets. A similar effect has been observed in other diffractive topologies in dijet events with two rapidity gaps by the CDF Collaboration at $\sqrt{s} = 1.8\TeV$~\cite{PhysRevLett.85.4215}. In the CDF measurement, comparisons are made of: (i) the ratio of yields of single-diffractive dijet events to nondiffractive dijet events, $R^\text{SD}_\text{ND}$, and (ii) the ratio of double-pomeron exchange dijet events to single-diffractive dijet events, $R^\text{DPE}_\text{SD}$. CDF finds that the double ratio has a value of $R^\text{SD}_\text{ND}/R^\text{DPE}_\text{SD} = 0.19 \pm 0.07$~\cite{PhysRevLett.85.4215}. An analogous double ratio for the present measurement is $f_\text{CSE} \text{(jet-gap-jet)}/f_\text{CSE}\text{(p-gap-jet-gap-jet)} = 0.34 \pm 0.08\stat ^{+0.12}_{-0.13}\syst$, which is similar to that for the double-pomeron exchange dijet topology reported by CDF. The present results further suggest that a gap is more likely to form or survive in the presence of another gap.

\section{Summary}\label{sec:summary}

Events with two leading jets separated by a large pseudorapidity ($\eta$) gap have been studied in proton-proton ($\Pp\Pp$) collisions at $\sqrt{s}=13\TeV$ with the CMS and TOTEM experiments at the CERN LHC in 2015. The pseudorapidity gap is defined by the absence of charged particles with transverse momentum $\pt>200\MeV$ in the $\abs{\eta}<1$ region. Each of the two leading $\pt$ jets has $1.4 < \abs{\eta^\text{jet}} < 4.7$ and $\pt^\text{jet} > 40\GeV$, with $\eta^\text{jet1}   \eta^\text{jet2} < 0$, where $\text{jet1}$ and $\text{jet2}$ are the leading and subleading jets in $\pt$. The pseudorapidity gap signature is assumed to be caused by hard color-singlet exchange, which is described in terms of two-gluon exchange in perturbative quantum chromodynamics. Color-singlet exchange events appear as an excess of events over the expected charged particle multiplicity contribution from color-exchange dijet events at the lowest charged particle multiplicity. The ratio of color-singlet exchange events to all dijet events, $f_\text{CSE}$, has been measured as a function of $\pt^\text{jet2}$, the $\eta$ difference between the two leading jets, $\Delta\eta_\text{jj} \equiv \abs{\eta^\text{jet1} - \eta^\text{jet2}}$, and the azimuthal angular separation between the two leading jets, $\Delta\phi_\text{jj} \equiv \abs{\phi^\text{jet1} - \phi^\text{jet2}}$.

The measured $f_\text{CSE}$ values are in the range of $0.4$--$1.0$\%. The ratio $f_\text{CSE}$ increases with $\Delta\eta_\text{jj}$, has a weak dependence on $\pt^\text{jet2}$, and increases as $\Delta\phi_\text{jj}$ approaches $\pi$. No significant difference in $f_\text{CSE}$ is observed between the $13\TeV$ results and those presented by the CMS Collaboration at $7\TeV$. This is in contrast to the trend found at lower energies of $0.63$ and $1.8\TeV$ by the D0 and CDF Collaborations, where a significant decrease of $f_\text{CSE}$ with increasing $\sqrt{s}$ was observed, as illustrated in Fig.~\ref{fig:fcse_ptaverage_comparison}. The results are compared with calculations based on the Balitsky--Fadin--Kuraev--Lipatov framework \cite{Kuraev:1977fs, Balitsky:1978ic,Lipatov:1985uk} with resummation of large logarithms of energy at next-to-leading logarithmic accuracy using leading order impact factors, and various treatments of gap survival probability effects. The implementation by Royon, Marquet, and Kepka \cite{Chevallier:2009cu, Kepka:2010hu} describes some features of the data, but is not able to simultaneously describe all aspects of the measurements. The implementation by Ekstedt, Enberg, Ingelman, and Motyka \cite{csp,cspLHC} gives a fair description of the data in $\Delta\eta_\text{jj}$ and $\pt^\text{jet2}$ within the uncertainties only when considering survival probability effects based on multiple-parton interactions and their soft color interaction model.

In addition, a sample of dijet events with intact protons collected by the CMS and TOTEM experiments is used to study jet-gap-jet events with intact protons, which correspond to proton-gap-jet-gap-jet topologies. This is the first analysis of this diffractive event topology. The $f_\text{CSE}$ value extracted in this sample is $2.91 \pm 0.70\stat ^{+ 1.08}_{- 1.01}\syst$ times larger than that found in inclusive dijet production, suggesting a larger abundance of jets with central gaps in events with detected intact protons. This can be interpreted in terms of a lower spectator parton activity in events with intact protons, which decreases the likelihood of the central gap signature being spoiled.

\begin{acknowledgments}

We thank Andreas Ekstedt, Rikard Enberg, Gunnar Ingelman, Leszek Motyka and Cyrille Marquet, Oldrich Kepka for providing the BFKL predictions of their respective models.

We congratulate our colleagues in the CERN accelerator departments for the excellent performance of the LHC and thank the technical and administrative staffs at CERN and at other CMS and TOTEM institutes for their contributions to the success of the CMS-TOTEM effort. In addition, we gratefully acknowledge the computing centers and personnel of the Worldwide LHC Computing Grid and other centers for delivering so effectively the computing infrastructure essential to our analyses. Finally, we acknowledge the enduring support for the construction and operation of the LHC, the CMS and TOTEM detectors, and the supporting computing infrastructure provided by the following funding agencies: BMBWF and FWF (Austria); FNRS and FWO (Belgium); CNPq, CAPES, FAPERJ, FAPERGS, and FAPESP (Brazil); MES (Bulgaria); CERN; CAS, MoST, and NSFC (China); COLCIENCIAS (Colombia); MSES and CSF (Croatia); RIF (Cyprus); SENESCYT (Ecuador); MoER, ERC PUT and ERDF (Estonia); Academy of Finland, Finnish Academy of Science and Letters (The Vilho Yrjö and Kalle V\"ais\"al\"a Fund), MEC, Magnus Ehrnrooth Foundation, HIP, and Waldemar von Frenckell Foundation (Finland); CEA and CNRS/IN2P3 (France); BMBF, DFG, and HGF (Germany); GSRT (Greece); the Circles of Knowledge Club and NKFIA (Hungary); DAE and DST (India); IPM (Iran); SFI (Ireland); INFN (Italy); MSIP and NRF (Republic of Korea); MES (Latvia); LAS (Lithuania); MOE and UM (Malaysia); BUAP, CINVESTAV, CONACYT, LNS, SEP, and UASLP-FAI (Mexico); MOS (Montenegro); MBIE (New Zealand); PAEC (Pakistan); MSHE and NSC (Poland); FCT (Portugal); JINR (Dubna); MON, RosAtom, RAS, RFBR, and NRC KI (Russia); MESTD (Serbia); SEIDI, CPAN, PCTI, and FEDER (Spain); MOSTR (Sri Lanka); Swiss Funding Agencies (Switzerland); MST (Taipei); ThEPCenter, IPST, STAR, and NSTDA (Thailand); TUBITAK and TAEK (Turkey); NASU (Ukraine); STFC (United Kingdom); DOE and NSF (USA).

\hyphenation{Rachada-pisek} Individuals have received support from the Marie-Curie program and the European Research Council and Horizon 2020 Grant, contract Nos.\ 675440, 724704, 752730, and 765710 (European Union); the Leventis Foundation; the Alfred P.\ Sloan Foundation; the Alexander von Humboldt Foundation; the Belgian Federal Science Policy Office; the Fonds pour la Formation \`a la Recherche dans l'Industrie et dans l'Agriculture (FRIA-Belgium); the Agentschap voor Innovatie door Wetenschap en Technologie (IWT-Belgium); the F.R.S.-FNRS and FWO (Belgium) under the ``Excellence of Science -- EOS" -- be.h project n.\ 30820817; the Beijing Municipal Science \& Technology Commission, No. Z191100007219010; the Ministry of Education, Youth and Sports (MEYS) and MSMT CR of the Czech Republic; the Nylands nation vid Helsingfors universitet (Finland); the Deutsche Forschungsgemeinschaft (DFG), under Germany's Excellence Strategy -- EXC 2121 ``Quantum Universe" -- 390833306, and under project number 400140256 - GRK2497; the Lend\"ulet (``Momentum") Program and the J\'anos Bolyai Research Scholarship of the Hungarian Academy of Sciences, the New National Excellence Program \'UNKP, the NKFIA research grants 123842, 123959, 124845, 124850, 125105, 128713, 128786, 129058, K 133046, and EFOP-3.6.1-16-2016-00001  (Hungary); the Council of Science and Industrial Research, India; the HOMING PLUS program of the Foundation for Polish Science, cofinanced from European Union, Regional Development Fund, the Mobility Plus program of the Ministry of Science and Higher Education, including Grant No. MNiSW DIR/WK/2018/13, the National Science Center (Poland), contracts Harmonia 2014/14/M/ST2/00428, Opus 2014/13/B/ST2/02543, 2014/15/B/ST2/03998, and 2015/19/B/ST2/02861, Sonata-bis 2012/07/E/ST2/01406; the National Priorities Research Program by Qatar National Research Fund; the Ministry of Science and Higher Education, project no. 0723-2020-0041 (Russia); the Tomsk Polytechnic University Competitiveness Enhancement Program; the Programa Estatal de Fomento de la Investigaci{\'o}n Cient{\'i}fica y T{\'e}cnica de Excelencia Mar\'{\i}a de Maeztu, grant MDM-2015-0509 and the Programa Severo Ochoa del Principado de Asturias; the Thalis and Aristeia programs cofinanced by EU-ESF and the Greek NSRF; the Rachadapisek Sompot Fund for Postdoctoral Fellowship, Chulalongkorn University and the Chulalongkorn Academic into Its 2nd Century Project Advancement Project (Thailand); the Kavli Foundation; the Nvidia Corporation; the SuperMicro Corporation; the Welch Foundation, contract C-1845; and the Weston Havens Foundation (USA).\end{acknowledgments}

\bibliography{auto_generated}

\providecommand{\href}[2]{#2}\begingroup\raggedright\begin{thebibliography}{100}%
\makeatletter
\providecommand{\hrefCMSnoop }[0]{\@secondoftwo}%
\makeatother
\providecommand{\doi}{\texttt{doi:}\begingroup \urlstyle{tt}\Url}

\bibitem{PDG}
\hrefCMSnoop {}{{Particle Data Group}, P.~A. Zyla {et~al.}, ``Review of
  particle physics'',} \textit{ Prog. Theor. Exp. Phys.} \textbf{ 2020} (2020)
  083C01,
  \href{http://dx.doi.org/10.1093/ptep/ptaa104}{\doi{10.1093/ptep/ptaa104}}.

\bibitem{Akiba:2016ofq}
K.~Akiba\hrefCMSnoop {}{ {et~al.}, ``{LHC forward physics}'',} \textit{ J.
  Phys. G} \textbf{ 43} (2016) 110201,
  \href{http://dx.doi.org/10.1088/0954-3899/43/11/110201}{\doi{10.1088/0954-3899/43/11/110201}},
  \href{http://www.arXiv.org/abs/1611.05079}{\texttt{arXiv:1611.05079}}.

\bibitem{Kuraev:1977fs}
\hrefCMSnoop {}{E.~A. Kuraev, L.~N. Lipatov, and V.~S. Fadin, ``{The
  Pomeranchuk singularity in nonabelian gauge theories}'',} \textit{ Sov. Phys.
  JETP} \textbf{ 45} (1977)
199.

\bibitem{Balitsky:1978ic}
\hrefCMSnoop {}{I.~I. Balitsky and L.~N. Lipatov, ``{The Pomeranchuk
  singularity in quantum chromodynamics}'',} \textit{ Sov. J. Nucl. Phys.}
  \textbf{ 28} (1978)
822.

\bibitem{Lipatov:1985uk}
\hrefCMSnoop {}{L.~N. Lipatov, ``{The bare pomeron in quantum
  chromodynamics}'',} \textit{ Sov. Phys. JETP} \textbf{ 63} (1986)
904.

\bibitem{Fadin:1998py}
\hrefCMSnoop {}{V.~S. Fadin and L.~N. Lipatov, ``{BFKL pomeron in the
  next-to-leading approximation}'',} \textit{ Phys. Lett. B} \textbf{ 429}
  (1998) 127,
  \href{http://dx.doi.org/10.1016/S0370-2693(98)00473-0}{\doi{10.1016/S0370-2693(98)00473-0}},
\href{http://www.arXiv.org/abs/hep-ph/9802290}{\texttt{arXiv:hep-ph/9802290}}.

\bibitem{Ciafaloni:1998gs}
\hrefCMSnoop {}{M.~Ciafaloni and G.~Camici, ``{Energy scale(s) and
  next-to-leading BFKL equation}'',} \textit{ Phys. Lett. B} \textbf{ 430}
  (1998) 349,
  \href{http://dx.doi.org/10.1016/S0370-2693(98)00551-6}{\doi{10.1016/S0370-2693(98)00551-6}},
\href{http://www.arXiv.org/abs/hep-ph/9803389}{\texttt{arXiv:hep-ph/9803389}}.

\bibitem{Khachatryan:2016udy}
\hrefCMSnoop {}{{CMS Collaboration}, ``{Azimuthal decorrelation of jets widely
  separated in rapidity in pp collisions at $ \sqrt{s}=7 $ TeV}'',} \textit{
  JHEP} \textbf{ 08} (2016) 139,
  \href{http://dx.doi.org/10.1007/JHEP08(2016)139}{\doi{10.1007/JHEP08(2016)139}},
\href{http://www.arXiv.org/abs/1601.06713}{\texttt{arXiv:1601.06713}}.

\bibitem{Aad:2011jz}
\hrefCMSnoop {}{{ATLAS Collaboration}, ``{Measurement of dijet production with
  a veto on additional central jet activity in pp collisions at {$\sqrt{s}=7$
  TeV} using the {ATLAS} detector}'',} \textit{ JHEP} \textbf{ 09} (2011) 053,
  \href{http://dx.doi.org/10.1007/JHEP09(2011)053}{\doi{10.1007/JHEP09(2011)053}},
\href{http://www.arXiv.org/abs/1107.1641}{\texttt{arXiv:1107.1641}}.

\bibitem{Chatrchyan:2012pb}
\hrefCMSnoop {}{{CMS Collaboration}, ``Ratios of dijet production cross
  sections as a function of the absolute difference in rapidity between jets in
  proton-proton collisions at {$\sqrt{s}=7$ TeV}'',} \textit{ Eur. Phys. J. C}
  \textbf{ 72} (2012) 2216,
  \href{http://dx.doi.org/10.1140/epjc/s10052-012-2216-6}{\doi{10.1140/epjc/s10052-012-2216-6}},
\href{http://www.arXiv.org/abs/1204.0696}{\texttt{arXiv:1204.0696}}.

\bibitem{Aaij:2014iea}
\hrefCMSnoop {}{{LHCb Collaboration}, ``{Updated measurements of exclusive
  \JPsi and $\psi$(2S) production cross sections in pp collisions at
  $\sqrt{s}=7$ TeV}'',} \textit{ J. Phys. G} \textbf{ 41} (2014) 055002,
  \href{http://dx.doi.org/10.1088/0954-3899/41/5/055002}{\doi{10.1088/0954-3899/41/5/055002}},
\href{http://www.arXiv.org/abs/1401.3288}{\texttt{arXiv:1401.3288}}.

\bibitem{Aaij:2015kea}
\hrefCMSnoop {}{{LHCb Collaboration}, ``{Measurement of the exclusive
  $\Upsilon$ production cross section in pp collisions at $ {\sqrt{s}=7} $ TeV
  and 8 TeV}'',} \textit{ JHEP} \textbf{ 09} (2015) 084,
  \href{http://dx.doi.org/10.1007/JHEP09(2015)084}{\doi{10.1007/JHEP09(2015)084}},
\href{http://www.arXiv.org/abs/1505.08139}{\texttt{arXiv:1505.08139}}.

\bibitem{Abelev:2012ba}
\hrefCMSnoop {}{{ALICE Collaboration}, ``{Coherent \JPsi photoproduction in
  ultraperipheral Pb-Pb collisions at $\sqrt{s_\text{NN}} = 2.76$ TeV}'',}
  \textit{ Phys. Lett. B} \textbf{ 718} (2013) 1273,
  \href{http://dx.doi.org/10.1016/j.physletb.2012.11.059}{\doi{10.1016/j.physletb.2012.11.059}},
\href{http://www.arXiv.org/abs/1209.3715}{\texttt{arXiv:1209.3715}}.

\bibitem{TheALICE:2014dwa}
\hrefCMSnoop {}{{ALICE Collaboration}, ``{Exclusive \JPsi photoproduction off
  protons in ultraperipheral p-Pb collisions at $\sqrt{s_{\rm NN}}=5.02$
  TeV}'',} \textit{ Phys. Rev. Lett.} \textbf{ 113} (2014) 232504,
  \href{http://dx.doi.org/10.1103/PhysRevLett.113.232504}{\doi{10.1103/PhysRevLett.113.232504}},
\href{http://www.arXiv.org/abs/1406.7819}{\texttt{arXiv:1406.7819}}.

\bibitem{Adam:2015gsa}
\hrefCMSnoop {}{{ALICE Collaboration}, ``{Coherent $\rho^{0}$ photoproduction
  in ultraperipheral Pb-Pb collisions at $ \sqrt{s_{\mathrm{NN}}}=2.76 $
  TeV}'',} \textit{ JHEP} \textbf{ 09} (2015) 095,
  \href{http://dx.doi.org/10.1007/JHEP09(2015)095}{\doi{10.1007/JHEP09(2015)095}},
\href{http://www.arXiv.org/abs/1503.09177}{\texttt{arXiv:1503.09177}}.

\bibitem{Sirunyan:2018sav}
\hrefCMSnoop {}{{CMS Collaboration}, ``{Measurement of exclusive $\Upsilon$
  photoproduction from protons in pPb collisions at $\sqrt{s_\mathrm{NN}} =$
  5.02 TeV}'',} \textit{ Eur. Phys. J. C} \textbf{ 79} (2019) 277,
  \href{http://dx.doi.org/10.1140/epjc/s10052-019-6774-8}{\doi{10.1140/epjc/s10052-019-6774-8}},
\href{http://www.arXiv.org/abs/1809.11080}{\texttt{arXiv:1809.11080}}.

\bibitem{Sirunyan:2019nog}
\hrefCMSnoop {}{{CMS Collaboration}, ``{Measurement of exclusive $\rho(770)^0$
  photoproduction in ultraperipheral pPb collisions at $\sqrt{s_\mathrm{NN}} =
  5.02$ {TeV}}'',} \textit{ Eur. Phys. J. C} \textbf{ 79} (2019) 702,
  \href{http://dx.doi.org/10.1140/epjc/s10052-019-7202-9}{\doi{10.1140/epjc/s10052-019-7202-9}},
\href{http://www.arXiv.org/abs/1902.01339}{\texttt{arXiv:1902.01339}}.

\bibitem{Bautista:2016xnp}
\hrefCMSnoop {}{I.~Bautista, A.~Fern\'{a}ndez~T\'{e}llez, and M.~Hentschinski,
  ``{BFKL evolution and the growth with energy of exclusive $J/\psi$ and
  $\Upsilon$ photoproduction cross sections}'',} \textit{ Phys. Rev. D}
  \textbf{ 94} (2016) 054002,
  \href{http://dx.doi.org/10.1103/PhysRevD.94.054002}{\doi{10.1103/PhysRevD.94.054002}},
  \href{http://www.arXiv.org/abs/1607.05203}{\texttt{arXiv:1607.05203}}.

\bibitem{Garcia:2019tne}
\hrefCMSnoop {}{A.~Arroyo~Garc\'{i}a, M.~Hentschinski, and K.~Kutak, ``{QCD
  evolution based evidence for the onset of gluon saturation in exclusive
  photo-production of vector mesons}'',} \textit{ Phys. Lett. B} \textbf{ 795}
  (2019) 569,
  \href{http://dx.doi.org/10.1016/j.physletb.2019.06.061}{\doi{10.1016/j.physletb.2019.06.061}},
  \href{http://www.arXiv.org/abs/1904.04394}{\texttt{arXiv:1904.04394}}.

\bibitem{dglap1}
\hrefCMSnoop {}{V.~N. Gribov and L.~N. Lipatov, ``{Deep inelastic ep scattering
  in perturbation theory}'',} \textit{ Sov. J. Nucl. Phys.} \textbf{ 15} (1972)
438.

\bibitem{dglap2}
\hrefCMSnoop {}{G.~Altarelli and G.~Parisi, ``{Asymptotic freedom in parton
  language}'',} \textit{ Nucl. Phys. B} \textbf{ 126} (1977) 298,
\href{http://dx.doi.org/10.1016/0550-3213(77)90384-4}{\doi{10.1016/0550-3213(77)90384-4}}.

\bibitem{dglap3}
\hrefCMSnoop {}{Y.~L. Dokshitzer, ``{Calculation of the structure functions for
  deep inelastic scattering and e$^{+}$ e$^{-}$ annihilation by perturbation
  theory in quantum chromodynamics}'',} \textit{ Sov. Phys. JETP} \textbf{ 46}
  (1977)
641.

\bibitem{Chatrchyan:2011qta}
\hrefCMSnoop {}{{CMS Collaboration}, ``{Measurement of the differential dijet
  production cross section in proton-proton collisions at {$\sqrt{s}=7$
  TeV}}'',} \textit{ Phys. Lett. B} \textbf{ 700} (2011) 187,
  \href{http://dx.doi.org/10.1016/j.physletb.2011.05.027}{\doi{10.1016/j.physletb.2011.05.027}},
\href{http://www.arXiv.org/abs/1104.1693}{\texttt{arXiv:1104.1693}}.

\bibitem{Chatrchyan:2012gwa}
\hrefCMSnoop {}{{CMS Collaboration}, ``{Measurement of the inclusive production
  cross sections for forward jets and for dijet events with one forward and one
  central jet in pp collisions at $\sqrt{s}=7$ TeV}'',} \textit{ JHEP} \textbf{
  06} (2012) 036,
  \href{http://dx.doi.org/10.1007/JHEP06(2012)036}{\doi{10.1007/JHEP06(2012)036}},
\href{http://www.arXiv.org/abs/1202.0704}{\texttt{arXiv:1202.0704}}.

\bibitem{Khachatryan:2015luy}
\hrefCMSnoop {}{{CMS Collaboration}, ``{Measurement of the inclusive jet cross
  section in pp collisions at $\sqrt{s} = 2.76\,\text {TeV}$}'',} \textit{ Eur.
  Phys. J. C} \textbf{ 76} (2016) 265,
  \href{http://dx.doi.org/10.1140/epjc/s10052-016-4083-z}{\doi{10.1140/epjc/s10052-016-4083-z}},
\href{http://www.arXiv.org/abs/1512.06212}{\texttt{arXiv:1512.06212}}.

\bibitem{Khachatryan:2016hkr}
\hrefCMSnoop {}{{CMS Collaboration}, ``Measurement of dijet azimuthal
  decorrelation in pp collisions at {$\sqrt{s}=8$ TeV}'',} \textit{ Eur. Phys.
  J. C} \textbf{ 76} (2016) 536,
  \href{http://dx.doi.org/10.1140/epjc/s10052-016-4346-8}{\doi{10.1140/epjc/s10052-016-4346-8}},
\href{http://www.arXiv.org/abs/1602.04384}{\texttt{arXiv:1602.04384}}.

\bibitem{Khachatryan:2016wdh}
\hrefCMSnoop {}{{CMS Collaboration}, ``{Measurement of the double-differential
  inclusive jet cross section in proton-proton collisions at $\sqrt{s} =
  13\,\text {TeV} $}'',} \textit{ Eur. Phys. J. C} \textbf{ 76} (2016) 451,
  \href{http://dx.doi.org/10.1140/epjc/s10052-016-4286-3}{\doi{10.1140/epjc/s10052-016-4286-3}},
\href{http://www.arXiv.org/abs/1605.04436}{\texttt{arXiv:1605.04436}}.

\bibitem{Khachatryan:2016mlc}
\hrefCMSnoop {}{{CMS Collaboration}, ``{Measurement and QCD analysis of
  double-differential inclusive jet cross sections in pp collisions at $
  \sqrt{s}=8 $ TeV and cross section ratios to 2.76 and 7 TeV}'',} \textit{
  JHEP} \textbf{ 03} (2017) 156,
  \href{http://dx.doi.org/10.1007/JHEP03(2017)156}{\doi{10.1007/JHEP03(2017)156}},
\href{http://www.arXiv.org/abs/1609.05331}{\texttt{arXiv:1609.05331}}.

\bibitem{Sirunyan:2017jnl}
\hrefCMSnoop {}{{CMS Collaboration}, ``{Azimuthal correlations for inclusive
  2-jet, 3-jet, and 4-jet events in pp collisions at $\sqrt{s}= $ 13 TeV}'',}
  \textit{ Eur. Phys. J. C} \textbf{ 78} (2018) 566,
  \href{http://dx.doi.org/10.1140/epjc/s10052-018-6033-4}{\doi{10.1140/epjc/s10052-018-6033-4}},
\href{http://www.arXiv.org/abs/1712.05471}{\texttt{arXiv:1712.05471}}.

\bibitem{Sirunyan:2018ffo}
\hrefCMSnoop {}{{CMS Collaboration}, ``{Measurement of inclusive very forward
  jet cross sections in proton-lead collisions at $ \sqrt{s_{\mathrm{NN}}} $ =
  5.02 TeV}'',} \textit{ JHEP} \textbf{ 05} (2019) 043,
  \href{http://dx.doi.org/10.1007/JHEP05(2019)043}{\doi{10.1007/JHEP05(2019)043}},
\href{http://www.arXiv.org/abs/1812.01691}{\texttt{arXiv:1812.01691}}.

\bibitem{Aad:2010ad}
\hrefCMSnoop {}{{ATLAS Collaboration}, ``{Measurement of inclusive jet and
  dijet cross sections in proton-proton collisions at {7 TeV} centre-of-mass
  energy with the {ATLAS} detector}'',} \textit{ Eur. Phys. J. C} \textbf{ 71}
  (2011) 1512,
  \href{http://dx.doi.org/10.1140/epjc/s10052-010-1512-2}{\doi{10.1140/epjc/s10052-010-1512-2}},
\href{http://www.arXiv.org/abs/1009.5908}{\texttt{arXiv:1009.5908}}.

\bibitem{Aad:2011fc}
\hrefCMSnoop {}{{ATLAS Collaboration}, ``{Measurement of inclusive jet and
  dijet production in pp collisions at {$\sqrt{s}=7$ TeV} using the {ATLAS}
  detector}'',} \textit{ Phys. Rev. D} \textbf{ 86} (2012) 014022,
  \href{http://dx.doi.org/10.1103/PhysRevD.86.014022}{\doi{10.1103/PhysRevD.86.014022}},
\href{http://www.arXiv.org/abs/1112.6297}{\texttt{arXiv:1112.6297}}.

\bibitem{Aad:2013tea}
\hrefCMSnoop {}{{ATLAS Collaboration}, ``{Measurement of dijet cross sections
  in pp collisions at {7 TeV} centre-of-mass energy using the {ATLAS}
  detector}'',} \textit{ JHEP} \textbf{ 05} (2014) 059,
  \href{http://dx.doi.org/10.1007/JHEP05(2014)059}{\doi{10.1007/JHEP05(2014)059}},
\href{http://www.arXiv.org/abs/1312.3524}{\texttt{arXiv:1312.3524}}.

\bibitem{Aad:2014pua}
\hrefCMSnoop {}{{ATLAS Collaboration}, ``Measurements of jet vetoes and
  azimuthal decorrelations in dijet events produced in pp collisions at
  {$\sqrt{s}=7$ TeV} using the {ATLAS} detector'',} \textit{ Eur. Phys. J. C}
  \textbf{ 74} (2014) 3117,
  \href{http://dx.doi.org/10.1140/epjc/s10052-014-3117-7}{\doi{10.1140/epjc/s10052-014-3117-7}},
\href{http://www.arXiv.org/abs/1407.5756}{\texttt{arXiv:1407.5756}}.

\bibitem{Aaboud:2017dvo}
\hrefCMSnoop {}{{ATLAS Collaboration}, ``{Measurement of the inclusive jet
  cross sections in proton-proton collisions at $\sqrt{s}=8$ TeV with the
  {ATLAS} detector}'',} \textit{ JHEP} \textbf{ 09} (2017) 020,
  \href{http://dx.doi.org/10.1007/JHEP09(2017)020}{\doi{10.1007/JHEP09(2017)020}},
\href{http://www.arXiv.org/abs/1706.03192}{\texttt{arXiv:1706.03192}}.

\bibitem{Aaboud:2018hie}
\hrefCMSnoop {}{{ATLAS Collaboration}, ``{Measurement of dijet azimuthal
  decorrelations in pp collisions at $\sqrt{s}=8$ TeV with the ATLAS detector
  and determination of the strong coupling}'',} \textit{ Phys. Rev. D} \textbf{
  98} (2018) 092004,
  \href{http://dx.doi.org/10.1103/PhysRevD.98.092004}{\doi{10.1103/PhysRevD.98.092004}},
\href{http://www.arXiv.org/abs/1805.04691}{\texttt{arXiv:1805.04691}}.

\bibitem{Ball:2017otu}
R.~D. Ball\hrefCMSnoop {}{ {et~al.}, ``{Parton distributions with small-$x$
  resummation: evidence for BFKL dynamics in HERA data}'',} \textit{ Eur. Phys.
  J. C} \textbf{ 78} (2018) 321,
  \href{http://dx.doi.org/10.1140/epjc/s10052-018-5774-4}{\doi{10.1140/epjc/s10052-018-5774-4}},
\href{http://www.arXiv.org/abs/1710.05935}{\texttt{arXiv:1710.05935}}.

\bibitem{MUELLER1992123}
\hrefCMSnoop {}{A.~H. Mueller and W.~K. Tang, ``{H}igh energy parton-parton
  elastic scattering in {QCD}'',} \textit{ Phys. Lett. B} \textbf{ 284} (1992)
  123,
  \href{http://dx.doi.org/10.1016/0370-2693(92)91936-4}{\doi{10.1016/0370-2693(92)91936-4}}.

\bibitem{h1}
\hrefCMSnoop {}{{H1} Collaboration, ``{Energy flow and rapidity gaps between
  jets in photoproduction at HERA}'',} \textit{ Eur. Phys. J. C} \textbf{ 24}
  (2002) 517,
  \href{http://dx.doi.org/10.1007/s10052-002-0988-9}{\doi{10.1007/s10052-002-0988-9}},
\href{http://www.arXiv.org/abs/hep-ex/0203011}{\texttt{arXiv:hep-ex/0203011}}.

\bibitem{zeus}
\hrefCMSnoop {}{{ZEUS} Collaboration, ``{Rapidity gaps between jets in
  photoproduction at HERA}'',} \textit{ Phys. Lett. B} \textbf{ 369} (1996) 55,
  \href{http://dx.doi.org/10.1016/0370-2693(95)01588-4}{\doi{10.1016/0370-2693(95)01588-4}},
\href{http://www.arXiv.org/abs/hep-ex/9510012}{\texttt{arXiv:hep-ex/9510012}}.

\bibitem{d01}
\hrefCMSnoop {}{{D0} Collaboration, ``{Rapidity gaps between jets in
  p$\bar{\mathrm{p}}$ collisions at $\sqrt{s} = 1.8$ TeV}'',} \textit{ Phys.
  Rev. Lett.} \textbf{ 72} (1994) 2332,
\href{http://dx.doi.org/10.1103/PhysRevLett.72.2332}{\doi{10.1103/PhysRevLett.72.2332}}.

\bibitem{d02}
\hrefCMSnoop {}{{D0} Collaboration, ``{Jet production via strongly-interacting
  color-singlet exchange in p$\bar{\mathrm{p}}$ collisions}'',} \textit{ Phys.
  Rev. Lett.} \textbf{ 76} (1996) 734,
  \href{http://dx.doi.org/10.1103/PhysRevLett.76.734}{\doi{10.1103/PhysRevLett.76.734}},
\href{http://www.arXiv.org/abs/hep-ex/9509013}{\texttt{arXiv:hep-ex/9509013}}.

\bibitem{d03}
\hrefCMSnoop {}{{D0} Collaboration, ``{Probing hard color-singlet exchange in
  p$\bar{\mathrm{p}}$ collisions at $\sqrt{s} = 630$ GeV and 1800 GeV}'',}
  \textit{ Phys. Lett. B} \textbf{ 440} (1998) 189,
  \href{http://dx.doi.org/10.1016/S0370-2693(98)01238-6}{\doi{10.1016/S0370-2693(98)01238-6}},
\href{http://www.arXiv.org/abs/hep-ex/9809016}{\texttt{arXiv:hep-ex/9809016}}.

\bibitem{cdf1}
\hrefCMSnoop {}{{CDF} Collaboration, ``{Observation of rapidity gaps in
  $\bar{\mathrm{p}}$p collisions at 1.8 TeV}'',} \textit{ Phys. Rev. Lett.}
  \textbf{ 74} (1995) 855,
\href{http://dx.doi.org/10.1103/PhysRevLett.74.855}{\doi{10.1103/PhysRevLett.74.855}}.

\bibitem{cdf2}
\hrefCMSnoop {}{{CDF} Collaboration, ``{Dijet production by color-singlet
  exchange at the {Fermilab} {Tevatron}}'',} \textit{ Phys. Rev. Lett.}
  \textbf{ 80} (1998) 1156,
\href{http://dx.doi.org/10.1103/PhysRevLett.80.1156}{\doi{10.1103/PhysRevLett.80.1156}}.

\bibitem{cdf3}
\hrefCMSnoop {}{{CDF} Collaboration, ``{Events with a rapidity gap between jets
  in $\bar{\mathrm{p}}$p collisions at $\sqrt{s} = 630$ GeV}'',} \textit{ Phys.
  Rev. Lett.} \textbf{ 81} (1998) 5278,
\href{http://dx.doi.org/10.1103/PhysRevLett.81.5278}{\doi{10.1103/PhysRevLett.81.5278}}.

\bibitem{jgjCMS}
\hrefCMSnoop {}{{CMS Collaboration}, ``{Study of dijet events with a large
  rapidity gap between the two leading jets in pp collisions at $\sqrt{s}=7$
  $\,\text {TeV}$}'',} \textit{ Eur. Phys. J. C} \textbf{ 78} (2018) 242,
  \href{http://dx.doi.org/10.1140/epjc/s10052-018-5691-6}{\doi{10.1140/epjc/s10052-018-5691-6}},
\href{http://www.arXiv.org/abs/1710.02586}{\texttt{arXiv:1710.02586}}.

\bibitem{Bjorken}
\hrefCMSnoop {}{J.~D. Bjorken, ``Rapidity gaps and jets as a new-physics
  signature in very-high-energy hadron-hadron collisions'',} \textit{ Phys.
  Rev. D} \textbf{ 47} (1993) 101,
  \href{http://dx.doi.org/10.1103/PhysRevD.47.101}{\doi{10.1103/PhysRevD.47.101}}.

\bibitem{barone}
V.~Barone and E.~Predazzi, ``High-energy particle diffraction''.
\newblock Springer, Berlin, 2002.
\newblock
  \href{http://dx.doi.org/10.1007/978-3-662-04724-8}{\doi{10.1007/978-3-662-04724-8}}.

\bibitem{donnachie}
S.~Donnachie, G.~Dosch, P.~Landshoff, and O.~Nachtmann, ``Pomeron physics and
  QCD''.
\newblock Cambridge Monographs on Particle Physics, Nuclear Physics and
  Cosmology. Cambridge University Press, 2002.
\newblock
  \href{http://dx.doi.org/10.1017/CBO9780511534935}{\doi{10.1017/CBO9780511534935}}.

\bibitem{forshaw}
J.~R. Forshaw and D.~A. Ross, ``Quantum chromodynamics and the pomeron''.
\newblock Cambridge Lecture Notes in Physics. Cambridge University Press, 1997.
\newblock
  \href{http://dx.doi.org/10.1017/CBO9780511524387}{\doi{10.1017/CBO9780511524387}}.

\bibitem{Cox:1999dw}
\hrefCMSnoop {}{B.~Cox, J.~R. Forshaw, and L.~L{\"o}nnblad, ``{Hard
  color-singlet exchange at the Tevatron}'',} \textit{ JHEP} \textbf{ 10}
  (1999) 023,
  \href{http://dx.doi.org/10.1088/1126-6708/1999/10/023}{\doi{10.1088/1126-6708/1999/10/023}},
\href{http://www.arXiv.org/abs/hep-ph/9908464}{\texttt{arXiv:hep-ph/9908464}}.

\bibitem{csp}
\hrefCMSnoop {}{R.~Enberg, G.~Ingelman, and L.~Motyka, ``{Hard color-singlet
  exchange and gaps between jets at the {T}evatron}'',} \textit{ Phys. Lett. B}
  \textbf{ 524} (2002) 273,
  \href{http://dx.doi.org/10.1016/S0370-2693(01)01379-X}{\doi{10.1016/S0370-2693(01)01379-X}},
\href{http://www.arXiv.org/abs/hep-ph/0111090}{\texttt{arXiv:hep-ph/0111090}}.

\bibitem{Chevallier:2009cu}
\hrefCMSnoop {}{F.~Chevallier, O.~Kepka, C.~Marquet, and C.~Royon, ``{Gaps
  between jets at hadron colliders in the next-to-leading BFKL framework}'',}
  \textit{ Phys. Rev. D} \textbf{ 79} (2009) 094019,
  \href{http://dx.doi.org/10.1103/PhysRevD.79.094019}{\doi{10.1103/PhysRevD.79.094019}},
\href{http://www.arXiv.org/abs/0903.4598}{\texttt{arXiv:0903.4598}}.

\bibitem{Kepka:2010hu}
\hrefCMSnoop {}{O.~Kepka, C.~Marquet, and C.~Royon, ``{Gaps between jets in
  hadronic collisions}'',} \textit{ Phys. Rev. D} \textbf{ 83} (2011) 034036,
  \href{http://dx.doi.org/10.1103/PhysRevD.83.034036}{\doi{10.1103/PhysRevD.83.034036}},
\href{http://www.arXiv.org/abs/1012.3849}{\texttt{arXiv:1012.3849}}.

\bibitem{cspLHC}
\hrefCMSnoop {}{A.~Ekstedt, R.~Enberg, and G.~Ingelman, ``{Hard color-singlet
  BFKL exchange and gaps between jets at the {LHC}}'',} 2017.
\href{http://www.arXiv.org/abs/1703.10919}{\texttt{arXiv:1703.10919}}.

\bibitem{bjorken_survival}
\hrefCMSnoop {}{J.~D. Bjorken, ``{A full-acceptance detector for {SSC} physics
  at low and intermediate mass scales: an expression of interest to the
  {SSC}}'',} \textit{ \it Int. J. Mod. Phys. A} \textbf{ 07} (1992) 4189,
  \href{http://dx.doi.org/10.1142/S0217751X92001885}{\doi{10.1142/S0217751X92001885}}.

\bibitem{survival_levin}
\hrefCMSnoop {}{E.~Gotsman, E.~Levin, and U.~Maor, ``{Energy dependence of the
  survival probability of large rapidity gaps}'',} \textit{ Phys. Lett. B}
  \textbf{ 438} (1998) 229,
  \href{http://dx.doi.org/10.1016/S0370-2693(98)00972-1}{\doi{10.1016/S0370-2693(98)00972-1}},
\href{http://www.arXiv.org/abs/hep-ph/9804404}{\texttt{arXiv:hep-ph/9804404}}.

\bibitem{khoze_survival_2013}
\hrefCMSnoop {}{V.~A. Khoze, A.~D. Martin, and M.~G. Ryskin, ``{{D}iffraction
  at the {LHC}}'',} \textit{ Eur. Phys. J. C} \textbf{ 73} (2013) 2503,
  \href{http://dx.doi.org/10.1140/epjc/s10052-013-2503-x}{\doi{10.1140/epjc/s10052-013-2503-x}},
  \href{http://www.arXiv.org/abs/1306.2149}{\texttt{arXiv:1306.2149}}.

\bibitem{gotsman_survival_2015}
\hrefCMSnoop {}{E.~Gotsman, E.~Levin, and U.~Maor, ``{{CGC}/saturation approach
  for soft interactions at high energy: survival probability of central
  exclusive production}'',} \textit{ Eur. Phys. J. C} \textbf{ 76} (2016) 177,
  \href{http://dx.doi.org/10.1140/epjc/s10052-016-4014-z}{\doi{10.1140/epjc/s10052-016-4014-z}},
  \href{http://www.arXiv.org/abs/1510.07249}{\texttt{arXiv:1510.07249}}.

\bibitem{Khoze:2017sdd}
\hrefCMSnoop {}{V.~A. Khoze, A.~D. Martin, and M.~G. Ryskin, ``{Multiple
  interactions and rapidity gap survival}'',} \textit{ J. Phys. G} \textbf{ 45}
  (2018) 053002,
  \href{http://dx.doi.org/10.1088/1361-6471/aab1bf}{\doi{10.1088/1361-6471/aab1bf}},
\href{http://www.arXiv.org/abs/1710.11505}{\texttt{arXiv:1710.11505}}.

\bibitem{Babiarz:2017jxc}
\hrefCMSnoop {}{I.~Babiarz, R.~Staszewski, and A.~Szczurek, ``{Multi-parton
  interactions and rapidity gap survival probability in jet-gap-jet
  processes}'',} \textit{ Phys. Lett. B} \textbf{ 771} (2017) 532,
  \href{http://dx.doi.org/10.1016/j.physletb.2017.05.095}{\doi{10.1016/j.physletb.2017.05.095}},
\href{http://www.arXiv.org/abs/1704.00546}{\texttt{arXiv:1704.00546}}.

\bibitem{PhysRevD.87.034010}
\hrefCMSnoop {}{C.~Marquet, C.~Royon, M.~Trzebi\ifmmode~\acute{n}\else
  \'{n}\fi{}ski, and R.~\ifmmode \check{Z}\else
  \v{Z}\fi{}leb\ifmmode~\check{c}\else \v{c}\fi{}\'{\i}k, ``{Gaps between jets
  in double-pomeron-exchange processes at the {LHC}}'',} \textit{ Phys. Rev. D}
  \textbf{ 87} (2013) 034010,
  \href{http://dx.doi.org/10.1103/PhysRevD.87.034010}{\doi{10.1103/PhysRevD.87.034010}},
  \href{http://www.arXiv.org/abs/1212.2059}{\texttt{arXiv:1212.2059}}.

\bibitem{totem1}
\hrefCMSnoop {}{{TOTEM} Collaboration, ``The {TOTEM} experiment at the {CERN}
  {L}arge {H}adron {C}ollider'',} \textit{ JINST} \textbf{ 3} (2008) S08007,
  \href{http://dx.doi.org/10.1088/1748-0221/3/08/s08007}{\doi{10.1088/1748-0221/3/08/s08007}}.

\bibitem{LHC_TDR}
O.~S. Br{\"o}ning {et~al.}, ``{LHC design report}''.
\newblock CERN-2004-003-v-1. 2004.
\newblock
  \href{http://dx.doi.org/10.1088/CERN-2004-003-V-1}{\doi{10.1088/CERN-2004-003-V-1}}.

\bibitem{TRK-11-001}
\hrefCMSnoop {}{{CMS Collaboration}, ``{Description and performance of track
  and primary-vertex reconstruction with the CMS tracker}'',} \textit{ JINST}
  \textbf{ 9} (2014) P10009,
  \href{http://dx.doi.org/10.1088/1748-0221/9/10/P10009}{\doi{10.1088/1748-0221/9/10/P10009}},
\href{http://www.arXiv.org/abs/1405.6569}{\texttt{arXiv:1405.6569}}.

\bibitem{PFnew}
\hrefCMSnoop {}{{CMS Collaboration}, ``{Particle-flow reconstruction and global
  event description with the CMS detector}'',} \textit{ JINST} \textbf{ 12}
  (2017) P10003,
  \href{http://dx.doi.org/10.1088/1748-0221/12/10/P10003}{\doi{10.1088/1748-0221/12/10/P10003}},
\href{http://www.arXiv.org/abs/1706.04965}{\texttt{arXiv:1706.04965}}.

\bibitem{Cacciari:2008gp}
\hrefCMSnoop {}{M.~Cacciari, G.~P. Salam, and G.~Soyez, ``{The anti-\kt jet
  clustering algorithm}'',} \textit{ JHEP} \textbf{ 04} (2008) 063,
  \href{http://dx.doi.org/10.1088/1126-6708/2008/04/063}{\doi{10.1088/1126-6708/2008/04/063}},
  \href{http://www.arXiv.org/abs/0802.1189}{\texttt{arXiv:0802.1189}}.

\bibitem{Cacciari:2011ma}
\hrefCMSnoop {}{M.~Cacciari, G.~P. Salam, and G.~Soyez, ``{\FASTJET user
  manual}'',} \textit{ Eur. Phys. J. C} \textbf{ 72} (2012) 1896,
  \href{http://dx.doi.org/10.1140/epjc/s10052-012-1896-2}{\doi{10.1140/epjc/s10052-012-1896-2}},
\href{http://www.arXiv.org/abs/1111.6097}{\texttt{arXiv:1111.6097}}.

\bibitem{Khachatryan:2016kdb}
\hrefCMSnoop {}{{CMS Collaboration}, ``{Jet energy scale and resolution in the
  CMS experiment in pp collisions at 8 TeV}'',} \textit{ JINST} \textbf{ 12}
  (2017) P02014,
  \href{http://dx.doi.org/10.1088/1748-0221/12/02/P02014}{\doi{10.1088/1748-0221/12/02/P02014}},
\href{http://www.arXiv.org/abs/1607.03663}{\texttt{arXiv:1607.03663}}.

\bibitem{Chatrchyan:2008zzk}
\hrefCMSnoop {}{{CMS Collaboration}, ``The {CMS} experiment at the {CERN}
  {LHC}'',} \textit{ JINST} \textbf{ 3} (2008) S08004,
  \href{http://dx.doi.org/10.1088/1748-0221/3/08/S08004}{\doi{10.1088/1748-0221/3/08/S08004}}.

\bibitem{optics}
\hrefCMSnoop {}{{TOTEM} Collaboration, ``{{LHC} optics measurement with proton
  tracks detected by the {R}oman pots of the {TOTEM} experiment}'',} \textit{
  New J. Phys.} \textbf{ 16} (2014) 103041,
  \href{http://dx.doi.org/10.1088/1367-2630/16/10/103041}{\doi{10.1088/1367-2630/16/10/103041}},
\href{http://www.arXiv.org/abs/1406.0546}{\texttt{arXiv:1406.0546}}.

\bibitem{totem2}
\hrefCMSnoop {}{{TOTEM} Collaboration, ``{Performance of the {TOTEM} detectors
  at the {LHC}}'',} \textit{ Int. J. Mod. Phys. A} \textbf{ 28} (2013) 1330046,
  \href{http://dx.doi.org/10.1142/S0217751X13300469}{\doi{10.1142/S0217751X13300469}},
\href{http://www.arXiv.org/abs/1310.2908}{\texttt{arXiv:1310.2908}}.

\bibitem{Niewiadomski:2008zz}
\href {http://cds.cern.ch/record/1131825}{H.~Niewiadomski, ``{Reconstruction of
  protons in the {TOTEM} {R}oman pot detectors at the {LHC}}''}.
\newblock PhD thesis, Manchester U., 2008.
\newblock {CERN-THESIS-2008-080}.

\bibitem{Sirunyan:2020ifc}
\hrefCMSnoop {}{{CMS and TOTEM Collaborations}, ``{{M}easurement of
  {single-diffractive} dijet production in proton-proton collisions at
  $\sqrt{s} =$ 8 {TeV} with the {CMS} and {TOTEM} experiments}'',} \textit{
  Eur. Phys. J. C} \textbf{ 80} (2020) 1164,
  \href{http://dx.doi.org/10.1140/epjc/s10052-020-08562-y}{\doi{10.1140/epjc/s10052-020-08562-y}},
  \href{http://www.arXiv.org/abs/2002.12146}{\texttt{arXiv:2002.12146}}.

\bibitem{Khachatryan:2016bia}
\hrefCMSnoop {}{{CMS Collaboration}, ``{The CMS trigger system}'',} \textit{
  JINST} \textbf{ 12} (2017) P01020,
  \href{http://dx.doi.org/10.1088/1748-0221/12/01/P01020}{\doi{10.1088/1748-0221/12/01/P01020}},
\href{http://www.arXiv.org/abs/1609.02366}{\texttt{arXiv:1609.02366}}.

\bibitem{Chatrchyan:2012vc}
\hrefCMSnoop {}{{CMS Collaboration}, ``{Observation of a diffractive
  contribution to dijet production in proton-proton collisions at $\sqrt{s}=7$
  TeV}'',} \textit{ Phys. Rev. D} \textbf{ 87} (2013) 012006,
  \href{http://dx.doi.org/10.1103/PhysRevD.87.012006}{\doi{10.1103/PhysRevD.87.012006}},
  \href{http://www.arXiv.org/abs/1209.1805}{\texttt{arXiv:1209.1805}}.

\bibitem{hentschinski1}
\hrefCMSnoop {}{M.~Hentschinski, J.~D. Madrigal Mart\ifmmode~\acute{i}\else
  \'{i}\fi{}nez, B.~Murdaca, and A.~Sabio~Vera, ``{The next-to-leading order
  vertex for a forward jet plus a rapidity gap at high energies}'',} \textit{
  Phys. Lett. B} \textbf{ 735} (2014) 168,
  \href{http://dx.doi.org/10.1016/j.physletb.2014.06.022}{\doi{10.1016/j.physletb.2014.06.022}},
  \href{http://www.arXiv.org/abs/1404.2937}{\texttt{arXiv:1404.2937}}.

\bibitem{hentschinski2}
\hrefCMSnoop {}{M.~Hentschinski, J.~D. Madrigal Mart\ifmmode~\acute{i}\else
  \'{i}\fi{}nez, B.~Murdaca, and A.~Sabio~Vera, ``{The gluon-induced
  {M}ueller-{T}ang jet impact factor at next-to-leading order}'',} \textit{
  {Nucl. Phys. B}} \textbf{ 889} (2014) 549,
  \href{http://dx.doi.org/10.1016/j.nuclphysb.2014.10.026}{\doi{10.1016/j.nuclphysb.2014.10.026}},
  \href{http://www.arXiv.org/abs/1409.6704}{\texttt{arXiv:1409.6704}}.

\bibitem{Giovannini1986}
\hrefCMSnoop {}{A.~Giovannini and L.~Van~Hove, ``Negative binomial multiplicity
  distributions in high energy hadron collisions'',} \textit{ Z. Phys. C}
  \textbf{ 30} (1986) 391,
  \href{http://dx.doi.org/10.1007/BF01557602}{\doi{10.1007/BF01557602}}.

\bibitem{Ghosh:2012xh}
\hrefCMSnoop {}{P.~Ghosh, ``{Negative binomial multiplicity distribution in
  proton-proton collisions in limited pseudorapidity intervals at {LHC} up to
  $\sqrt{s} = 7$ {TeV} and the clan model}'',} \textit{ Phys. Rev. D} \textbf{
  85} (2012) 054017,
  \href{http://dx.doi.org/10.1103/PhysRevD.85.054017}{\doi{10.1103/PhysRevD.85.054017}},
\href{http://www.arXiv.org/abs/1202.4221}{\texttt{arXiv:1202.4221}}.

\bibitem{ua5}
\hrefCMSnoop {}{{UA5} Collaboration, ``Multiplicity distributions in different
  pseudorapidity intervals at a center-of-mass energy of {540 GeV}'',} \textit{
  Phys. Lett. B} \textbf{ 160} (1985) 193,
\href{http://dx.doi.org/10.1016/0370-2693(85)91491-1}{\doi{10.1016/0370-2693(85)91491-1}}.

\bibitem{ua5failstart}
\hrefCMSnoop {}{{UA5} Collaboration, ``Charged particle multiplicity
  distributions at {200 GeV} and {900 GeV} center-of-mass energy'',} \textit{
  Z. Phys. C} \textbf{ 43} (1989) 357,
\href{http://dx.doi.org/10.1007/BF01506531}{\doi{10.1007/BF01506531}}.

\bibitem{alice}
\hrefCMSnoop {}{{ALICE Collaboration}, ``{Charged-particle multiplicities in
  proton-proton collisions at $\sqrt{s} = 0.9$ to 8 {TeV}}'',} \textit{ Eur.
  Phys. J. C} \textbf{ 77} (2017) 33,
  \href{http://dx.doi.org/10.1140/epjc/s10052-016-4571-1}{\doi{10.1140/epjc/s10052-016-4571-1}},
\href{http://www.arXiv.org/abs/1509.07541}{\texttt{arXiv:1509.07541}}.

\bibitem{pythia8}
T.~Sj{\"o}strand\hrefCMSnoop {}{ {et~al.}, ``{An introduction to PYTHIA
  8.2}'',} \textit{ Comput. Phys. Commun.} \textbf{ 191} (2015) 159,
  \href{http://dx.doi.org/10.1016/j.cpc.2015.01.024}{\doi{10.1016/j.cpc.2015.01.024}},
\href{http://www.arXiv.org/abs/1410.3012}{\texttt{arXiv:1410.3012}}.

\bibitem{NNPDF1}
\hrefCMSnoop {}{{NNPDF} Collaboration, ``{Unbiased global determination of
  parton distributions and their uncertainties at NNLO and at LO}'',} \textit{
  Nucl. Phys. B} \textbf{ 855} (2012) 153,
  \href{http://dx.doi.org/10.1016/j.nuclphysb.2011.09.024}{\doi{10.1016/j.nuclphysb.2011.09.024}},
  \href{http://www.arXiv.org/abs/1107.2652}{\texttt{arXiv:1107.2652}}.

\bibitem{NNPDF2}
\hrefCMSnoop {}{{NNPDF} Collaboration, ``{Parton distributions with QED
  corrections}'',} \textit{ Nucl. Phys. B} \textbf{ 877} (2013) 290,
  \href{http://dx.doi.org/10.1016/j.nuclphysb.2013.10.010}{\doi{10.1016/j.nuclphysb.2013.10.010}},
  \href{http://www.arXiv.org/abs/1308.0598}{\texttt{arXiv:1308.0598}}.

\bibitem{LundString}
\hrefCMSnoop {}{B.~Andersson, G.~Gustafson, G.~Ingelman, and T.~Sj{\"o}strand,
  ``{Parton fragmentation and string dynamics}'',} \textit{ Phys. Rept.}
  \textbf{ 97} (1983) 31,
  \href{http://dx.doi.org/10.1016/0370-1573(83)90080-7}{\doi{10.1016/0370-1573(83)90080-7}}.

\bibitem{CUETP8M1}
\hrefCMSnoop {}{{CMS Collaboration}, ``{Event generator tunes obtained from
  underlying event and multiparton scattering measurements}'',} \textit{ Eur.
  Phys. J. C} \textbf{ 76} (2016) 155,
  \href{http://dx.doi.org/10.1140/epjc/s10052-016-3988-x}{\doi{10.1140/epjc/s10052-016-3988-x}},
  \href{http://www.arXiv.org/abs/1512.00815}{\texttt{arXiv:1512.00815}}.

\bibitem{herwig}
G.~Corcella\hrefCMSnoop {}{ {et~al.}, ``{HERWIG 6: an event generator for
  hadron emission reactions with interfering gluons (including supersymmetric
  processes)}'',} \textit{ JHEP} \textbf{ 01} (2001) 010,
  \href{http://dx.doi.org/10.1088/1126-6708/2001/01/010}{\doi{10.1088/1126-6708/2001/01/010}},
\href{http://www.arXiv.org/abs/hep-ph/0011363}{\texttt{arXiv:hep-ph/0011363}}.

\bibitem{cteq6l1}
J.~Pumplin\hrefCMSnoop {}{ {et~al.}, ``{New generation of parton distributions
  with uncertainties from global QCD analysis}'',} \textit{ JHEP} \textbf{ 07}
  (2002) 012,
  \href{http://dx.doi.org/10.1088/1126-6708/2002/07/012}{\doi{10.1088/1126-6708/2002/07/012}},
\href{http://www.arXiv.org/abs/hep-ph/0201195}{\texttt{arXiv:hep-ph/0201195}}.

\bibitem{clusterFragmentation}
\hrefCMSnoop {}{B.~R. Webber, ``{A QCD model for jet fragmentation including
  soft gluon interference}'',} \textit{ Nucl. Phys. B} \textbf{ 238} (1984)
  492,
\href{http://dx.doi.org/10.1016/0550-3213(84)90333-X}{\doi{10.1016/0550-3213(84)90333-X}}.

\bibitem{jimmy}
\hrefCMSnoop {}{J.~M. Butterworth, J.~R. Forshaw, and M.~H. Seymour,
  ``Multiparton interactions in photoproduction at {HERA}'',} \textit{ Z. Phys.
  C} \textbf{ 72} (1996) 637,
  \href{http://dx.doi.org/10.1007/s002880050286}{\doi{10.1007/s002880050286}},
  \href{http://www.arXiv.org/abs/hep-ph/9601371}{\texttt{arXiv:hep-ph/9601371}}.

\bibitem{geant}
\hrefCMSnoop {}{{GEANT4} Collaboration, ``{{\GEANTfour}---A simulation
  toolkit}'',} \textit{ Nucl. Instrum. Meth. A} \textbf{ 506} (2003) 250,
  \href{http://dx.doi.org/10.1016/S0168-9002(03)01368-8}{\doi{10.1016/S0168-9002(03)01368-8}}.

\bibitem{Abbott:1999km}
\hrefCMSnoop {}{{D0} Collaboration, ``{Hard single diffraction in \PAp{}\Pp
  collisions at $\sqrt{s} = 630$ GeV and 1800 {GeV}}'',} \textit{ Phys. Lett.
  B} \textbf{ 531} (2002) 52,
  \href{http://dx.doi.org/10.1016/S0370-2693(02)01364-3}{\doi{10.1016/S0370-2693(02)01364-3}},
\href{http://www.arXiv.org/abs/hep-ex/9912061}{\texttt{arXiv:hep-ex/9912061}}.

\bibitem{Affolder:2000vb}
\hrefCMSnoop {}{{CDF} Collaboration, ``{Diffractive dijets with a leading
  antiproton in \PAp{}\Pp collisions at $\sqrt{s} = 1800$ GeV}'',} \textit{
  Phys. Rev. Lett.} \textbf{ 84} (2000) 5043,
\href{http://dx.doi.org/10.1103/PhysRevLett.84.5043}{\doi{10.1103/PhysRevLett.84.5043}}.

\bibitem{Affolder:2001zn}
\hrefCMSnoop {}{{CDF} Collaboration, ``{Diffractive dijet production at
  $\sqrt{s} = 630$ GeV and 1800 GeV at the {F}ermilab {T}evatron}'',} \textit{
  Phys. Rev. Lett.} \textbf{ 88} (2002) 151802,
  \href{http://dx.doi.org/10.1103/PhysRevLett.88.151802}{\doi{10.1103/PhysRevLett.88.151802}},
\href{http://www.arXiv.org/abs/hep-ex/0109025}{\texttt{arXiv:hep-ex/0109025}}.

\bibitem{PhysRevD.77.052004}
\hrefCMSnoop {}{{CDF} Collaboration, ``{Observation of exclusive dijet
  production at the Fermilab Tevatron \PAp{}\Pp Collider}'',} \textit{ Phys.
  Rev. D} \textbf{ 77} (2008) 052004,
  \href{http://dx.doi.org/10.1103/PhysRevD.77.052004}{\doi{10.1103/PhysRevD.77.052004}},
  \href{http://www.arXiv.org/abs/0712.0604}{\texttt{arXiv:0712.0604}}.

\bibitem{Aaltonen:2012tha}
\hrefCMSnoop {}{{CDF} Collaboration, ``{Diffractive dijet production in
  \PAp{}\Pp collisions at $\sqrt{s}=1.96$ TeV}'',} \textit{ Phys. Rev. D}
  \textbf{ 86} (2012) 032009,
  \href{http://dx.doi.org/10.1103/PhysRevD.86.032009}{\doi{10.1103/PhysRevD.86.032009}},
\href{http://www.arXiv.org/abs/1206.3955}{\texttt{arXiv:1206.3955}}.

\bibitem{Andreev:2015cwa}
\hrefCMSnoop {}{{H1} Collaboration, ``{Diffractive dijet production with a
  leading proton in $\rm{e}\Pp$ collisions at {HERA}}'',} \textit{ JHEP}
  \textbf{ 05} (2015) 056,
  \href{http://dx.doi.org/10.1007/JHEP05(2015)056}{\doi{10.1007/JHEP05(2015)056}},
\href{http://www.arXiv.org/abs/1502.01683}{\texttt{arXiv:1502.01683}}.

\bibitem{Trzebinski:2015dna}
\hrefCMSnoop {}{P.~\ifmmode~\acute{S}\else \'{S}\fi{}wierska and
  M.~Trzebi\ifmmode~\acute{n}\else \'{n}\fi{}ski, ``{BFKL amplitude
  parametrization for the jet-gap-jet events at the {LHC} energies}'',}
  \textit{ Acta Phys. Polon. B} \textbf{ 46} (2015) 2477,
  \href{http://dx.doi.org/10.5506/APhysPolB.46.2477}{\doi{10.5506/APhysPolB.46.2477}},
\href{http://www.arXiv.org/abs/1504.06271}{\texttt{arXiv:1504.06271}}.

\bibitem{Colferai:2010wu}
\hrefCMSnoop {}{D.~Colferai, F.~Schwennsen, L.~Szymanowski, and S.~Wallon,
  ``{{M}ueller--{N}avelet jets at {LHC} - complete {NLL} {BFKL}
  calculation}'',} \textit{ JHEP} \textbf{ 12} (2010) 026,
  \href{http://dx.doi.org/10.1007/JHEP12(2010)026}{\doi{10.1007/JHEP12(2010)026}},
  \href{http://www.arXiv.org/abs/1002.1365}{\texttt{arXiv:1002.1365}}.

\bibitem{PhysRevLett.85.4215}
\hrefCMSnoop {}{{CDF} Collaboration, ``Dijet production by double-pomeron
  exchange at the {Fermilab} {Tevatron}'',} \textit{ Phys. Rev. Lett.} \textbf{
  85} (2000) 4215,
  \href{http://dx.doi.org/10.1103/PhysRevLett.85.4215}{\doi{10.1103/PhysRevLett.85.4215}}.

\end{thebibliography}\endgroup
\cleardoublepage \appendix\section{The CMS Collaboration \label{app:collab}}\begin{sloppypar}\hyphenpenalty=5000\widowpenalty=500\clubpenalty=5000\vskip\cmsinstskip
\textbf{Yerevan Physics Institute, Yerevan, Armenia}\\*[0pt]
A.M.~Sirunyan$^{\textrm{\dag}}$, A.~Tumasyan
\vskip\cmsinstskip
\textbf{Institut f\"{u}r Hochenergiephysik, Wien, Austria}\\*[0pt]
W.~Adam, F.~Ambrogi, T.~Bergauer, M.~Dragicevic, J.~Er\"{o}, A.~Escalante~Del~Valle, R.~Fr\"{u}hwirth\cmsAuthorMark{1}, M.~Jeitler\cmsAuthorMark{1}, N.~Krammer, L.~Lechner, D.~Liko, T.~Madlener, I.~Mikulec, F.M.~Pitters, N.~Rad, J.~Schieck\cmsAuthorMark{1}, R.~Sch\"{o}fbeck, M.~Spanring, S.~Templ, W.~Waltenberger, C.-E.~Wulz\cmsAuthorMark{1}, M.~Zarucki
% ...
\vskip\cmsinstskip
\dag: Deceased\\
1:  Also at Vienna University of Technology, Vienna, Austria\\
% ...

\section{The TOTEM Collaboration\label{app:totem}}
\newcommand{\AddAuthor}[2]{#1$^{#2}$,\ }
\newcommand\AddAuthorLast[2]{#1$^{#2}$}
\noindent
\AddAuthor{G.~Antchev}{a}
\AddAuthor{P.~Aspell}{9}
\AddAuthor{I.~Atanassov}{a}
\AddAuthor{V.~Avati}{7,9}
\AddAuthor{J.~Baechler}{9}
\AddAuthor{C.~Baldenegro~Barrera}{11}
\AddAuthor{V.~Berardi}{4a,4b}
\AddAuthor{M.~Berretti}{2a}
\AddAuthor{V.~Borchsh}{8}
\AddAuthor{E.~Bossini}{6b,9}
\AddAuthor{U.~Bottigli}{6c}
\AddAuthor{M.~Bozzo}{5a,5b}
\AddAuthor{H.~Burkhardt}{9}
\AddAuthor{F.S.~Cafagna}{4a}
\AddAuthor{M.G.~Catanesi}{4a}
\AddAuthor{M.~Csan\'{a}d}{3a,b}
\AddAuthor{T.~Cs\"{o}rg\H{o}}{3a,3b}
\AddAuthor{M.~Deile}{9}
\AddAuthor{F.~De~Leonardis}{4a,4c}
\AddAuthor{M.~Doubek}{1c}
\AddAuthor{D.~Druzhkin}{8,9}
\AddAuthor{K.~Eggert}{10}
\AddAuthor{V.~Eremin}{d}
\AddAuthor{A.~Fiergolski}{9}
\AddAuthor{L.~Forthomme}{2a,2b}
\AddAuthor{F.~Garcia}{2a}
\AddAuthor{V.~Georgiev}{1a}
\AddAuthor{S.~Giani}{9}
\AddAuthor{L.~Grzanka}{7}
\AddAuthor{J.~Hammerbauer}{1a}
\AddAuthor{T.~Isidori}{11}
\AddAuthor{V.~Ivanchenko}{8}
\AddAuthor{M.~Janda}{1c}
\AddAuthor{A.~Karev}{9}
\AddAuthor{J.~Ka\v{s}par}{1b,9}
\AddAuthor{B.~Kaynak}{e}
\AddAuthor{J.~Kopal}{9}
\AddAuthor{V.~Kundr\'{a}t}{1b}
\AddAuthor{S.~Lami}{6a}
\AddAuthor{R.~Linhart}{1a}
\AddAuthor{C.~Lindsey}{11}
\AddAuthor{M.V.~Lokaj\'{i}\v{c}ek$^{\textrm{\dag,}}$}{1b}
\AddAuthor{L.~Losurdo}{6c}
\AddAuthor{F.~Lucas~Rodr\'{i}guez}{9}
\AddAuthor{M.~Macr\'{i}}{5a}
\AddAuthor{M.~Malawski}{7}
\AddAuthor{N.~Minafra}{11}
\AddAuthor{S.~Minutoli}{5a}
\AddAuthor{T.~Naaranoja}{2a,2b}
\AddAuthor{F.~Nemes}{3a,9}
\AddAuthor{H.~Niewiadomski}{10}
\AddAuthor{T.~Nov\'{a}k}{3b}
\AddAuthor{E.~Oliveri}{9}
\AddAuthor{F.~Oljemark}{2a,2b}
\AddAuthor{M.~Oriunno}{f}
\AddAuthor{K.~\"{O}sterberg}{2a,2b}
\AddAuthor{P.~Palazzi}{9}
\AddAuthor{V.~Passaro}{4a,4c}
\AddAuthor{Z.~Peroutka}{1a}
\AddAuthor{J.~Proch\'{a}zka}{1b}
\AddAuthor{M.~Quinto}{4a,4b}
\AddAuthor{E.~Radermacher}{9}
\AddAuthor{E.~Radicioni}{4a}
\AddAuthor{F.~Ravotti}{9}
\AddAuthor{C.~Royon}{11}
\AddAuthor{G.~Ruggiero}{9}
\AddAuthor{H.~Saarikko}{2a,2b}
\AddAuthor{V.D.~Samoylenko}{c}
\AddAuthor{A.~Scribano}{6a}
\AddAuthor{J.~\v{S}irok\'{y}}{1a}
\AddAuthor{J.~Smajek}{9}
\AddAuthor{W.~Snoeys}{9}
\AddAuthor{R.~Stefanovitch}{9}
\AddAuthor{J.~Sziklai}{3a}
\AddAuthor{C.~Taylor}{10}
\AddAuthor{E.~Tcherniaev}{8}
\AddAuthor{N.~Turini}{6c}
\AddAuthor{O.~Urban}{1a}
\AddAuthor{V.~Vacek}{1c}
\AddAuthor{O.~Vavroch}{1a}
\AddAuthor{J.~Welti}{2a,2b}
\AddAuthor{J.~Williams}{11}
\AddAuthor{J.~Zich}{1a}
\AddAuthor{K.~Zielinski}{7}

\vskip 4pt plus 4pt
\let\thefootnote\relax
\newcommand{\AddInstitute}[2]{${}^{#1}$#2\\}
\newcommand{\AddExternalInstitute}[2]{\footnote{${}^{#1}$ #2}}
\noindent
\dag Deceased\\
\AddInstitute{1a}{University of West Bohemia, Pilsen, Czech Republic}
\AddInstitute{1b}{Institute of Physics of the Academy of Sciences of the Czech Republic, Prague, Czech Republic}
\AddInstitute{1c}{Czech Technical University, Prague, Czech Republic}
\AddInstitute{2a}{Helsinki Institute of Physics, University of Helsinki, Helsinki, Finland}
\AddInstitute{2b}{Department of Physics, University of Helsinki, Helsinki, Finland}
\AddInstitute{3a}{Wigner Research Centre for Physics, RMKI, Budapest, Hungary}
\AddInstitute{3b}{Eszterhazy Karoly University KRC, Gy\"{o}ngy\"{o}s, Hungary}
\AddInstitute{4a}{INFN Sezione di Bari, Bari, Italy}
\AddInstitute{4b}{Dipartimento Interateneo di Fisica di Bari, University of Bari, Bari, Italy}
\AddInstitute{4c}{Dipartimento di Ingegneria Elettrica e dell'Informazione --- Politecnico di Bari, Bari, Italy}
\AddInstitute{5a}{INFN Sezione di Genova, Genova, Italy}
\AddInstitute{5b}{Universit\`{a} degli Studi di Genova, Genova, Italy}
\AddInstitute{6a}{INFN Sezione di Pisa, Pisa, Italy}
\AddInstitute{6b}{Universit\`{a} degli Studi di Pisa, Pisa, Italy}
\AddInstitute{6c}{Universit\`{a} degli Studi di Siena and Gruppo Collegato INFN di Siena, Siena, Italy}
\AddInstitute{7}{Akademia G\'{o}rniczo-Hutnicza (AGH) University of Science and Technology, Krakow, Poland}
\AddInstitute{8}{Tomsk State University, Tomsk, Russia}
\AddInstitute{9}{CERN, Geneva, Switzerland}
\AddInstitute{10}{Case Western Reserve University, Department of Physics, Cleveland, OH, USA}
\AddInstitute{11}{The University of Kansas, Lawrence, KS, USA}
\AddExternalInstitute{a}{INRNE-BAS, Institute for Nuclear Research and Nuclear Energy, Bulgarian Academy of Sciences, Sofia, Bulgaria}
\AddExternalInstitute{b}{Department of Atomic Physics, E\"{o}tv\"{o}s Lor\'{a}nd University, Budapest, Hungary}
\AddExternalInstitute{c}{NRC 'Kurchatov Institute'--IHEP, Protvino, Russia}
\AddExternalInstitute{d}{Ioffe Physical Technical Institute, Russian Academy of Sciences, St. Petersburg, Russian Federation}
\AddExternalInstitute{e}{Istanbul University, Istanbul, Turkey}
\AddExternalInstitute{f}{SLAC, Stanford University, California, USA}

\end{sloppypar}
\end{document}